\documentclass{aa}
\usepackage{graphicx}
\usepackage[varg]{txfonts}
\usepackage{hyperref}
\usepackage{booktabs}
\usepackage{xcolor}
\usepackage{soul}

\usepackage{xcolor}

\newcommand{\ppf}{\texttt{PreProFit}~}

\begin{document}
   \title{How Cluster Pressure Profiles Scale: Mass, Redshift, and Intrinsic Scatter}
   
\authorrunning{}
\titlerunning{}
   \author{S. Andreon
          \inst{1}
          \and 
          F. Castagna\inst{1}
          }

    \institute{INAF-Osservatorio Astronomico di Brera, via Brera 28, 20121 Milano, Italy \\ \email{andreon@inaf.it} 
   \date{xx
   }
 }
    \abstract{ 
The thermal pressure profile of the intracluster medium is a critical component for cluster cosmology and for understanding the astrophysics of galaxy clusters. We aim to determine the population-averaged pressure profile of Sunyaev-Zel'dovich (SZ) selected galaxy clusters, characterize its intrinsic scatter, and measure its scaling with cluster mass and redshift.  We select a high-quality sample of 60 clusters from the South Pole Telescope (SPT) SZ catalog with $0.08<z<0.6$, requiring high signal-to-noise and 5 element resolution on Planck-SPT Compton-$y$ maps. We extract Compton-$y$ maps and use a hierarchical Bayesian framework to model the  spherical three-dimensional pressure profiles using restricted cubic splines. Crucially, we employ an updated weak-lensing-calibrated Compton-$Y$--mass scaling relation, effectively removing hydrostatic mass bias from the inferred masses. Our model is internally self-consistent, accommodating cluster-to-cluster profile variations and allowing the foreground and background contributions to vary between lines of sight. Furthermore, we allow for outliers arising from both bona fide clusters with intrinsically distinct profiles and line-of-sight projection effects. The code is made public available with this paper. We derive the population-averaged pressure profile and confirm that the intrinsic scatter is minimal at intermediate radii ($0.4 \le r/r_{500} \le 0.7$). Due to the adoption of the unbiased mass scaling, our derived pressure profile is approximately 40\% lower than previous estimates at all radii. We find suggestive, though not conclusive, evidence that the currently adopted mass dependence may require revision specifically by canceling the additional $M^{0.12}$ mass scaling introduced by \citet{Arnaud2010}. We further constrain departures from self-similar evolution in the mean profile to be smaller than 0.1-0.2 dex and limit the evolution of the intrinsic scatter to less than 8 \% per $\Delta z=0.1$. Future work should assess the impact of this updated pressure-profile model on SZ observables, cluster mass calibration, survey completeness, thermal SZ power spectrum predictions, halo-model calculations, and cosmological parameter inference.
}


   \maketitle

\section{Introduction}

The thermal pressure profile of the intracluster medium (ICM) is a fundamental descriptor of the baryonic content of galaxy clusters. Because the Sunyaev-Zel'dovich (SZ) effect directly probes the line-of-sight integral of the electron pressure \citep{Sunyaev1972,Birkinshaw1999,Carlstrom2002}, the pressure distribution determines the morphology and amplitude of the SZ signal and therefore plays a central role in cluster cosmology and in studies of cluster astrophysics. The population-average pressure profile and its intrinsic scatter enter a wide range of quantities of astrophysical and cosmological interest, including the integrated Compton-$Y$ signal used as a mass proxy \citep{Motl2005,Nagai2006,Andersson2011,Planck2014_XX}, matched-filter cluster detection, selection functions of SZ surveys \citep{Melin2006,Bleem2015,Hilton2021}, hydrostatic mass estimates \citep{Ettori2013}, predictions for the thermal SZ power spectrum \citep{Komatsu2002,Shaw2010,Battaglia2012_tsz}, halo model calculations \citep{Cooray2002,Hill2014}, and cosmological analyses based on cluster counts \citep{Bocquet2019,Planck2020_clusters}. In addition, the radial dependence of the scatter provides insight into the relative importance of cooling, AGN feedback, mergers, accretion, and departures from hydrostatic equilibrium across the cluster population \citep{McDonald2014,Lau2015,Barnes2017}.

A robust characterization of the population-averaged pressure profile and its covariance structure is therefore required not only to understand the thermodynamic structure of clusters, but also to derive unbiased cosmological constraints from current and future surveys \citep{Pratt2019,Mantz2022}. This requirement has become increasingly important with the advent of high-quality SZ observations from the South Pole Telescope \citep{Carlstrom2011}, Planck \citep{Planck2011_overview}, and the Atacama Cosmology Telescope \citep{Hincks10}, together with forthcoming surveys such as the Simons Observatory \citep{Ade2019_SO} and CMB-S4 \citep{Abazajian2019_CMBS4}, whose statistical precision is now sufficiently high that systematic uncertainties in pressure modeling become limiting factors.

The pressure profile is commonly modeled through generalized Navarro-Frenk-White (gNFW) parameterizations or related functional forms \citep{Nagai2007,Arnaud2010}. Early analyses suggested the existence of a nearly universal pressure profile after appropriate scaling with mass and redshift \citep{Nagai2007,Arnaud2010,Planck2013_intermediate}. However, the degree of universality, the level and radial dependence of the intrinsic scatter, and the precise scaling with mass and cosmic epoch remain actively debated \citep{Sayers2013,McDonald2014,Ruppin2018,Ghirardini2019}. These questions are not merely descriptive. Even modest biases in the average pressure normalization propagate directly into biases in integrated Compton-$Y$ estimates and therefore into cluster mass calibration and cosmological inference \citep{Planck2014_XX,Andreon2014,Ruppin2018}. Likewise, an incorrect characterization of the intrinsic scatter modifies inferred selection functions and Eddington-type biases, affecting cluster abundance analyses \citep{Eddington1913,Stanek2006,AB12,Evrard2014,Farahi2018}.

Despite the importance of these issues, several past analyses suffer from statistical inconsistencies or simplifying assumptions that are difficult to justify with present-day data quality. A common approximation consists in fitting pressure profiles after scaling clusters by masses affected by hydrostatic bias, effectively propagating biases in mass calibration directly into the inferred pressure normalization \citep{vonderLinden2014,Hoekstra2015}. 
Many works rely on hydrostatic masses or on scaling relations calibrated with them \citep{Arnaud2010,Planck2013_intermediate}, but since hydrostatic masses are known to be biased low relative to weak-lensing masses \citep{Mahdavi2013,vonderLinden2014,Hoekstra2015,Andreon2025}, this choice propagates directly into an overestimate of the pressure normalization at fixed quoted mass. This issue is particularly relevant because pressure normalization, intrinsic scatter, and mass scaling are strongly coupled quantities in hierarchical analyses. %\citep{Mantz2016,Farahi2019}.

Another widespread simplification consists in adopting models that are unable to adequately describe the data, for example by assuming that all clusters share exactly the same pressure profile shape. This approach is adopted, for instance, in the recent analysis by \citet{Munoz-Echeverria25}, where all clusters are forced to share identical gNFW shape parameters. In practice, this implies that cool-core and non-cool-core systems are required to have the same inner pressure slope and overall profile shape, artificially suppressing genuine cluster-to-cluster variations. While such assumptions simplify the modeling, they also require a quantitative assessment of model adequacy. However, this aspect is rarely discussed, including in \citet{Munoz-Echeverria25}, where no evidence is provided that the adopted model yields an acceptable description of the profiles of individual clusters.
In fact, the inadequacy of the model is directly indicated by the posterior distribution of the additional variance term referred to by the authors as ``intrinsic scatter'', which is found to be significantly different from zero. Since the model explicitly assumes identical gNFW parameters for all clusters, this term cannot represent genuine cluster-to-cluster variations in profile shape. Instead, the non-zero variance term indicates that the adopted model fails to reproduce the observed diversity of the data. In this context, the additional variance term effectively compensates for model misspecification rather than describing astrophysical scatter. 

Another common simplification \citep[e.g.][]{Munoz-Echeverria25,Ghirardini2019,Sayers2023} is the assumption that the 
line-of-sight foreground and background contributions do not vary from one line of sight to another. This choice directly affects the inferred intrinsic scatter at large radii, because the intrinsic scatter is, by definition, the component of the observed variance not explained by measurement uncertainties. For example, \citet{Sayers2023} neglected line-of-sight  fore/background variations altogether, which is likely responsible for the large scatter they reported at large radii \citep{Castagna25}. In contrast, \citet{Ghirardini2019} inflated the fore/background uncertainties, rather than explicitly modeling the possibility that the fore/background along the cluster line of sight differs from the value measured in surrounding regions. Although the adopted error inflation may be reasonable in magnitude, a robust determination of the intrinsic scatter at large radii requires a generative description of fore/background fluctuations. This was implemented by \citet{Castagna25} through the pedestal parameter, inferred from the data, rather than through the introduction of ad hoc variance terms.

More generally, conclusions derived from models that do not provide an adequate description of the observations should be interpreted with caution.
Conversely, analyses performed independently at each radial bin often fail to enforce physically meaningful profile smoothness and may amplify noise-driven fluctuations. Flexible but statistically controlled approaches are therefore needed to capture the underlying population structure without overfitting the data \citep{Gelman2013,AndreonHurn2010,Castagna25}.

In this work, we revisit the determination of the population-averaged ICM pressure profile using a sample of high signal-to-noise SZ-selected clusters from the SPT survey. We model the three-dimensional pressure profiles within a hierarchical Bayesian framework using restricted cubic splines, allowing the data to determine the profile shape while preserving smoothness and statistical consistency \citep{Gelman2013}. Crucially, we adopt the updated weak-lensing calibrated Compton-$Y$--mass relation of \citet{Andreon2025}, effectively removing hydrostatic mass bias from the inferred pressure normalization. Our model is internally self-consistent, accommodating cluster-to-cluster profile variations and allowing line-of-sight foreground and background contributions to vary between lines of sight. Furthermore, we allow for outliers arising from both bona fide clusters with intrinsically distinct profiles and line-of-sight projection effects. In addition to deriving the mean profile, we characterize the intrinsic scatter as a function of radius and constrain the mass and redshift dependence of the population. The statistical code developed for this analysis is made publicly available on GitHub\footnote{\url{https://github.com/fcastagna/preprofit}}

Throughout our work, we adopted a flat $\Lambda$CDM cosmology with $H_0=70 \mathrm{~km~s^{-1}~Mpc^{-1}}$, $\Omega_M=0.3$, and $\Omega_\Lambda=0.7$ to convert between observed and physical quantities. Throughout the paper, we adopt as summary measures for the posterior distribution of a parameter its median and the 68\% uncertainty defined by the corresponding percentiles of the distribution, and we use instead $\sigma$'s when posteriors are Normal. All logarithms are in base 10.

\section{Methods} \label{sec:methods}

\subsection{Data}

\begin{figure}
\begin{center}
\begin{tabular}{c}
\includegraphics[width=.955\linewidth]{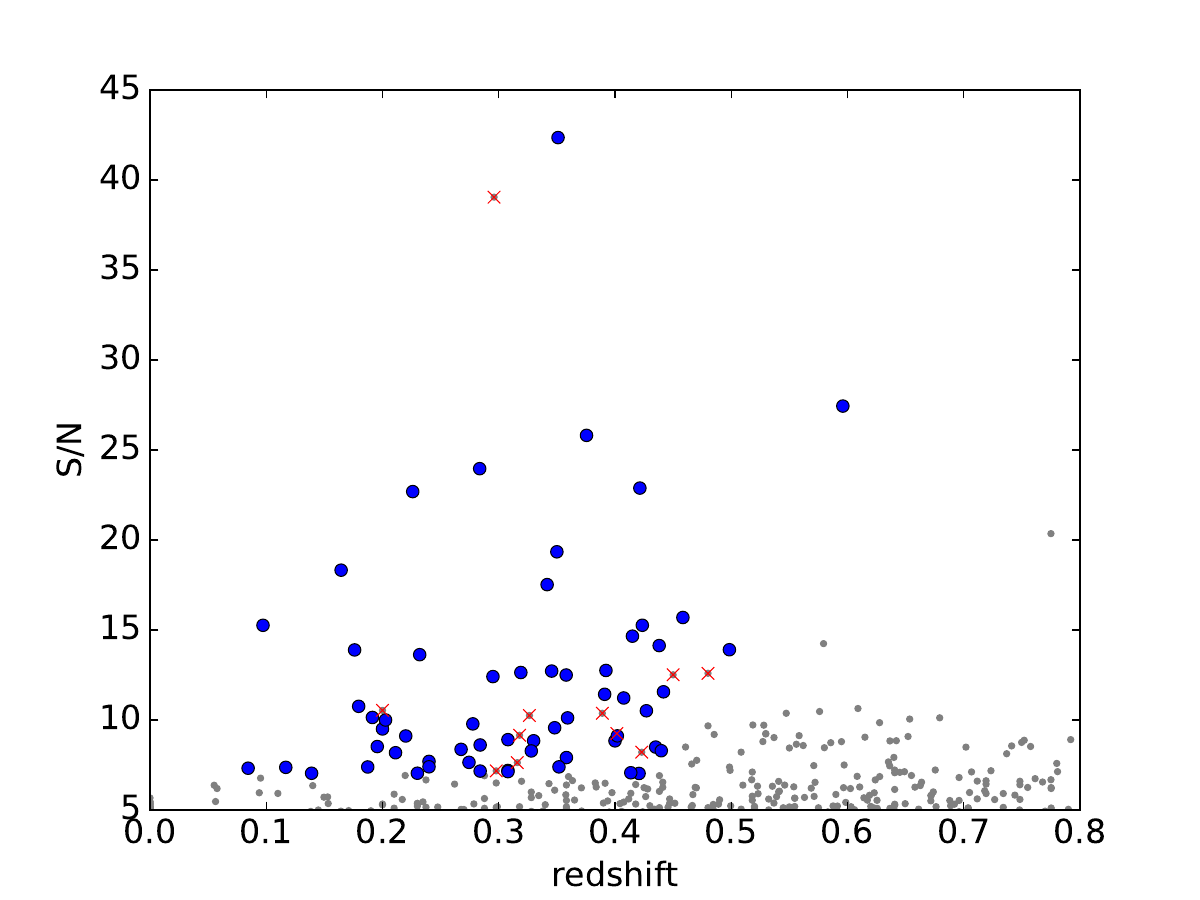}
\end{tabular}
\caption{Studied sample (blue points) and entire SPT catalog~\citep[gray points,][]{Bocquet2019}. Below $z\sim0.45$ the sample is limited by S/N, whereas at larger redshift our requirement on size dominates the selection. Crosses indicate galaxy clusters that are in low-quality regions, or even outside, of the footprint of the used Compton y map.} 
\label{fig:catalog}
\end{center}
\end{figure}

The sample selection was guided by our goal of focusing on clusters with well-resolved, high-quality data. From the South Pole Telescope (SPT) SZ galaxy cluster catalog~\citep{Bocquet2019}, we selected clusters with a signal-to-noise ratio (S/N) greater than 7 and an angular size at least five times the SPT Half Width at Half Maximum (HWHM), $\theta_{500} > 5 \times \rm{HWHM_{SPT}}$\footnote{The overdensity radius $r_{500}$ is defined as the radius within which the mean density is 500 times the critical density at the cluster's redshift.}. We further excluded clusters located near the boundary of the SPT map or in masked regions~\citep[][see Fig.~\ref{fig:catalog}]{Bleem2022}. One cluster, SPT-CLJ0405-4916, was removed because its background shows oscillations instead of flattening. The final studied sample consists of 60 clusters, as illustrated in Fig.~\ref{fig:catalog}. In practice, the size criterion corresponds to a redshift-dependent S/N limit due to its derivation from the cluster flux in \citet{Bleem2022} (see Fig.~\ref{fig:catalog}). Table~\ref{tab:individual} lists the full sample.

As in \citet{Castagna25}, 
the raw Compton-$y$ map of each cluster was extracted from the Sanson-Flamsteed projection minimum-variance Compton-$y$ maps based on both SPT and \textit{Planck} data~\citep{Bleem2022}. Radial profiles were computed in circular annuli with a width of 75 arcsec, which corresponds to the beam's full width at half maximum (FWHM), accounting for field boundaries and after flagging other sources, using the point source mask occasionaly supplemented by masking additional sources such as other galaxy cluster and regions of lower quality. The chosen width ensures that the data covariance between radial bins is negligible. We adopted as centers the coordinates derived by \citet{Bocquet2019}  and the point spread function (PSF) and transfer function provided with the Compton maps~\citep{Bleem2022}.
To estimate the errors of the profiles, we measured the scatter across profiles computed from random centers placed around each cluster. Radial profiles were extracted up to $r\sim17.5$ arcmin. Given the SPT resolution, the data are sampling the pressure profile with a $\approx 250$ kpc scale FWHM. Compton profiles are shown in Fig.~\ref{fig:Compton_prof1} to Fig.~\ref{fig:Compton_prof3}.

\subsection{Pressure profile model and fitting} \label{sec:press_prof}

To model each individual cluster, following the approach of \citet{Castagna25}, the three-dimensional spherical pressure profile was represented using a restricted cubic spline~\citep{durrleman1989} with $n_k$ knots. This approach defines a cubic spline between knots with linear extrapolation beyond the first and last knot, preventing unphysical oscillations in regions unconstrained by data. We adopted $n_k = 5$ knots, placed at fixed radii $\left[ 0.15, 0.4, 0.7, 1, 1.3 \right] \times r_{500}$. By scaling the knots with $r_{500}$, we assume that the scatter depends on cluster size, as in previous studies~\citep[e.g.,][]{Ghirardini2019}, rather than on absolute radii. The spacing between knots was chosen to exceed the radial bin width, minimizing covariances between adjacent bins. Modeling was performed in log-log space to avoid unphysical negative values for either radius or pressure. The innermost radius is larger than in \citet{Castagna25} to accommodate the higher-redshift clusters in our sample.

The three-dimensional pressure profile was numerically integrated along the line of sight using an Abel transform to produce a two-dimensional Compton-$y$ map, which was subsequently convolved with the beam and transfer function. Radial profiles were then extracted from the filtered Compton-$y$ maps, converted to surface brightness units, and compared to the observed profiles within a Bayesian framework to infer the pressure parameters and their uncertainties. To ensure a finite Compton-$y$ integral and distinguish a flat background from a flat cluster signal, we imposed an upper limit on the slope beyond the outermost knot, as in previous works~\citep{Romero2018, Andreon2021,Castagna25}, and required this slope to be negative, a less restrictive choice than in some earlier studies.

Since clusters have different masses and sizes, we modeled the scaled pressure profiles, $p(x)$, defined in \citet{Arnaud2010}, rather than the absolute pressure profile. A pedestal parameter was added to account for any nonzero background in the surface brightness profile, with a Gaussian prior centered at zero and $\sigma = 10^{-6}$ in Compton y. 

To model the full cluster population, we adopt a mean scaled pressure profile that depends on mass and redshift, with a Student's t distribution describing the scatter around it. Specifically, at each knot, the deviation of individual cluster profiles from the population mean is modeled with a Student's t distribution with 10 degrees of freedom that is allowed to be mass- and redshift-dependent and with a scatter that is redshift-dependent. The adoption of a Student's t distribution accounts for potential anomalies, including but not limited to binary clusters (which lack a well-defined center or the assumed spherical symmetry) or clusters that are close along the line of sight but not recognized as distinct objects. In detail,
the scatter of the individual  scaled pressures, $lgP_{k,i}$, around the population-averaged scaled pressure, $lgP_k$, is described as
\begin{align}
lgP_{k,i} &\sim t_{\nu=10} \Big( lgP_k + \gamma_z \mathcal{Z}_i + \gamma_M \mathcal{M}_i, \Big(\frac{1+z_i}{1+z_{\rm piv}}\Big)^{\gamma_\sigma} \sqrt{\frac{\nu-2}{\nu}}\sigma_{int,k}  \Big),\\
\mathcal{Z}_i &= \log \frac{1+z_i}{1+z_{\rm piv}},\\
\mathcal{M}_i &= \log \frac{M_{500,i}}{M_{500,{\rm med}}},
\end{align}

where $M_{500,{\rm med}} = 8 \times 10^{14} M_{\odot}$ and $z_{\rm piv} = 0.3$. The prefactor $\sqrt{(\nu-2)/\nu}$ ensures that the variance of the t-distribution is $\sigma_{{\rm int},k}^2$, whereas the $\Big(\frac{1+z_i}{1+z_{\rm piv}}\Big)^{\gamma_\sigma}$ factor models the redshift dependence of the intrinsic scatter. Here, $lgP$ is shorthand for $\log [p(x)]$. In short, we allow the mean model to change with both mass and redshift, and the scatter to be redshift-dependent and different from knot to knot.

\begin{figure}
\begin{center}
\begin{tabular}{c}
\includegraphics[width=.955\linewidth]{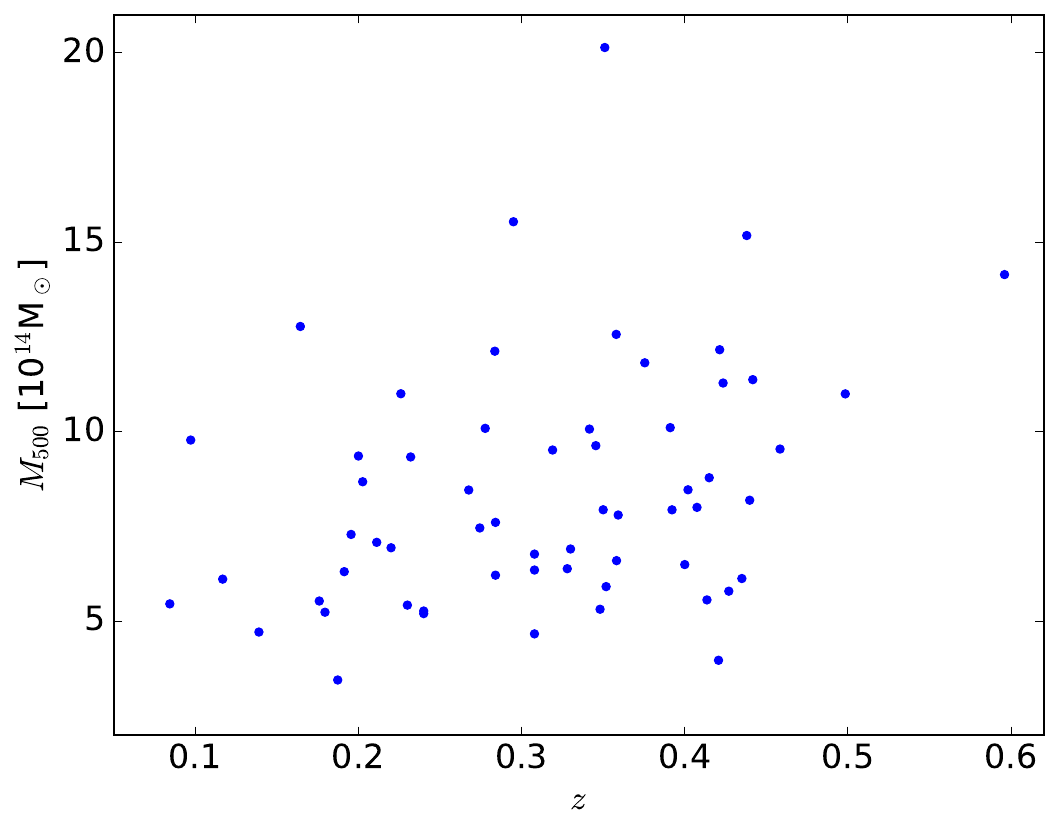}
\end{tabular}
\caption{Redshift vs mass for our sample. 
The average mass in our sample does not vary significantly with redshift, implying minimal covariance between the mass and redshift dependencies in the fit.} 
\label{fig:zP500}
\end{center}
\end{figure}

Fig.~\ref{fig:zP500} shows that the average mass in our sample does not vary significantly with redshift, implying minimal covariance between the mass and redshift dependencies in the fit.

We re-derived the cluster's $r_{500}$ from its SZ flux, $Y_{500}$, using the updated
Compton-Y mass scaling in \citet{Andreon2025} that, being based on weak-lensing masses, is
not affected by the hydrostatic mass bias.  Because of this choice (i.e. larger mass, or, equivalently, lower Compton Y at fixed mass), we expect that profiles derived using this scaling are lower at all radii than those using lower (biased) masses even for the same data and sample, as already verified in \citet{Castagna25}. More precisely, we use
\begin{eqnarray}
\log Y_{500}  = (1.7\pm0.2) (\log (M_{500}/M_\odot) -14.788+\log(1.5)) \nonumber \\
 -4.54\pm0.08  +2/3 (\log (E(z)/E(0.25)) -\log(1.34)\ ,
\label{eq:YM} 
\end{eqnarray}
where $E(z)$ is the usual Hubble ratio term with fixed self-similar evolutionary slope of 2/3, and the other numbers are for conversion from $\Delta=200$ to $\Delta=500$ assuming a NFW profile (the 1.5 value) or the Universal Pressure profile (the 1.34 value), resulting in a
minor offset of 0.17 dex compared to the relation originally derived at the overdensity $\Delta=500$.
$Y_{500}$ values are reported in Table~\ref{tab:individual}. %Derivation of $r_{500}$ and $M_{500}$ values is simple algebra.

Our prior for the population-averaged  scaled pressure profile at the $k$-th knot is a Gaussian with a large standard deviation, $\sigma = 0.5$ dex, centered on the universal pressure profile (UPP) of \citet{Arnaud2010}:
\begin{equation}
lgP_k \sim N\Big(\mu = lgP^{\rm UPP}(x_k), \sigma = 0.5 \Big).
\end{equation}
The prior on the intrinsic scatter at each knot is taken as a uniform distribution over a wide range, encompassing all plausible values:
\begin{equation}
\sigma_{{\rm int},k} \sim U(0,1).
\end{equation}
For the slopes $\gamma_M$, $\gamma_Z$, and $\gamma_\sigma$ we adopt a uniform prior on the angle, equivalent to a Student's t distribution with one degree of freedom on the slope. This choice avoids favoring any particular slope direction, following \citet{Andreon2010}.

The model contains 377 parameters (60 clusters with six parameters each, plus 10 population-level parameters, six redshift-dependence parameters, and one mass-dependence parameter). We sampled the posterior using 24 chains and 8000 iterations, discarding the first half of the samples as burn-in. After convergence assessment, the small number of chains that failed to converge were excluded from the final analysis.

\subsection{The Hierarchical \ppf Package}

All computations were performed using an updated version of the \texttt{PreProFit} package \citep{Castagna2019}.
%, which we publicly release together with this paper on GitHub\footnote{\url{https://github.com/fcastagna/preprofit}}.

Although the present analysis adopts restricted cubic splines to parameterize the pressure profile, hierarchical \ppf additionally supports: (a) the widely used generalized Navarro-Frenk-White profile \citep{Nagai2007}, which offers limited flexibility and contains parameters that are often difficult to interpret \citep[see also][]{Castagna25}; (b) the piecewise power-law model proposed by \citet{Romero2018}, whose main drawback is that the profile uncertainties tend to peak at the knot locations; and (c) cubic-spline interpolation between knots, which can produce undesirable oscillations near the boundaries of the radial range. While we adopt five radial knots in this work, users may freely specify both the number and the location of the knots according to their scientific requirements.

As in previous versions, hierarchical \ppf provides a flexible and automated framework for handling filtering effects arising from the point spread function (PSF) and transfer functions from different instruments. The implementation is compatible with a variety of instruments and supports both exact kernels and their Normal or Cumulated Normal approximations. Relative to earlier versions, the computational performance has been improved by combining beam and transfer-function convolutions into a single operation and by moving all redundant calculations to a pre-computation stage. The perhaps more remarkable new capability of hierarchical \ppf is the simultaneous fitting of all cluster profiles in a sample, enabling the direct inference of the population mean profile, intrinsic scatter, and their dependence on mass and redshift.

Hierarchical \ppf relies on \texttt{PyMC}\footnote{\url{https://www.pymc.io/welcome.html}} \citep{Patil2010} for Bayesian inference, replacing the \texttt{emcee} package used in previous versions. \texttt{PyMC} is particularly well suited for hierarchical Bayesian modeling and facilitates the implementation of complex dependency structures. Users can readily customize the statistical model to suit their specific applications; for example, prior distributions can be modified using standard \texttt{PyMC} syntax. As demonstrated in this work, analyses of cluster populations may include parameters describing mass and redshift trends, either through global dependencies or through radial-dependent parametrizations.

The GitHub repository\footnote{\url{https://github.com/fcastagna/preprofit}} includes two comprehensive examples illustrating these capabilities: a single-cluster analysis based on NIKA data using a gNFW profile, and a population-level analysis based on SPT data using the restricted cubic-spline model. To facilitate reproducibility and ease of use, both examples are configured through YAML files that centralize all relevant settings and can be readily adapted to new datasets.

\begin{table*}[]
\centering
\caption{Sample of galaxy clusters and estimated parameters in the population analysis. }
\footnotesize{
\begin{tabular}{lrrrrrrrrrr}
\toprule
Name & RA & Dec & $z$ & $M_{500}$ & $Y_{500}$ & $lgP_{0,i}$ & $lgP_{1,i}$ & $lgP_{2,i}$ & $lgP_{3,i}$ & $lgP_{4,i}$ \\ 
& [deg] & [deg] & & [10$^{14}M_\odot$] & [10$^{-4}$ arcmin$^2$] & & & & & \\
\midrule
SPT-CLJ0051-4834 & 12.7905 & -48.5776 & 0.187 & $3.45^{+0.39}_{-0.42}$ & $0.16^{+0.03}_{-0.03}$ & 0.57$^{+0.03}_{-0.03}$ & -0.05$^{+0.03}_{-0.03}$ & -0.66$^{+0.04}_{-0.05}$ & -1.10$^{+0.11}_{-0.12}$ & -1.59$^{+0.14}_{-0.18}$ \\
SPT-CLJ0145-5301 & 26.2645 & -53.0295 & 0.117 & $6.11^{+0.28}_{-0.28}$ & $0.41^{+0.03}_{-0.03}$ & 0.60$^{+0.01}_{-0.02}$ & -0.04$^{+0.02}_{-0.02}$ & -0.70$^{+0.04}_{-0.04}$ & -1.15$^{+0.10}_{-0.11}$ & -1.61$^{+0.13}_{-0.16}$ \\
SPT-CLJ0145-6033 & 26.2958 & -60.5594 & 0.179 & $5.24^{+0.41}_{-0.42}$ & $0.32^{+0.04}_{-0.05}$ & 0.52$^{+0.04}_{-0.05}$ & -0.03$^{+0.03}_{-0.03}$ & -0.67$^{+0.04}_{-0.05}$ & -1.05$^{+0.11}_{-0.11}$ & -1.58$^{+0.13}_{-0.16}$ \\
SPT-CLJ0232-4421 & 38.0701 & -44.3541 & 0.284 & $12.13^{+0.39}_{-0.39}$ & $1.38^{+0.07}_{-0.08}$ & 0.66$^{+0.04}_{-0.05}$ & -0.02$^{+0.03}_{-0.04}$ & -0.64$^{+0.05}_{-0.05}$ & -1.11$^{+0.12}_{-0.13}$ & -1.66$^{+0.15}_{-0.18}$ \\
SPT-CLJ0235-5121 & 38.9468 & -51.3516 & 0.278 & $10.09^{+0.54}_{-0.56}$ & $1.01^{+0.09}_{-0.10}$ & 0.55$^{+0.02}_{-0.02}$ & -0.07$^{+0.02}_{-0.02}$ & -0.65$^{+0.04}_{-0.04}$ & -1.04$^{+0.09}_{-0.09}$ & -1.76$^{+0.13}_{-0.18}$ \\
...\\
\bottomrule
\end{tabular}
\break 
Cluster Names, RA, Dec and z are from \citet{Bocquet2019}, the remaining quantities are computed in this work. The scaled pressure values reported in this table differ from those published in \citet{Castagna25} for clusters in common, primarily due to differences in $P_{500}$ and, in the case of the first radial bin, in its radial location.  The full Table is only available in electronic form at the CDS via \url{http://cdsweb.u-strasbg.fr/cgi-bin/qcat?J/A+A/}. URL TO BE ADDED WHEN AVAILABLE\hfill \break
\label{tab:individual}
}
\end{table*}

\section{Results} \label{sec:results}

\begin{figure}
\begin{center}
\begin{tabular}{c}
\includegraphics[width=.89\linewidth]{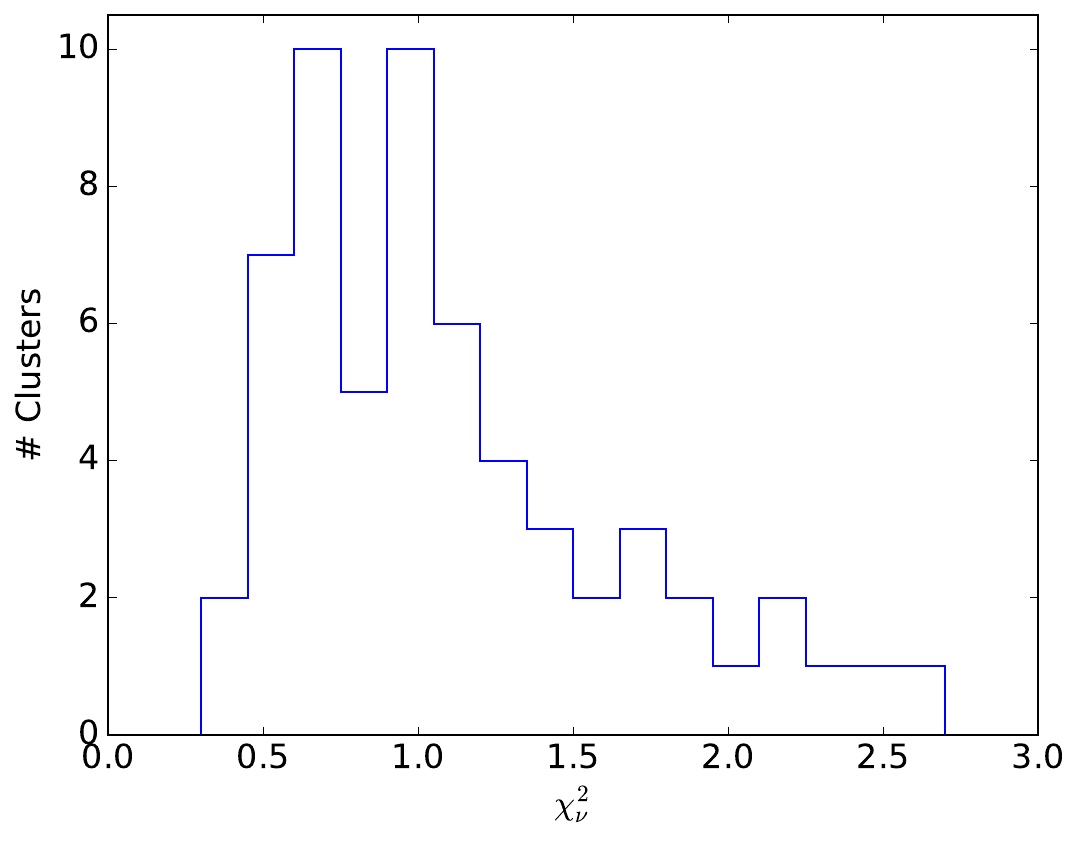}
\end{tabular}
\caption{Reduced chi-square $\chi^2_{\nu}$ of our fit. The adopted model provides a good description of the data, especially considering that the quoted $\chi^2_{\nu}$ is not that of the best-fit model, but corresponds instead to the median model.
} 
\label{fig:chisq}
\end{center}
\end{figure}

Our model simultaneously fits the 17 parameters describing the population-averaged scaled pressure profile and the 60 individual scaled pressure profiles. 
Before presenting the results, it is useful to summarize two lessons learned from running models with different settings.  

First, we explored replacing Eq.~3 with a model including an explicit dependence on $P_{500}$, as adopted in pressure-scaled analyses such as \citet{Arnaud2010}. Since $P_{500}\propto M^{2/3}E(z)^{8/3}$, this parameterization mixes mass and redshift dependences by construction. In our sample, nearly all clusters above the median redshift have $P_{500,i}>P_{500,\rm med}$, while the opposite holds for low-redshift systems. Consequently, introducing a free dependence on $P_{500}$ generates an unnecessary covariance between mass and redshift trends, complicating their interpretation. Similar covariance issues induced by parameterization choices were discussed by \citet{Andreon2014}.

Second, the innermost knot, set at $0.1 \, r_{500}$ in \citet{Castagna25}, is smaller than the beam for some clusters with $z \gtrsim 0.3$. This introduces additional covariances between cluster parameters through $\gamma_z$, because this parameter becomes poorly constrained at the first knot. For this reason, we adopted a larger innermost knot, at $0.15 \, r_{500}$

Figure~\ref{fig:chisq} shows how well our model fits the data, namely the distribution of the reduced chi-square $\chi_{\nu}^2$ of the median surface brightness profile, which is therefore not the minimal $\chi_\nu^2$. Our model adequately describes the data and does not require additional flexibility, as also confirmed by visual inspection of the predicted and observed Compton-$y$ profiles shown in Fig.~\ref{fig:Compton_prof1} to Fig.~\ref{fig:Compton_prof3}.

Table~\ref{tab:individual} lists the individual scaled pressure parameter estimates $lgP_{k,i}$, while
Table~\ref{tab:population} lists the population parameter estimates, including the population-averaged scaled pressures $lgP_k$ and the intrinsic scatters $\sigma_{{\rm int},k}$. These individual estimates are more accurate than those obtained from fitting each cluster independently, as the joint hierarchical model uses the population-level posteriors as informative priors for the individual $lgP_{k,i}$.  
For objects in common with \citet{Castagna25}, these values differ numerically from those reported there because the two works adopt different $r_{500}$ and $P_{500}$ values, due to the use of different $Y$--$M$ scalings.

\begin{table}[ht]
\centering\caption{Population-averaged scaled pressure, intrinsic scatter, and $\gamma$ values.}
\footnotesize{
\begin{tabular}{lrrrrr}
\toprule
r/r$_{500}$ & 0.15 & 0.40 & 0.70 & 1.00 & 1.30 \\
\midrule
$lgP_k$ & 0.547 & -0.047 & -0.614 & -1.056 & -1.628 \\ 
err & 0.028 & 0.015 & 0.024 & 0.037 & 0.051 \\ 
$\sigma_{int,k}$ & 0.176 & 0.043 & 0.054 & 0.154 & 0.206 \\ 
err & 0.024 & 0.015 & 0.024 & 0.032 & 0.046 \\ 
$\gamma_z$ & -0.272 & -0.019 & 0.834 & 0.691 & -0.113 \\ 
err & 0.573 & 0.302 & 0.691 & 0.779& 0.048 \\ 
\hline
$\gamma_M$ & & & -0.113 & & \\ 
err & & & 0.048 & & \\
$\gamma_\sigma$ & & & 1.061 & & \\ 
err & & & 0.984 & & \\
\hline
\bottomrule
\end{tabular}}
\break $\gamma_M$ and $\gamma_\sigma$ are assumed to be radius-independent. \hfill\break
\label{tab:population}
\end{table}

\begin{figure}
\begin{center}
\begin{tabular}{c}
\includegraphics[width=.955\linewidth]{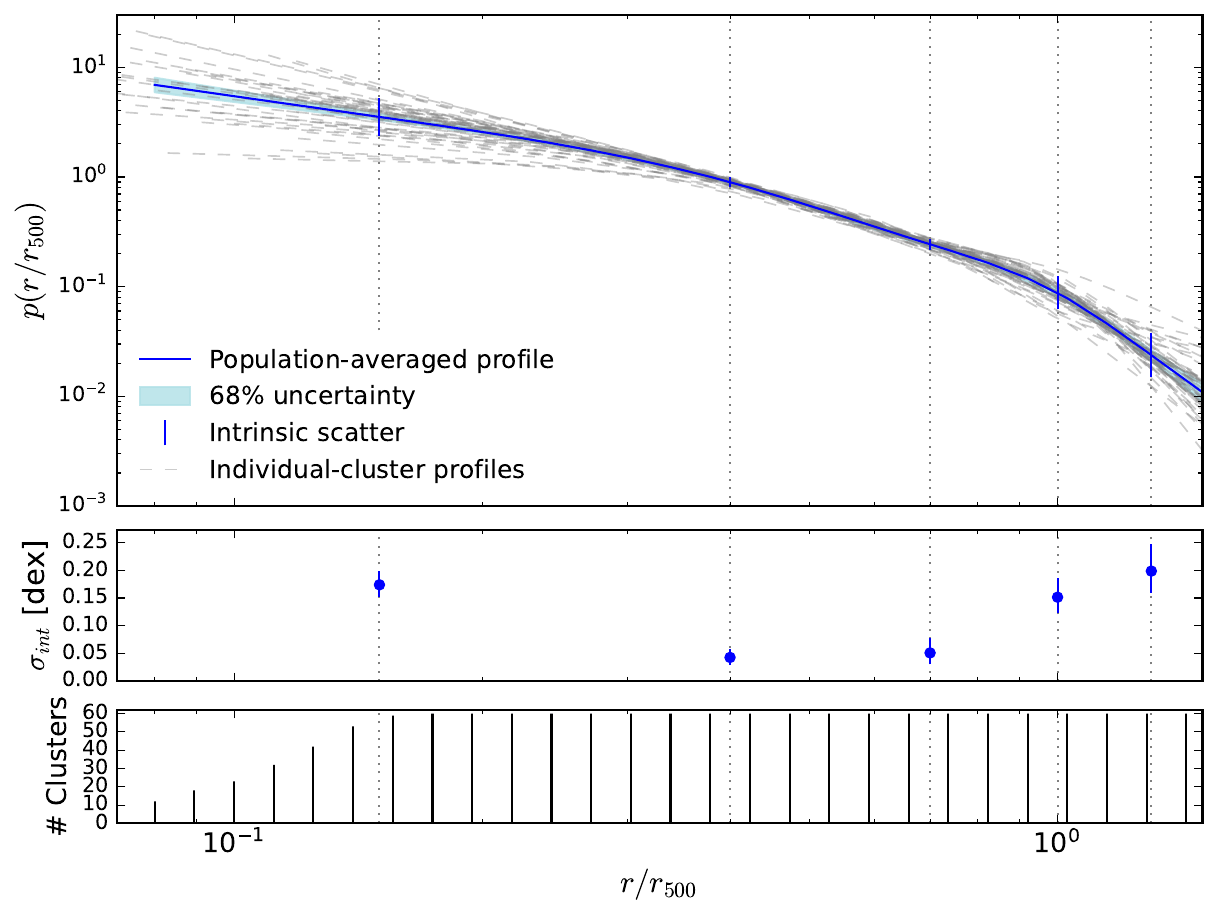}
\end{tabular}
\end{center}
\caption{Scaled pressure profiles (top panel) and intrinsic scatter (top and central panels). The scatter is minimal at $0.4r_{500}\leq r \leq0.7r_{500}$. The bottom panel shows the number of objects contributing at each radius. Nearly all clusters contribute to every knot, which is not true adopting a smaller knot, for example at $0.1r_{500}$.}
\label{fig:press_joint}
\end{figure}

\begin{figure}
\begin{center}
\begin{tabular}{c}
\includegraphics[width=.955\linewidth]{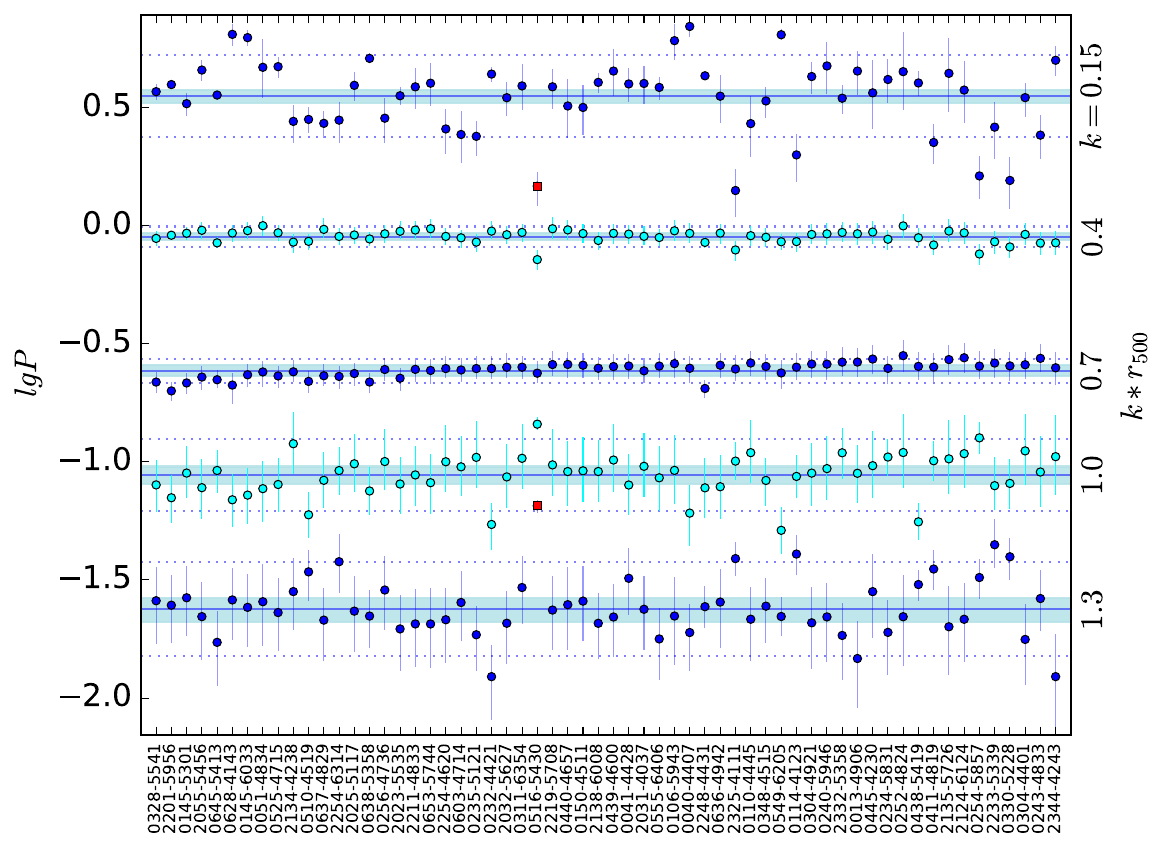}
\end{tabular}
\caption{Individual scaled pressure parameters $lgP_{k,i}$ (points with error bars) at the five knots (marked on the right y axis). Horizontal lines represent the population-averaged scaled pressure estimates $lgP_k$ (solid lines) with their errors (shaded areas), and intrinsic scatter estimates (dotted lines). The red squares indicates $>2$ combined $\sigma$ outliers.}
\label{fig:outliers}
\end{center}
\end{figure}

\begin{figure}
\begin{center}
\begin{tabular}{c}
\includegraphics[width=.955\linewidth]{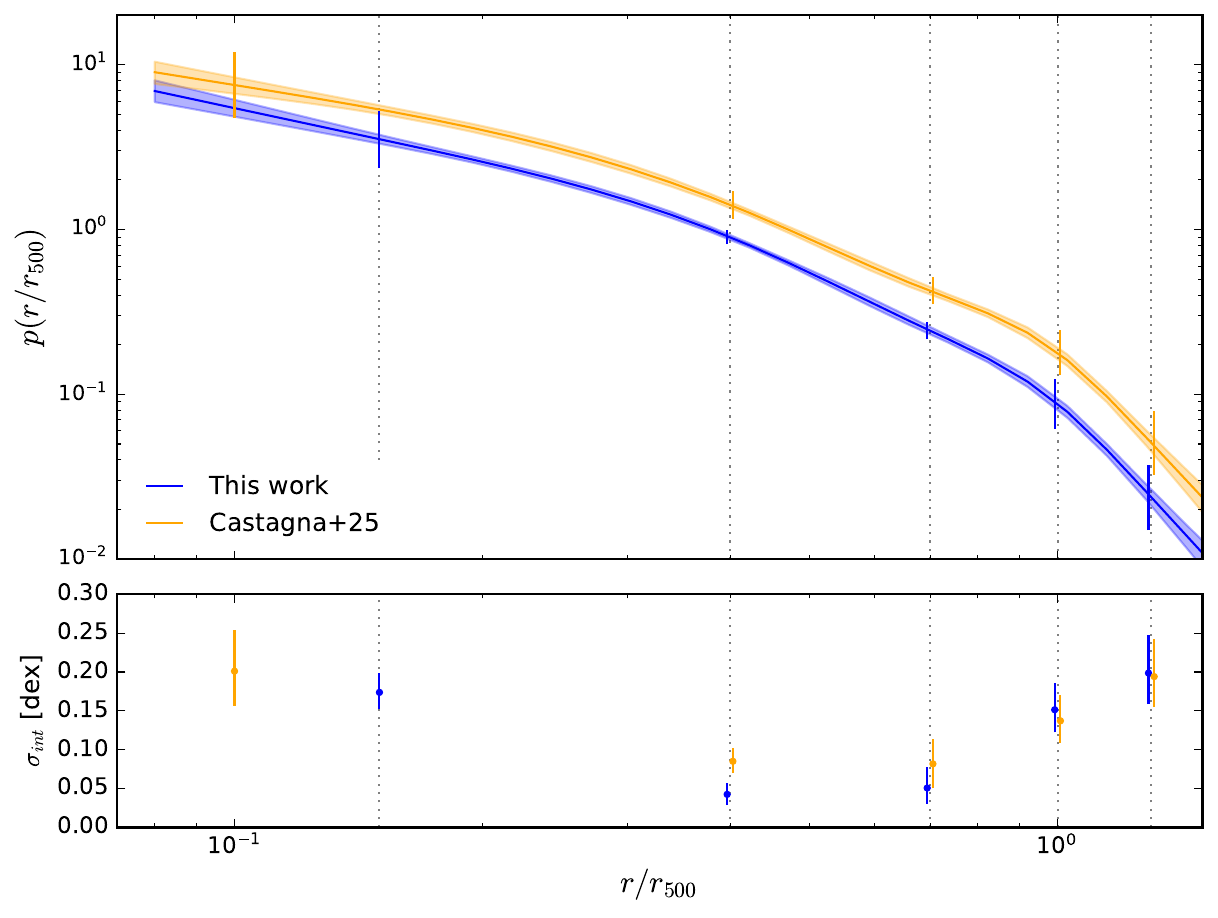}
\end{tabular}
\caption{ Scaled pressure profiles (top panel) and intrinsic scatters (bottom panel) for our sample and \citet{Castagna25} sample, adopting a different Y-M scaling (and a different sample). The scaled pressure profile is about 40 \% lower at all radii because clusters have about 40\% larger mass for they Compton-Y parameter (lower Y for their mass) with the revised scaling. }
\label{fig:lit_comp}
\end{center}
\end{figure}

The top panel of Figure~\ref{fig:press_joint} shows the population-averaged scaled pressure profile (solid blue line), its 68\% uncertainty (blue shaded area), the 68\% intrinsic scatter around it (error bars), and the individual cluster estimates from the population analysis (dashed gray lines)\footnote{As discussed in Castagna et al. for a previous sample, and by \citet{Andreon2013Understanding} in a similar hierarchical fit, we expect that with noisy data, more than 68\% of the individual posterior median profiles lie within the $1\sigma_{\rm int}$ interval because of Bayesian shrinkage.}. The central panel shows the radial distribution of the intrinsic scatter, estimated at the five knots. The dispersion is minimal at intermediate radii (second and third knots), in agreement with previous works, as already shown by Castagna et al. A similar set of information, emphasizing the level of homogeneity between clusters, is presented in Figure~\ref{fig:outliers}.  
Only one cluster, SPT-CLJ0516-5430, differs by more than $2\sigma$ from the mean. Visual inspection of Chandra data (OBSID=15099) shows that it is highly elongated and exhibits a tail. 

Our choice of modeling the scatter with a Student-t distribution is well justified: for example, SPT-CLJ0330-5228 is an 
intermediate-redshift cluster (with a prominent gravitational arc visible in the Legacy Survey) seen through a very low-redshift cluster, Abell 3128 at $z=0.06$. The cluster is not an outlier, being about at most $\sim 1.7\sigma$  away from the average. Removing or retaining this profile would be arbitrary because both the contamination and the outlier status are uncertain; the Student-t modeling provides a consistent approach.  

Figure~\ref{fig:lit_comp} compares our determination with the previous result by \citet{Castagna25} for a sample in a reduced redshift range and using a different $Y$--$M$ scaling. The scaled pressure profile is about 40\% lower at all radii. The change is due to the about  40\% larger mass for their Compton-$Y$ parameter (i.e., lower $Y$ at fixed mass) with the revised scaling. We verified that this difference is not due to sample selection by also comparing the average profile of the overlapping sample. Extensive comparison with previous works
is presented in \citet{Castagna25}.

\begin{figure}
\begin{center}
\begin{tabular}{c}
\includegraphics[width=.955\linewidth]{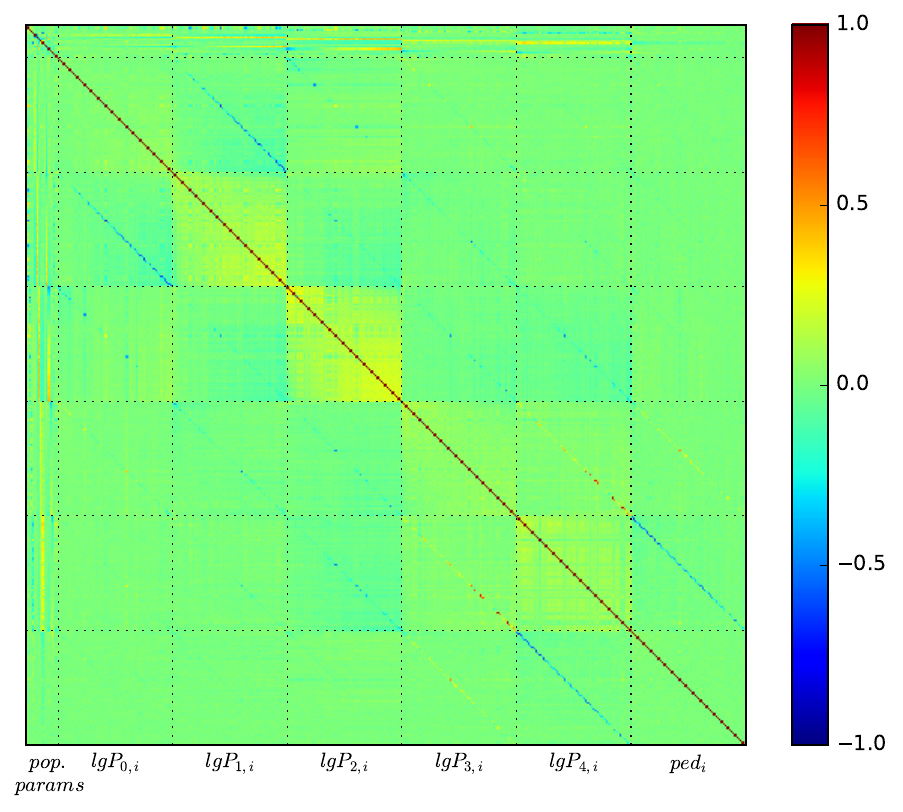}
\end{tabular}
\end{center}
\caption{Correlation matrix for all 377 fit parameters. Most parameters show little covariance, except where expected: between the scaled pressures of the two inner knots and between the scaled pressures at the last knot and the pedestal parameters (see text for details).}
\label{fig:corrmat}
\end{figure}

Figure~\ref{fig:cornerplot} shows the marginal and joint probability distributions for the population parameters. Almost all parameters exhibit little correlation, due to our choice of radial profile model and the spacing of the knots. This is further illustrated in Figure~\ref{fig:corrmat}, which shows the correlation matrix for all 377 parameters, where each entry $\rho_{i,j}$ represents the correlation between the parameter pair $(\theta_i, \theta_j)$.  Some correlations remain: between the  scaled pressures at the two innermost knots (where higher-resolution data would help break the degeneracy), and between the scaled pressures at the outermost knots and the pedestal parameter (because it is difficult to distinguish a flat profile from a flat background).

\begin{figure}
\begin{center}
\begin{tabular}{c}
\includegraphics[width=.8\linewidth]{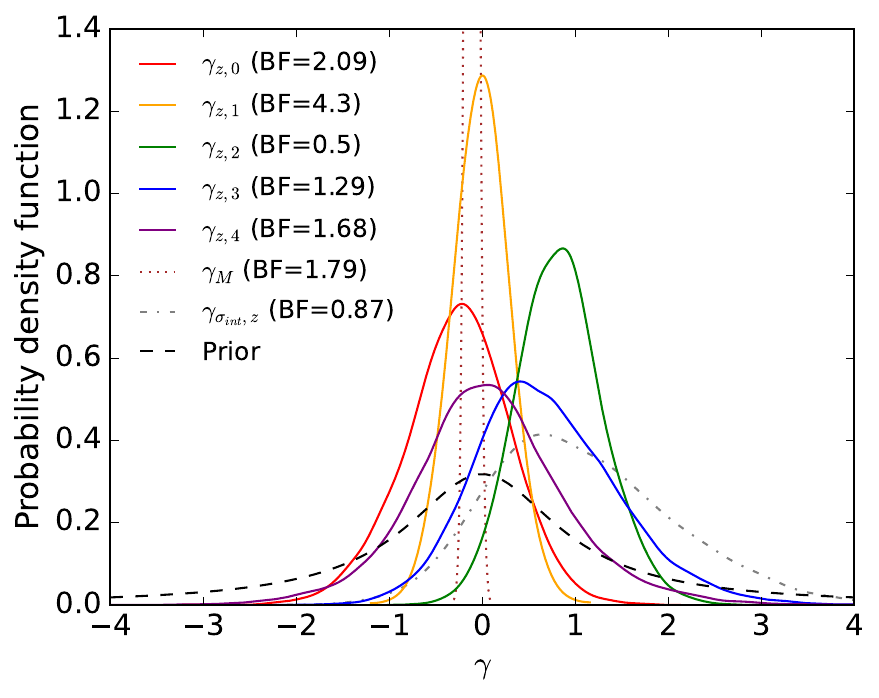}
\end{tabular}
\end{center}
\caption{Posterior and prior probability for the redshift parameter $\gamma_z$, the mass parameter $\gamma_M$ and scatter parameter $\gamma_\sigma$. Data are inconclusive to draw firm conclusions about the redshift dependency, although they provide the constraints given in Table~\ref{tab:population}.}
\label{fig:bayes_factor}
\end{figure}

The data are insufficient to draw firm conclusions on the redshift dependency based on the Bayes factor, as models with and without these dependencies have comparable probabilities. Since the models are nested, the Bayes factor can be derived directly from the available posterior samples using the marginal distribution shown in Fig.~\ref{fig:cornerplot}, as the ratio of the posterior density to the prior density evaluated at $\gamma_z = 0$, as illustrated in Fig.~\ref{fig:bayes_factor}. We also tested the hypothesis of self-similar evolution by replacing $(1+z)$ with $E(z)$ in Eq.~2. Recomputing the posteriors and Bayes factors yields similarly inconclusive results, which can also be seen visually in Fig.~\ref{fig:bayes_factor} at $\gamma_z \sim 0.7$ (the approximated $\gamma$ values corresponding to the self-similar prediction).
About mass dependency, we are not actually comparing two models, but basically comparing our and \citet{Arnaud2010} mass dependence parameters. Our data mildly favor a small negative $\gamma_M\sim -0.11$ (i.e., $M^{-0.11}$), with $\sim2.3 \sigma$ evidence against zero (Table~\ref{tab:population}). This factor almost exactly cancels the additional $M^{0.12}$ mass scaling introduced by \citet{Arnaud2010} to compute the scaled pressure profile. For completeness, Fig.~\ref{fig:bayes_factor} reports the Bayes factor, that as mentioned is partially misleading, resulting in a suggestive, but not strong, evidence.
Anyway, our data provide a $1\sigma$ sensitivity in $\gamma$ of 0.04 dex per unit mass for the mass dependence, and 0.1--0.2 dex per unit redshift for the redshift dependence (see Table~\ref{tab:population} for the individual values in their respective units).  Finally, about the redshift dependence of the intrinsic scatter, as for the redshift dependence of the mean scaled pressure, data are are insufficient to draw firm conclusions based on the Bayes factor (Fig.~\ref{fig:bayes_factor}), with a  
$1\sigma$ sensitivity of 8\% variation of $\sigma_{\rm intr}$ over a $\Delta z=0.1$.

As discussed in Andreon et al. (2021), redshift dependence should not be confused with evolution. The objects being compared at different redshifts are not related as ancestors and descendants. Moreover, while the average distance of gas particles from the cluster center remains constant or decreases due to infall, the comparison radius increases because it scales with $r_{500}$, which grows with time. Therefore, the chosen comparison radius implies that different physical regions are being compared at different redshifts. Unsurprisingly, the no-evolution prediction from the Magneticum simulations yields negative $\gamma_z$ at all radii except $r=0$ (Andreon et al. 2021).

\section{Conclusions} \label{sec:conclusion}

We have presented a comprehensive analysis of the thermal pressure profiles of a population of 60 clusters from the South Pole Telescope (SPT) SZ catalog with $0.08<z<0.6$, requiring high signal-to-noise and 5 element resolution on Planck-SPT Compton-$y$ maps. We extract Compton-$y$ maps and use a hierarchical Bayesian framework to model the three-dimensional spherical pressure profiles using restricted cubic splines. Crucially, we employ an updated weak-lensing-calibrated Compton-$Y$--mass scaling relation, effectively removing hydrostatic mass bias from the inferred masses. Our model is internally self-consistent, accommodating cluster-to-cluster profile variations and allowing the  foreground and background contributions {to vary between lines of sight. Furthermore, we allow for outliers arising from both bona fide clusters with intrinsically distinct profiles and line-of-sight projection effects. The code is made public available with this paper. We derive a new population-averaged pressure profile and confirm that the intrinsic scatter is minimal at intermediate radii ($0.4 \le r/r_{500} \le 0.7$). Due to the adoption of the unbiased mass scaling, our derived pressure profile is approximately 40\% lower  at all radii than previous estimates using biased hydrostatic masses. 
We find suggestive, though not conclusive, evidence that the currently adopted mass dependence may require revision, specifically by canceling the additional $M^{0.12}$ mass scaling introduced by \citet{Arnaud2010}. We further constrain departures from self-similar evolution in the mean profile to be smaller than 0.1-0.2 dex and limit the evolution of the intrinsic scatter to less than 8 \% per $\Delta z=0.1$. 

Given the broad role of pressure profiles in cluster astrophysics and cosmology, future work should incorporate the updated mean profile and scatter presented here and reassess their impact on SZ observables, cluster mass calibration, survey completeness, thermal SZ power spectrum predictions, halo-model calculations, and cosmological parameter inference.
Future high-resolution multi-wavelength data will be critical to breaking the remaining covariances at the inner and outer cluster boundaries, better identifying redshift and mass depedences and refining the scaling relations.

\section*{Data availability}
The statistical code developed for this analysis is made publicly available on GitHub\footnote{\url{https://github.com/fcastagna/preprofit}}
Full Table~\ref{tab:individual} is only available in electronic form at the CDS via \url{http://cdsweb.u-strasbg.fr/cgi-bin/qcat?J/A+A/} [INSERT URL WHEN AVAILABLE].

\begin{acknowledgements}
FC and SA acknowledge INAF grant “Characterizing the newly discovered clusters of low surface brightness” and PRIN-MIUR grant 20228B938N “Mass and selection biases of galaxy clusters: a multi-probe approach”, the latter funded by the European Union Next generation EU, Mission 4 Component 1 CUP C53D2300092 0006.
\end{acknowledgements}

\bibliographystyle{aa} 
\bibliography{biblio}

\begin{appendix}
\onecolumn
\section{Radial surface brightness profiles of the sample}
Fig.~\ref{fig:Compton_prof1} to \ref{fig:Compton_prof3} show surface brightness profiles of the studied 60 clusters.

\begin{figure}[h]
\centerline{\includegraphics[width=.25\textwidth]{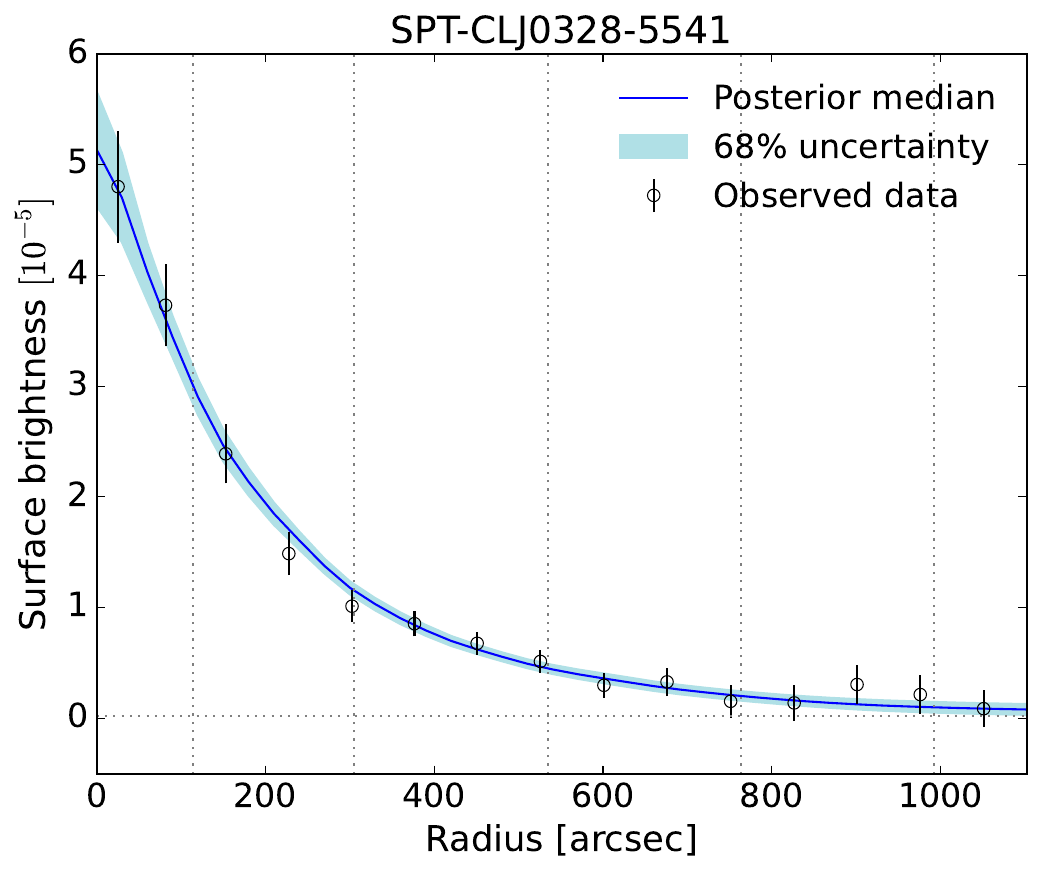} 
\includegraphics[width=.25\textwidth]{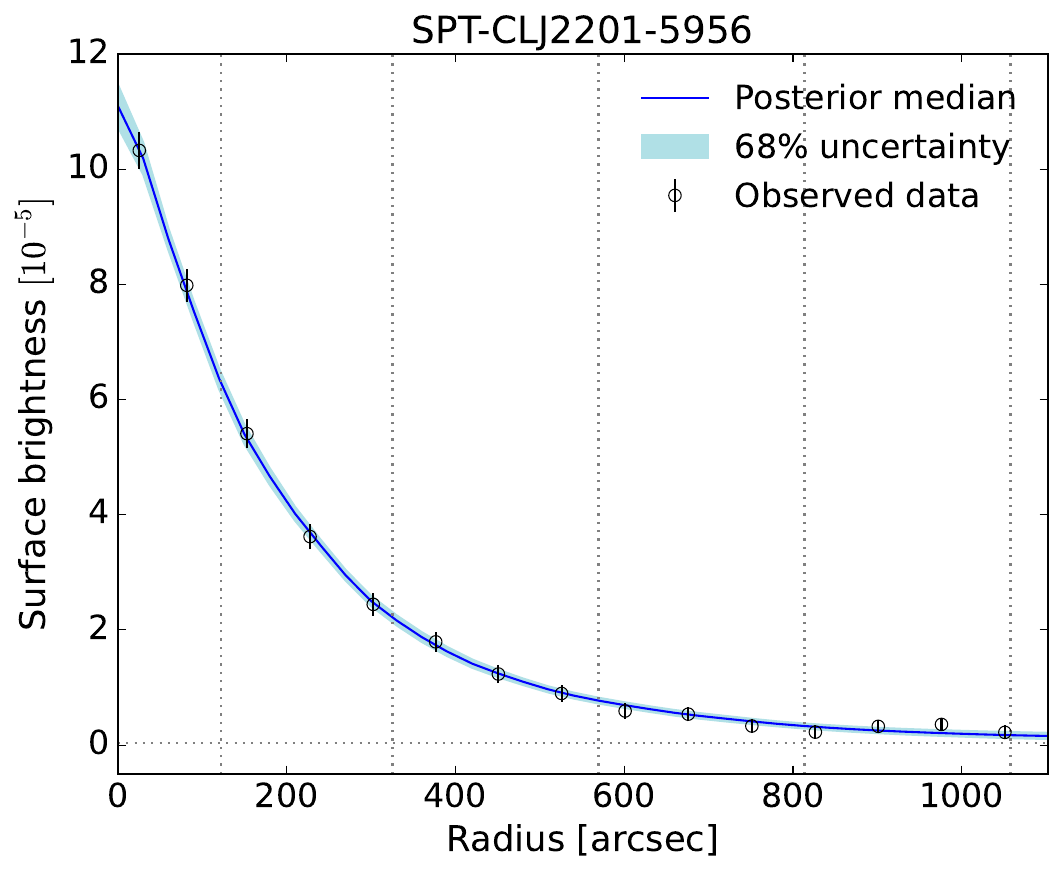} 
\includegraphics[width=.25\textwidth]{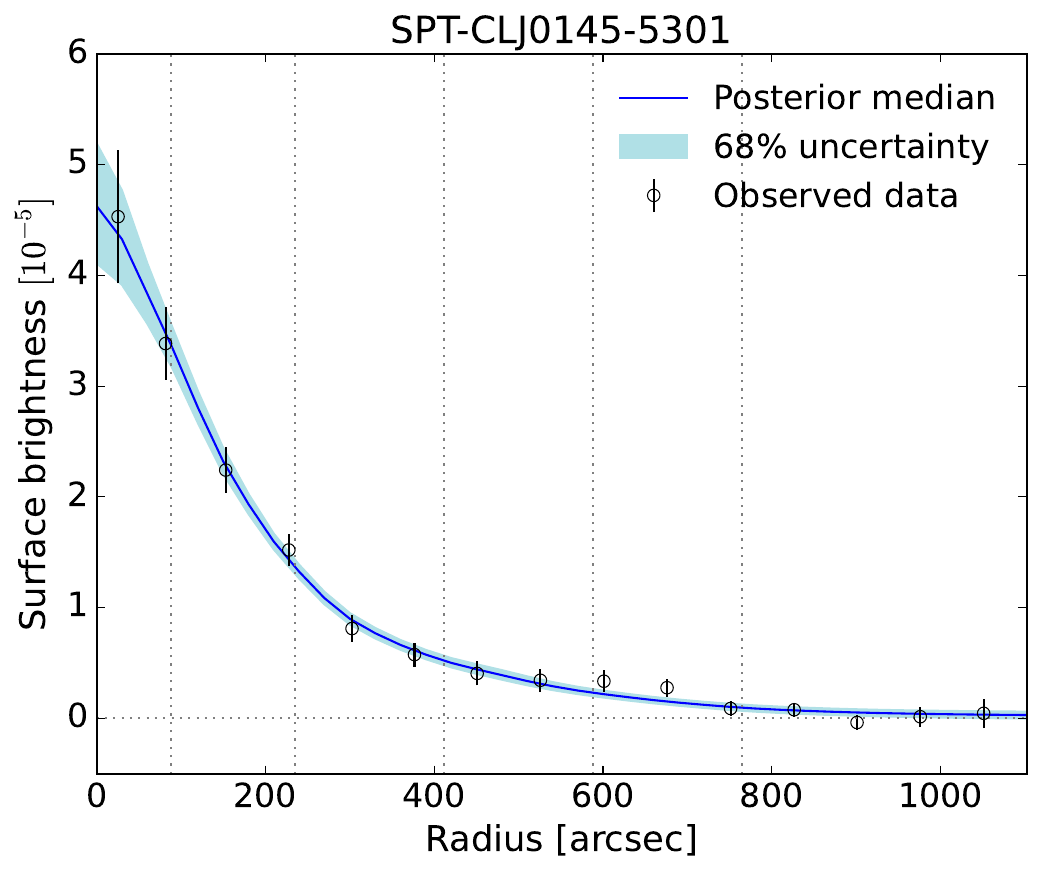} 
\includegraphics[width=.25\textwidth]{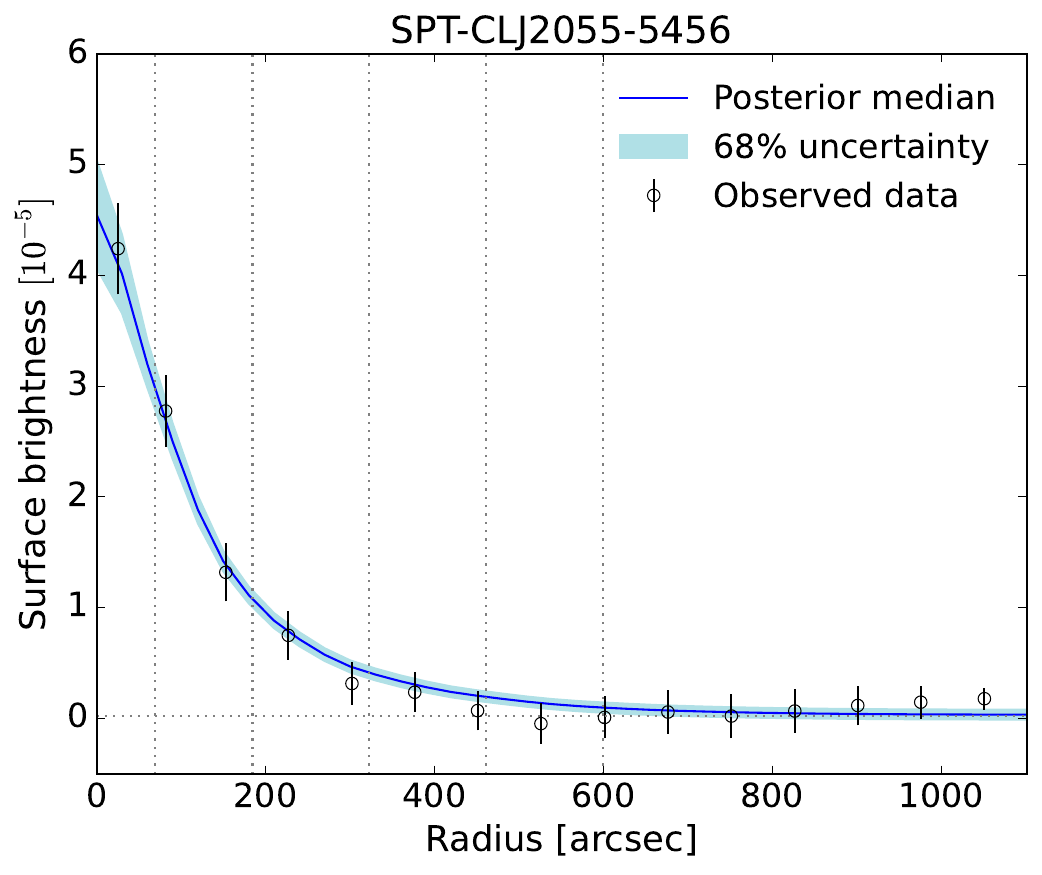}} 
\centerline{\includegraphics[width=.25\textwidth]{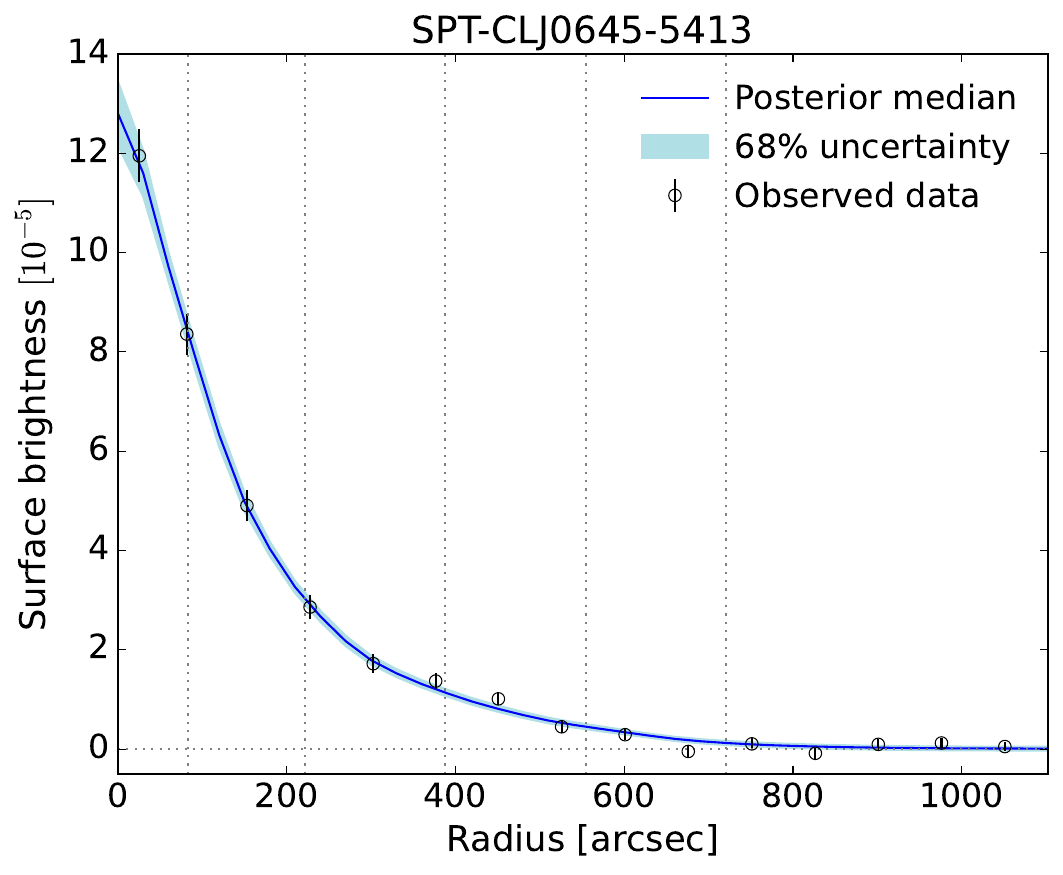} 
\includegraphics[width=.25\textwidth]{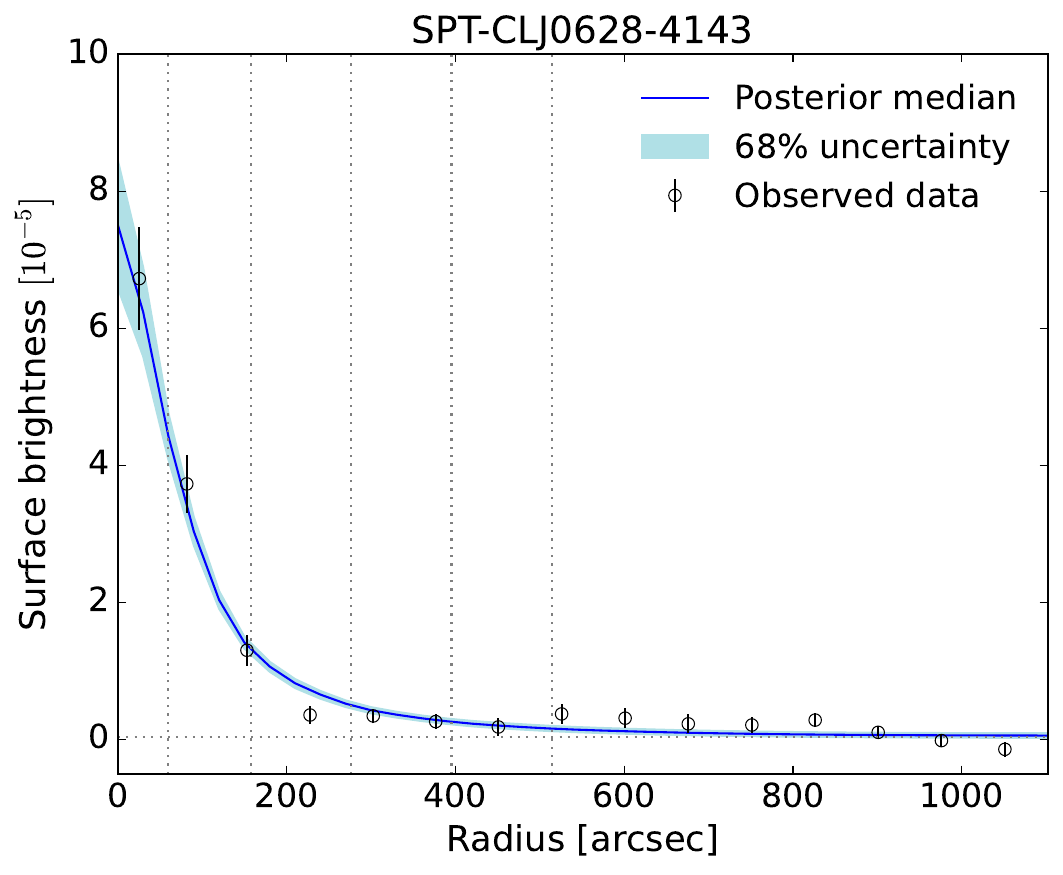} 
\includegraphics[width=.25\textwidth]{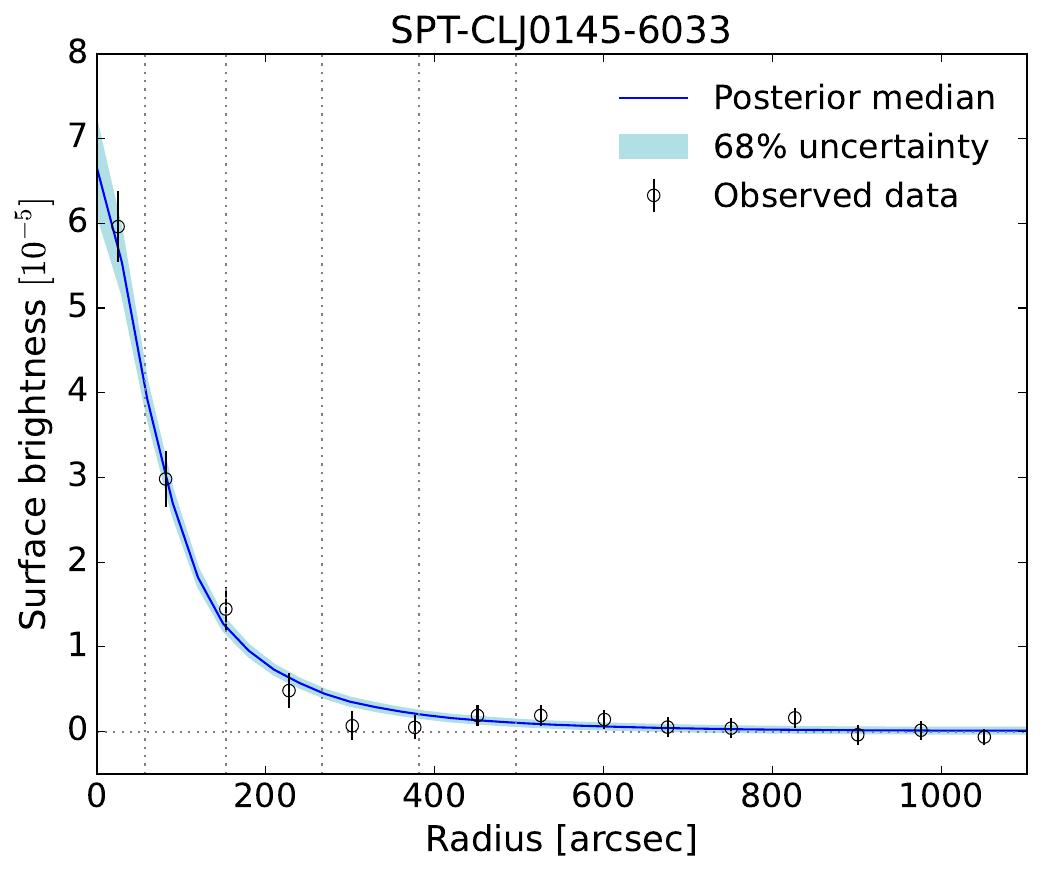} 
\includegraphics[width=.25\textwidth]{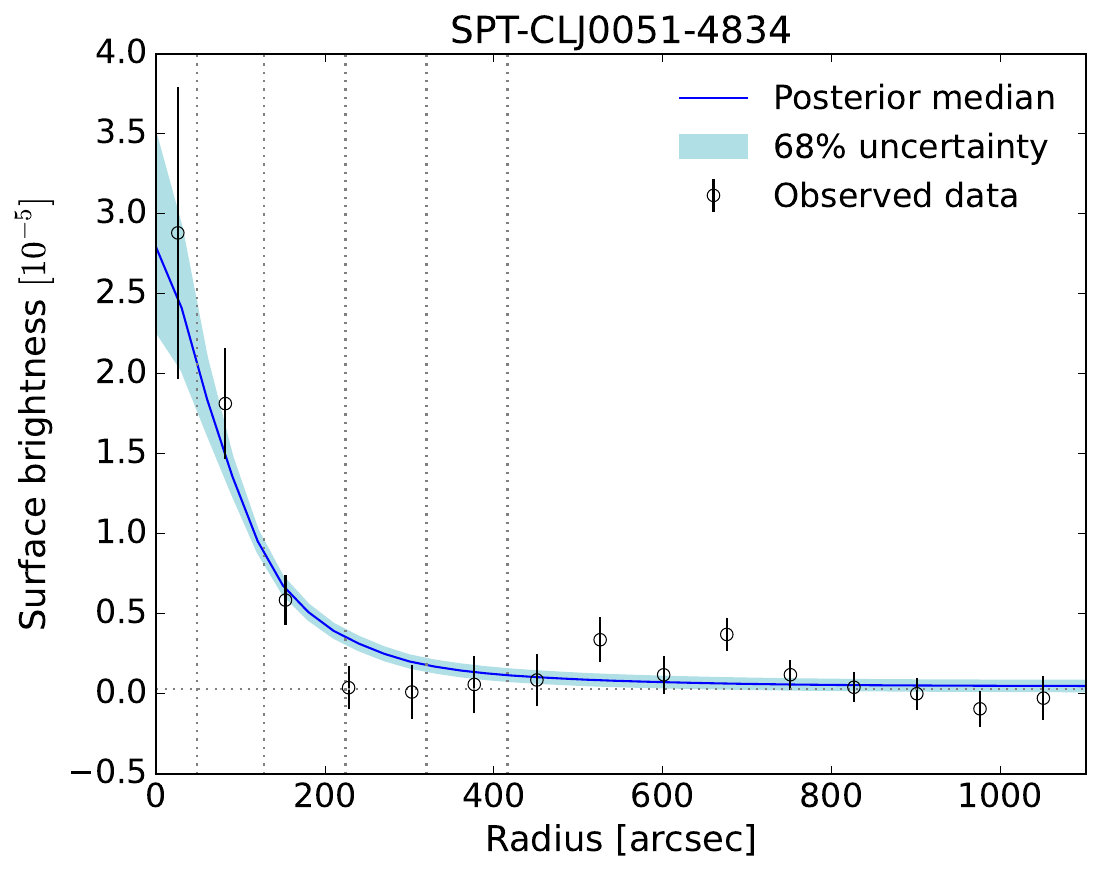}} 
\centerline{\includegraphics[width=.25\textwidth]{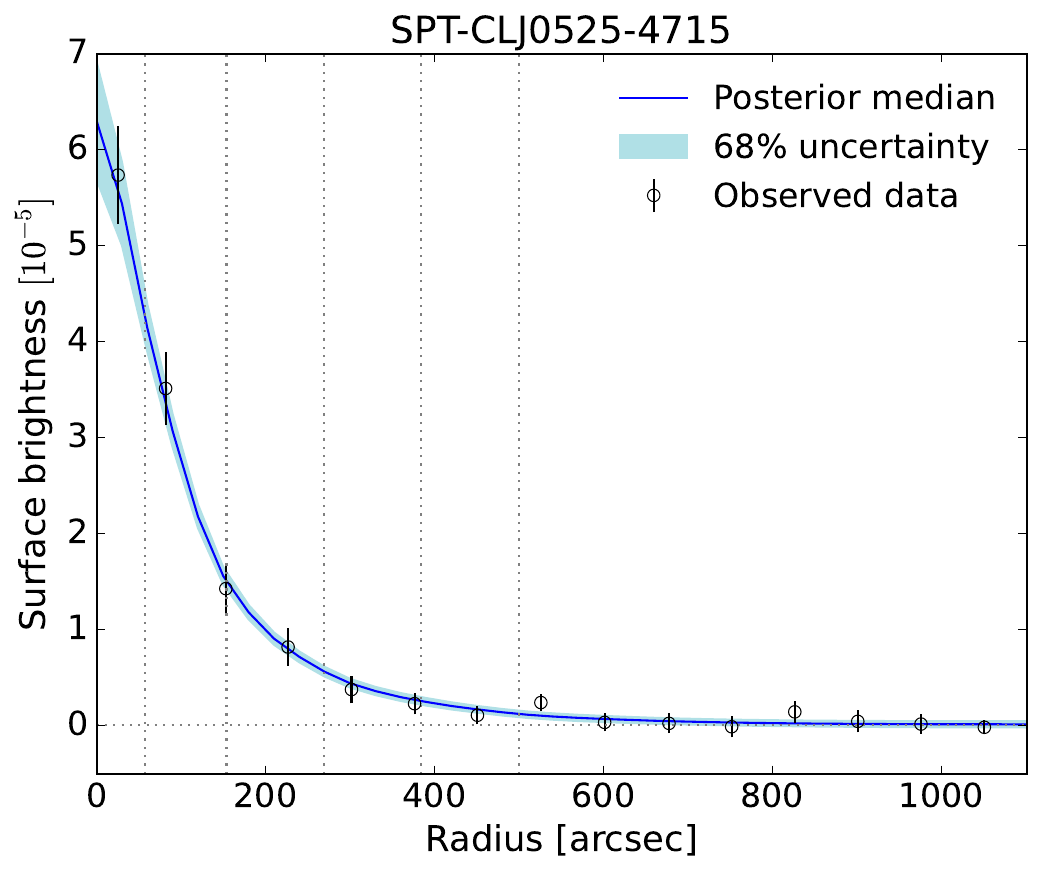} 
\includegraphics[width=.25\textwidth]{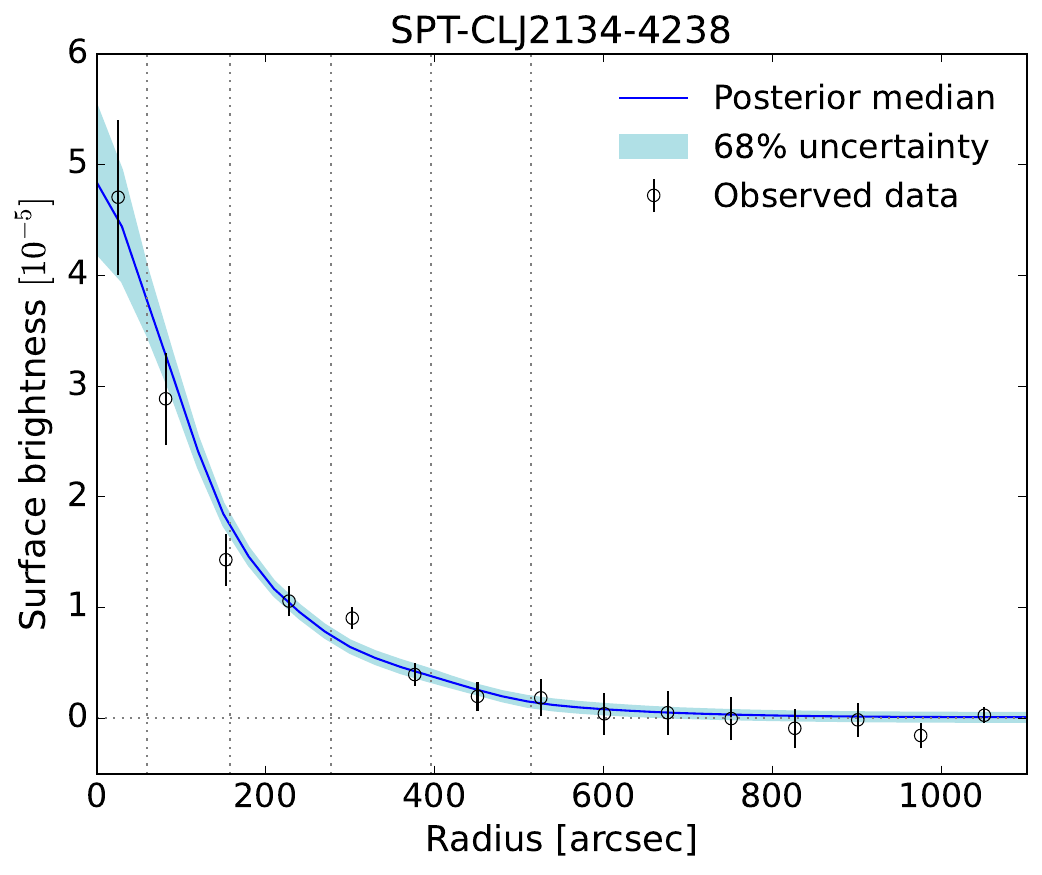} 
\includegraphics[width=.25\textwidth]{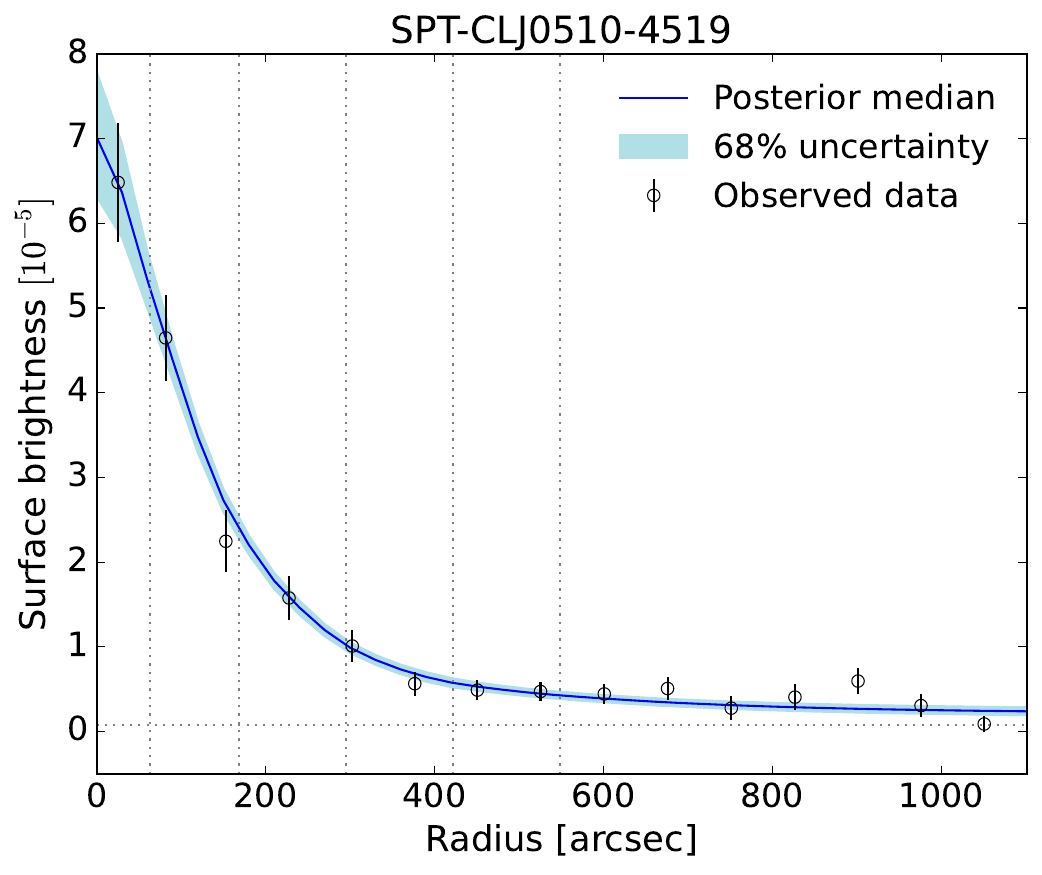} 
\includegraphics[width=.25\textwidth]{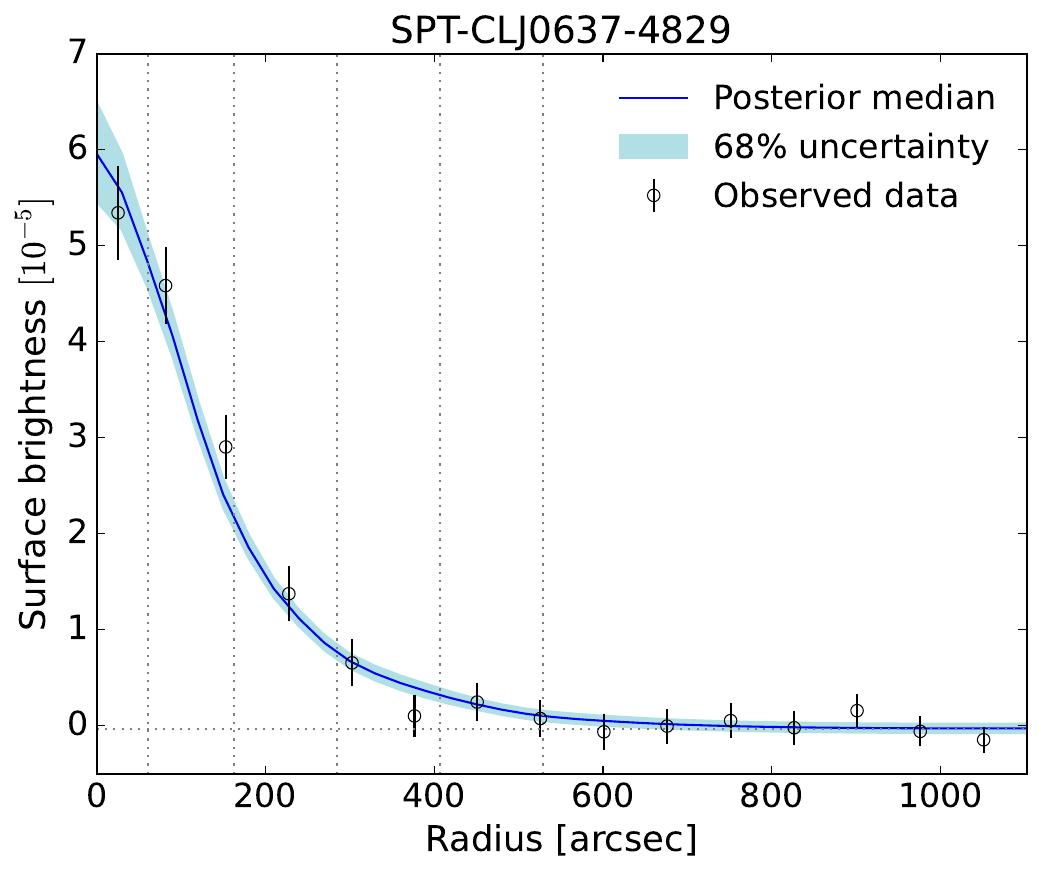}} 
\centerline{\includegraphics[width=.25\textwidth]{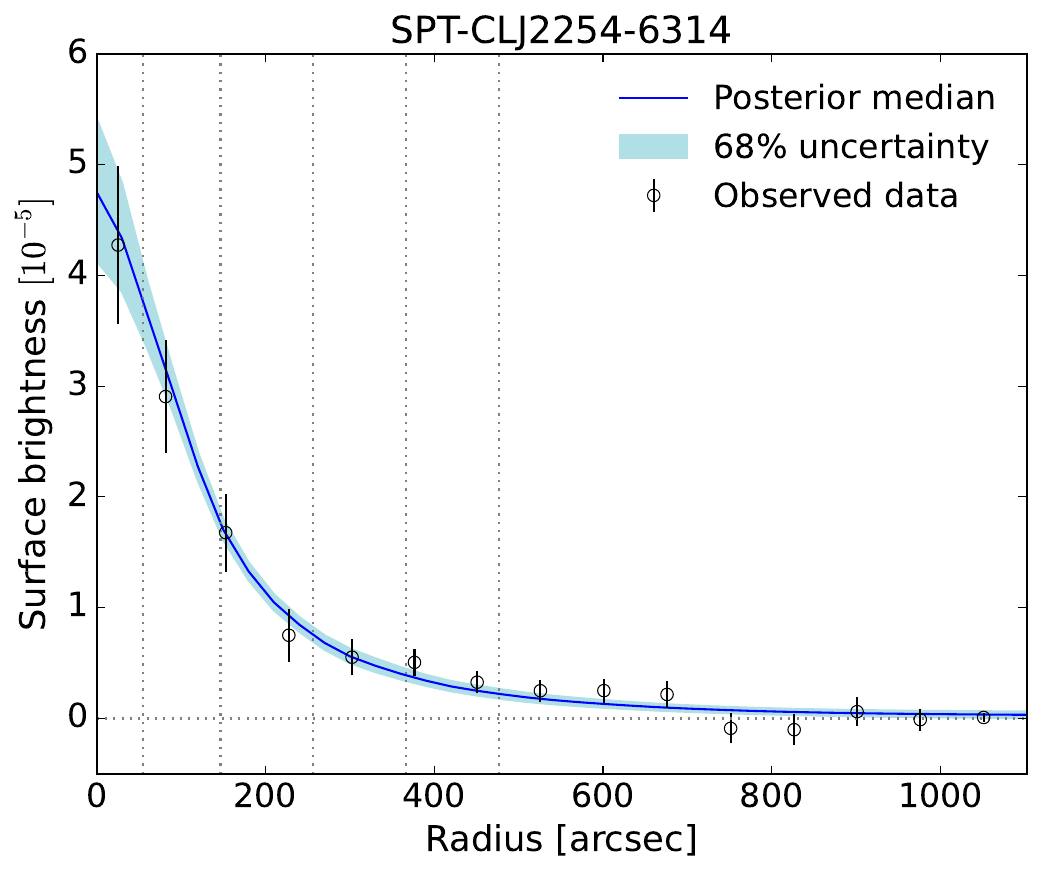} 
\includegraphics[width=.25\textwidth]{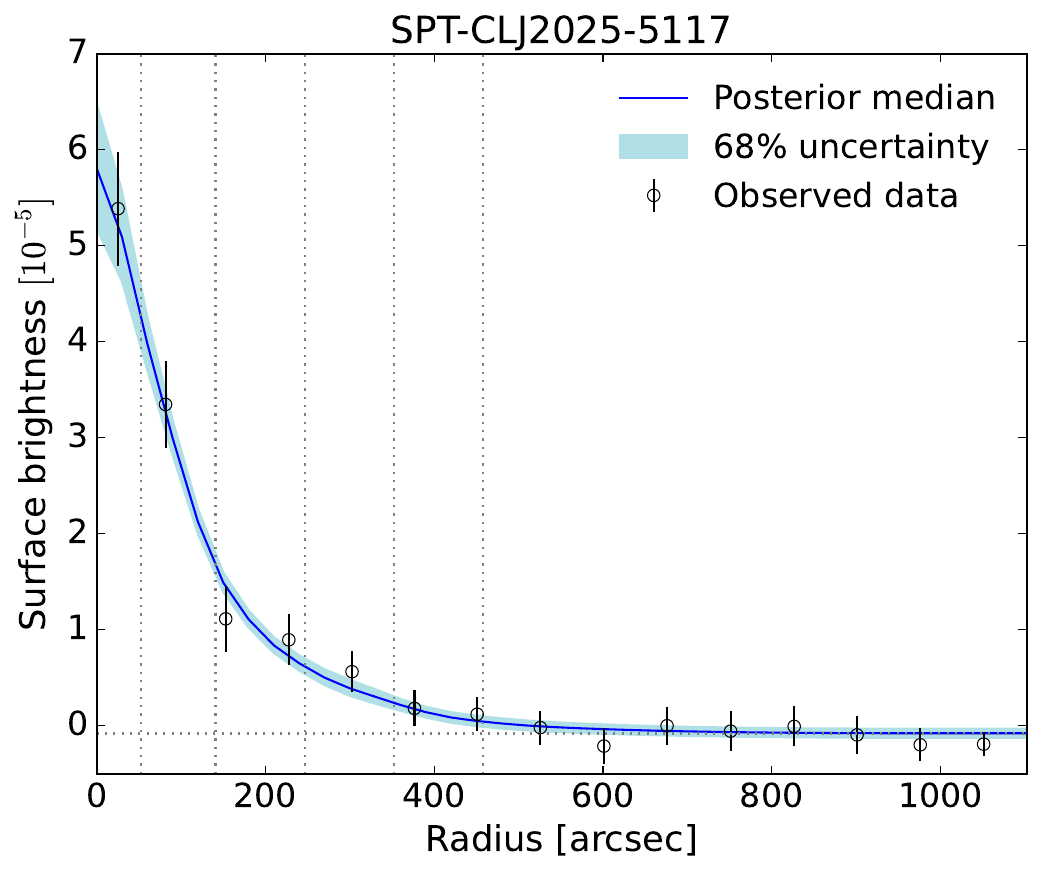} 
\includegraphics[width=.25\textwidth]{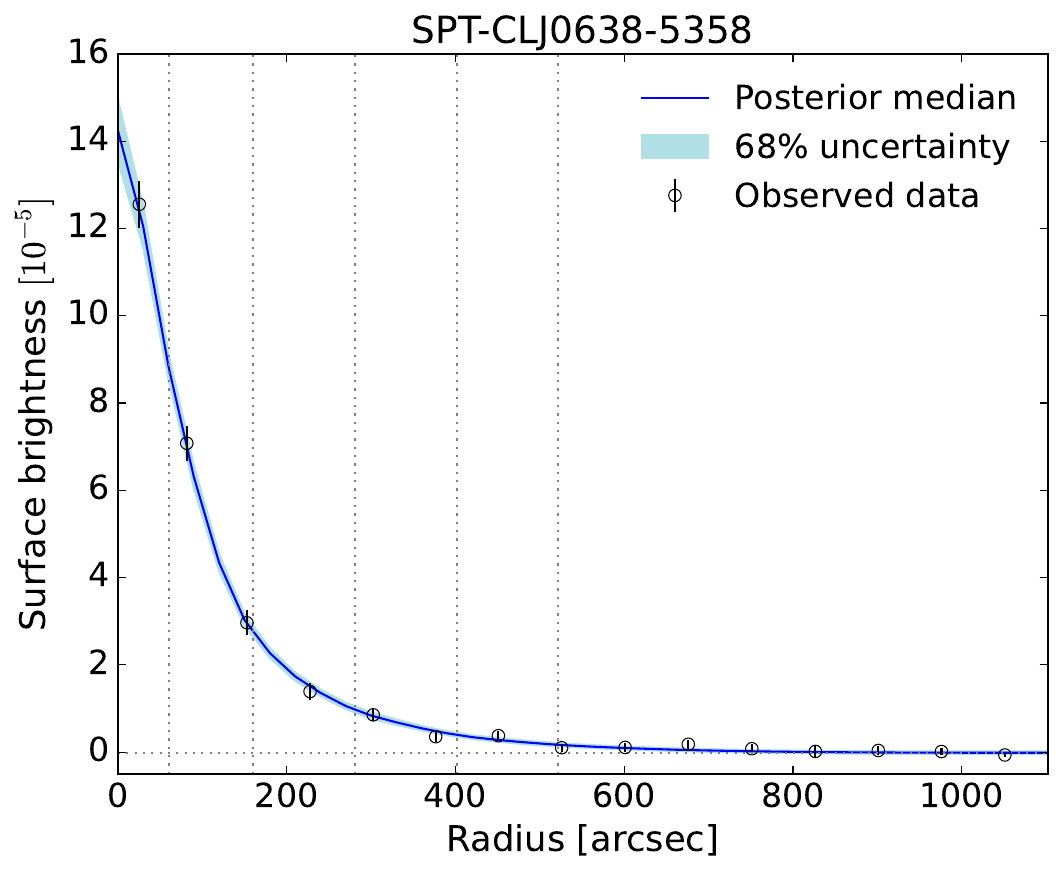} 
\includegraphics[width=.25\textwidth]{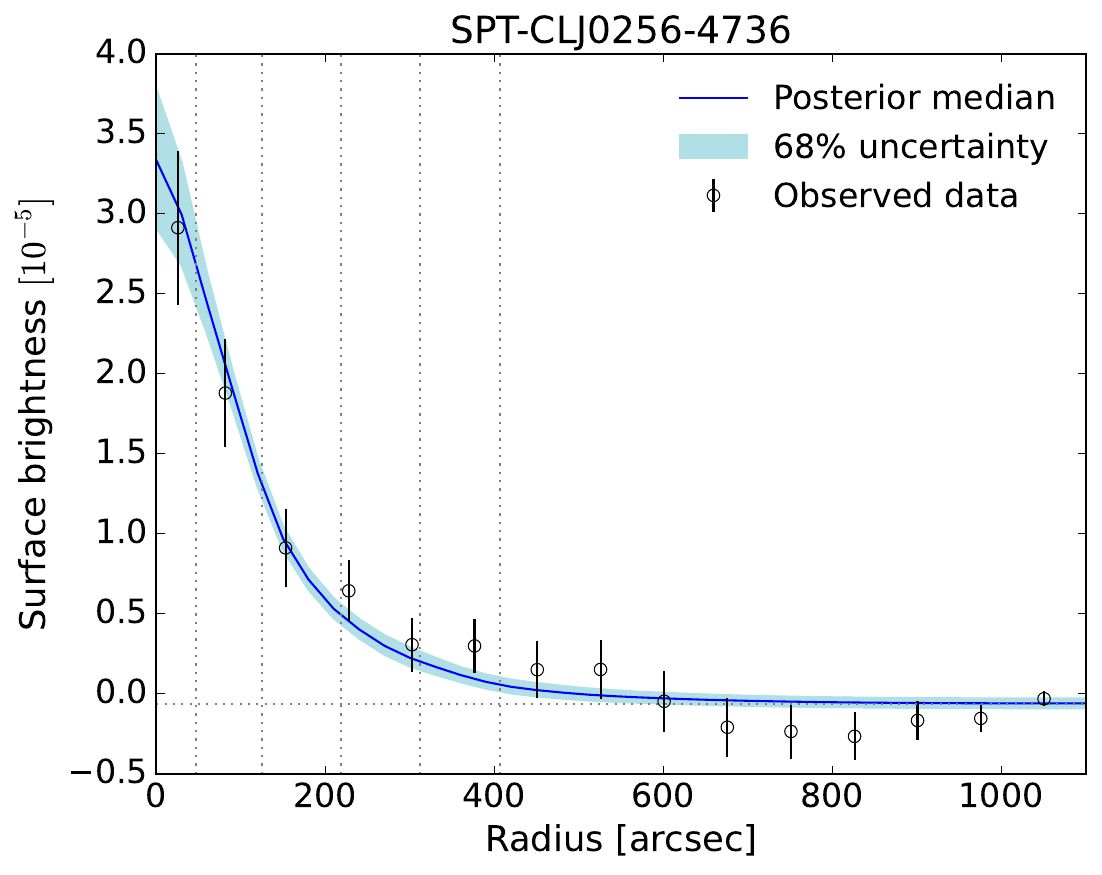}}
\centerline{\includegraphics[width=.25\textwidth]{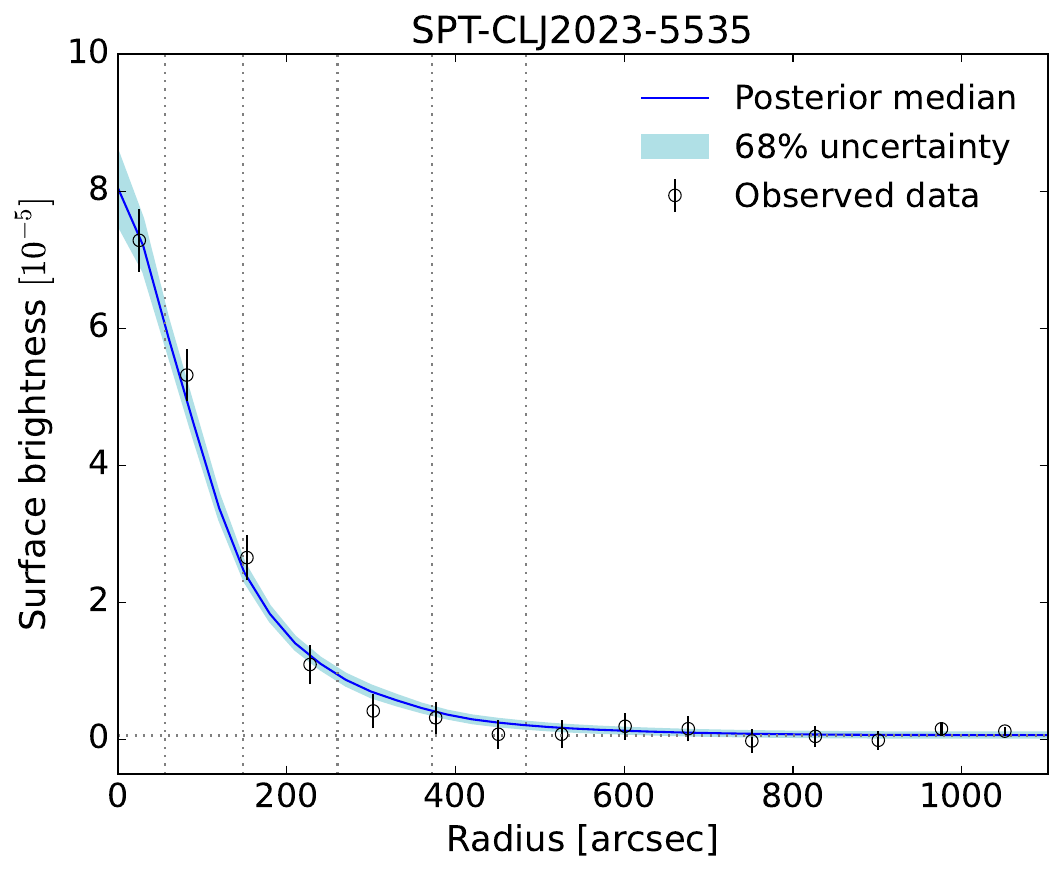} 
\includegraphics[width=.25\textwidth]{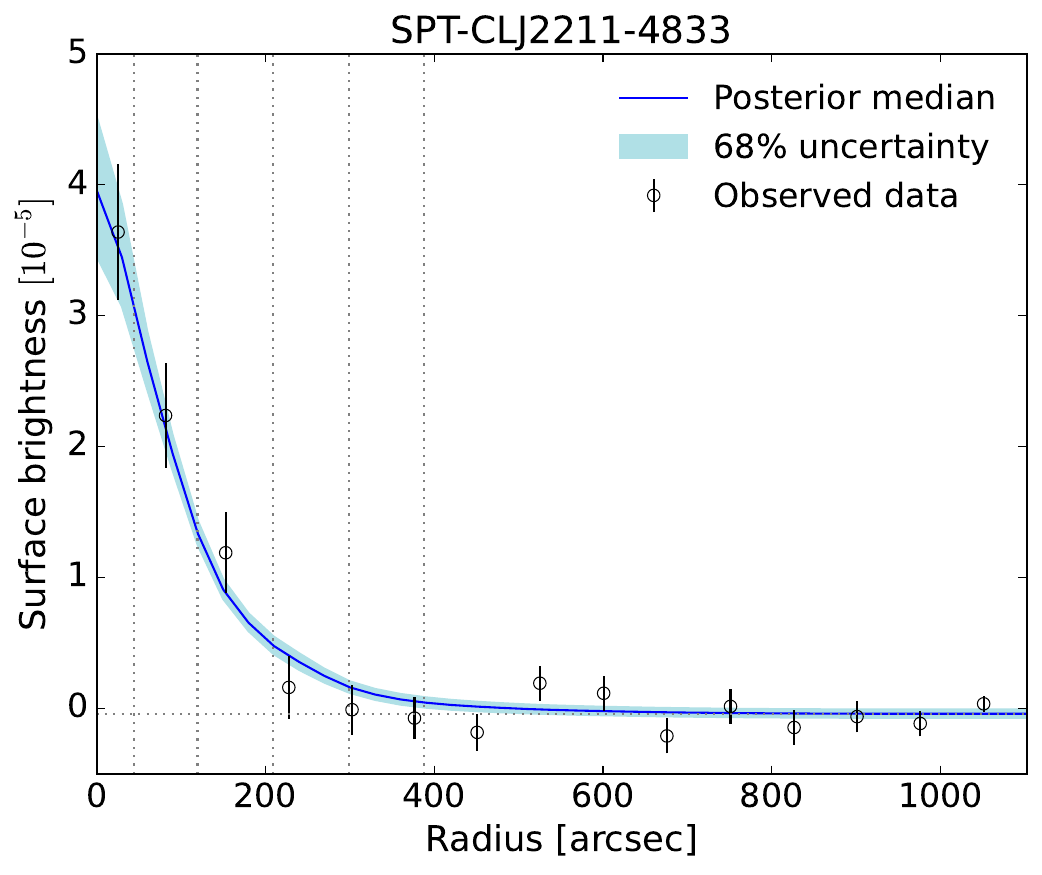} 
\includegraphics[width=.25\textwidth]{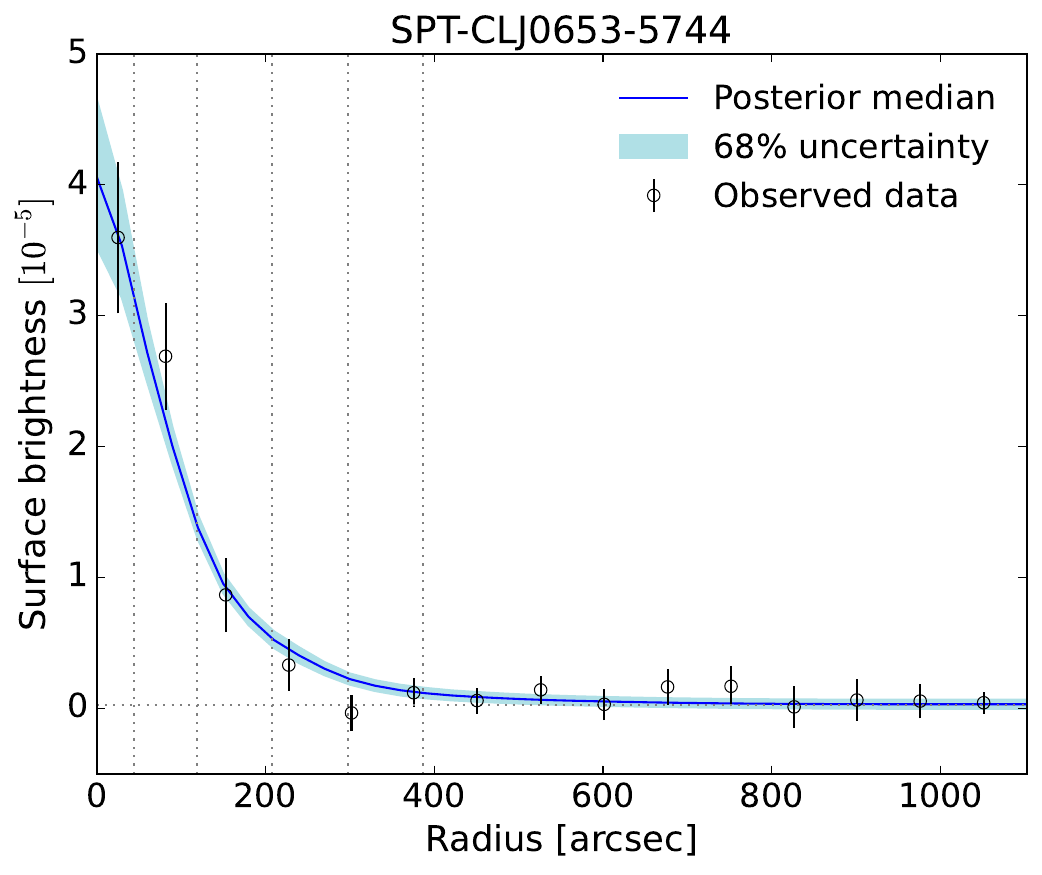} 
\includegraphics[width=.25\textwidth]{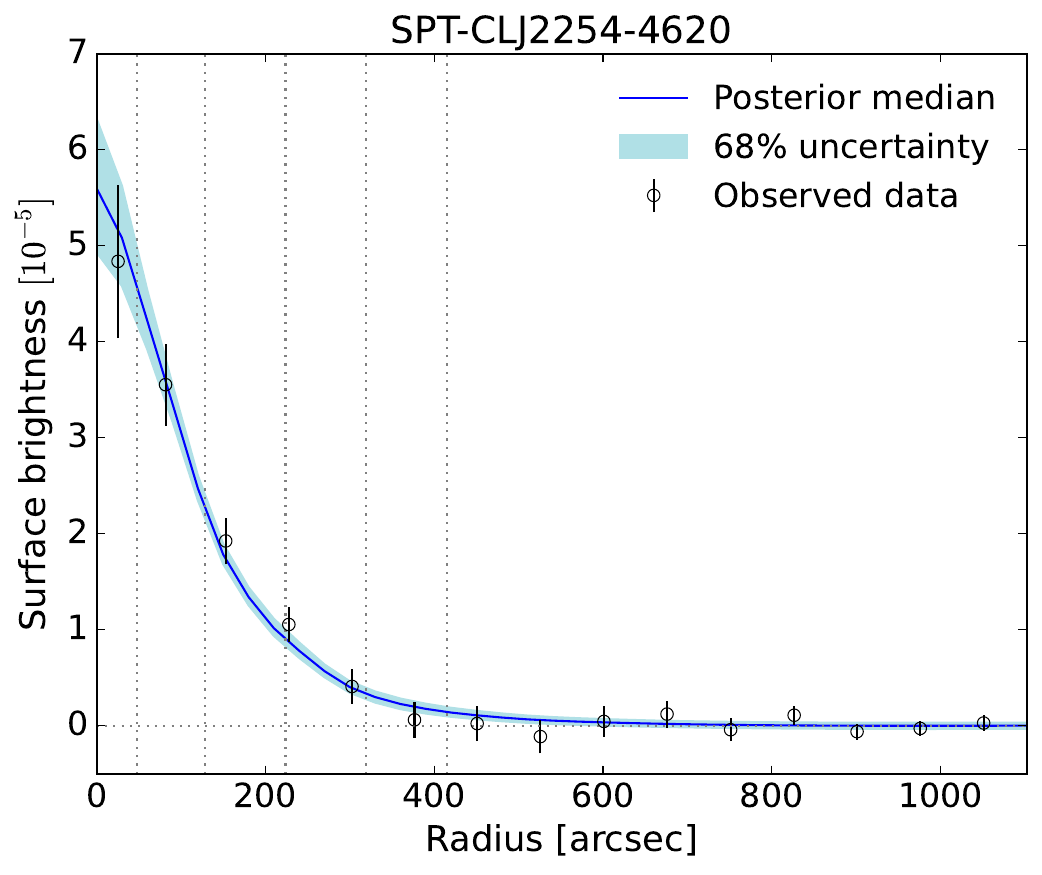}} 
\caption{Surface brightness profiles of the first 20 clusters. The ordinate is in Compton-y units. Vertical dotted gray lines show the placement of the
knots, whereas the horizontal one indicates the estimated pedestal level.}
\label{fig:Compton_prof1}
\end{figure} 

\begin{figure}[h]
\centerline{\includegraphics[width=.25\textwidth]{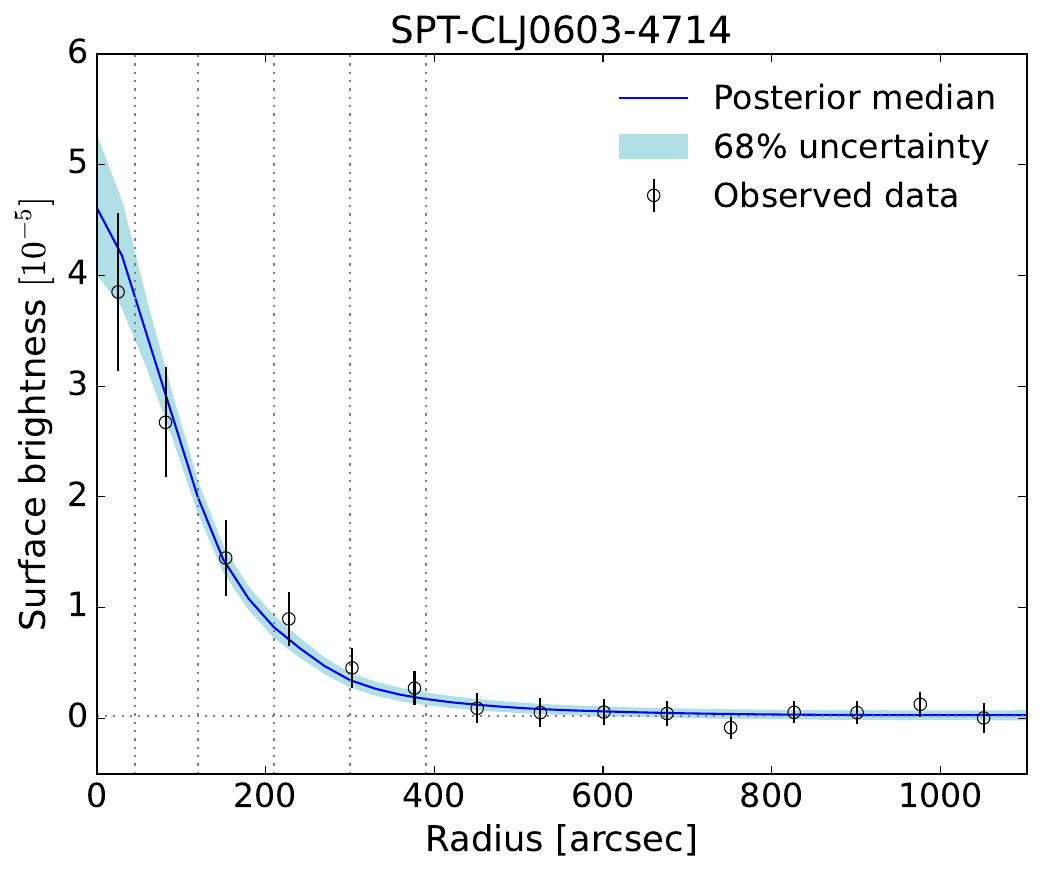} 
\includegraphics[width=.25\textwidth]{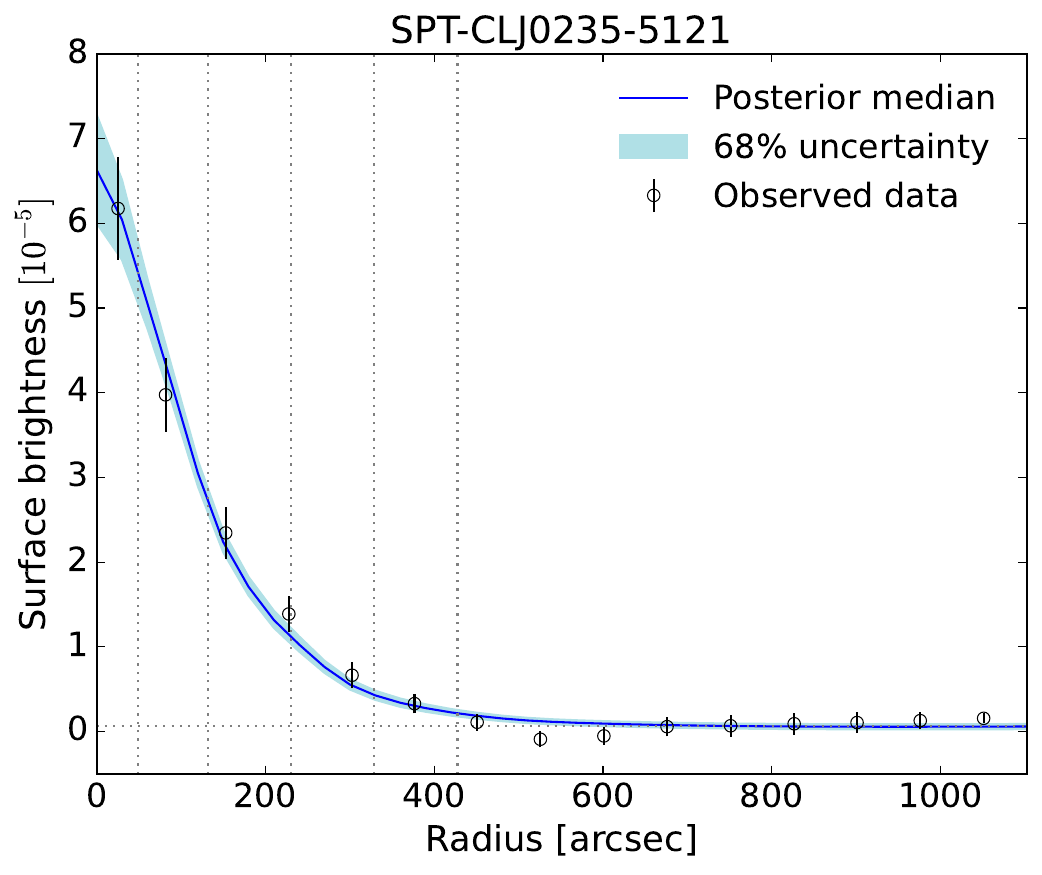} 
\includegraphics[width=.25\textwidth]{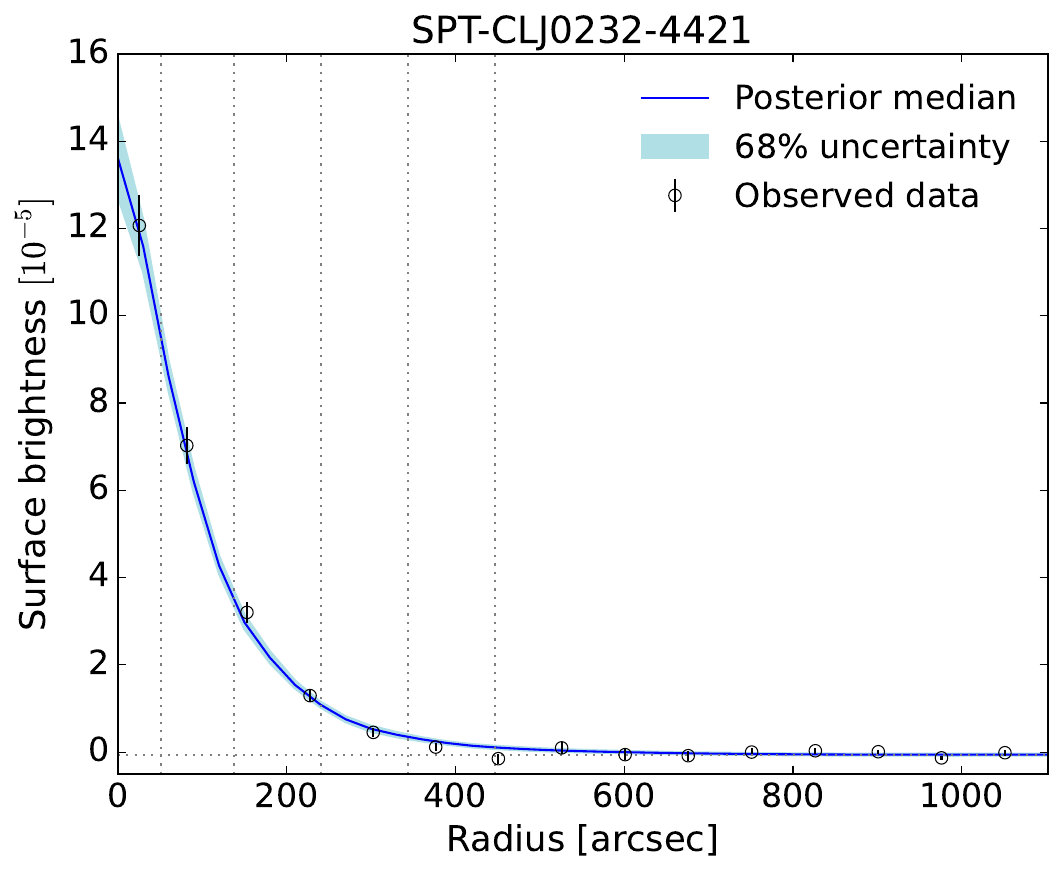} 
\includegraphics[width=.25\textwidth]{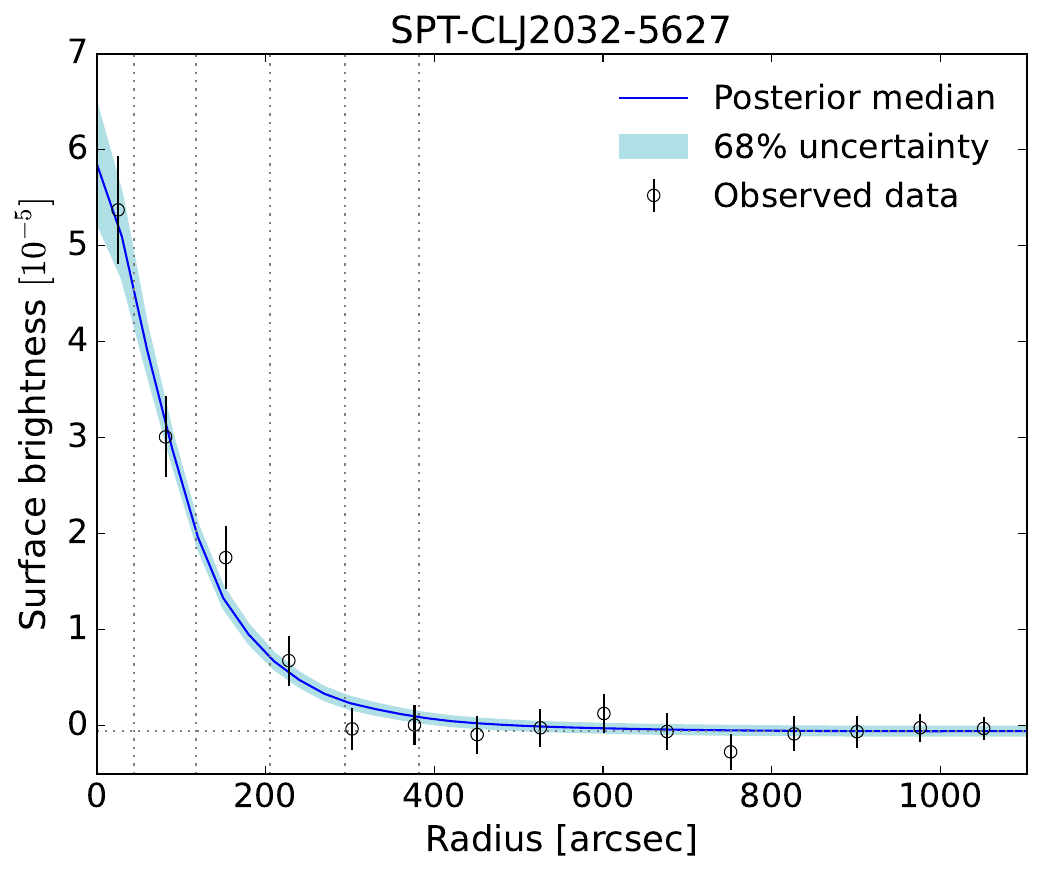}} 
\centerline{\includegraphics[width=.25\textwidth]{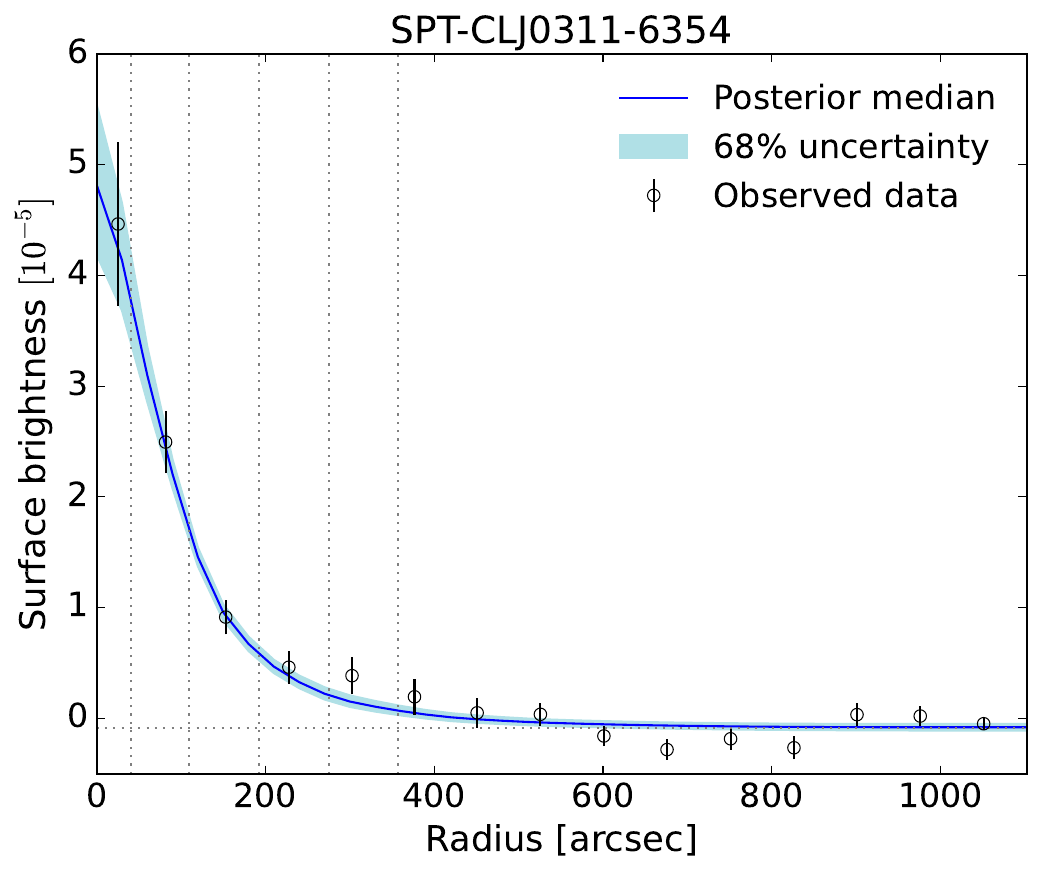} 
\includegraphics[width=.25\textwidth]{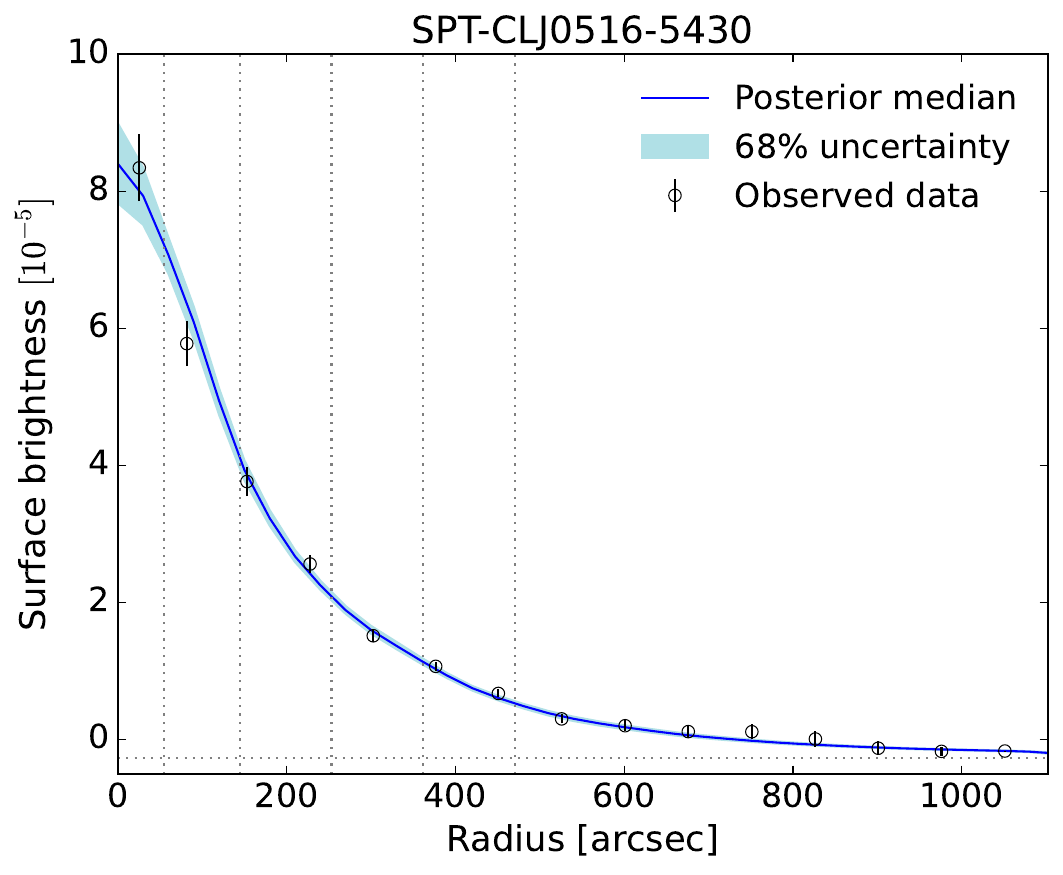} 
\includegraphics[width=.25\textwidth]{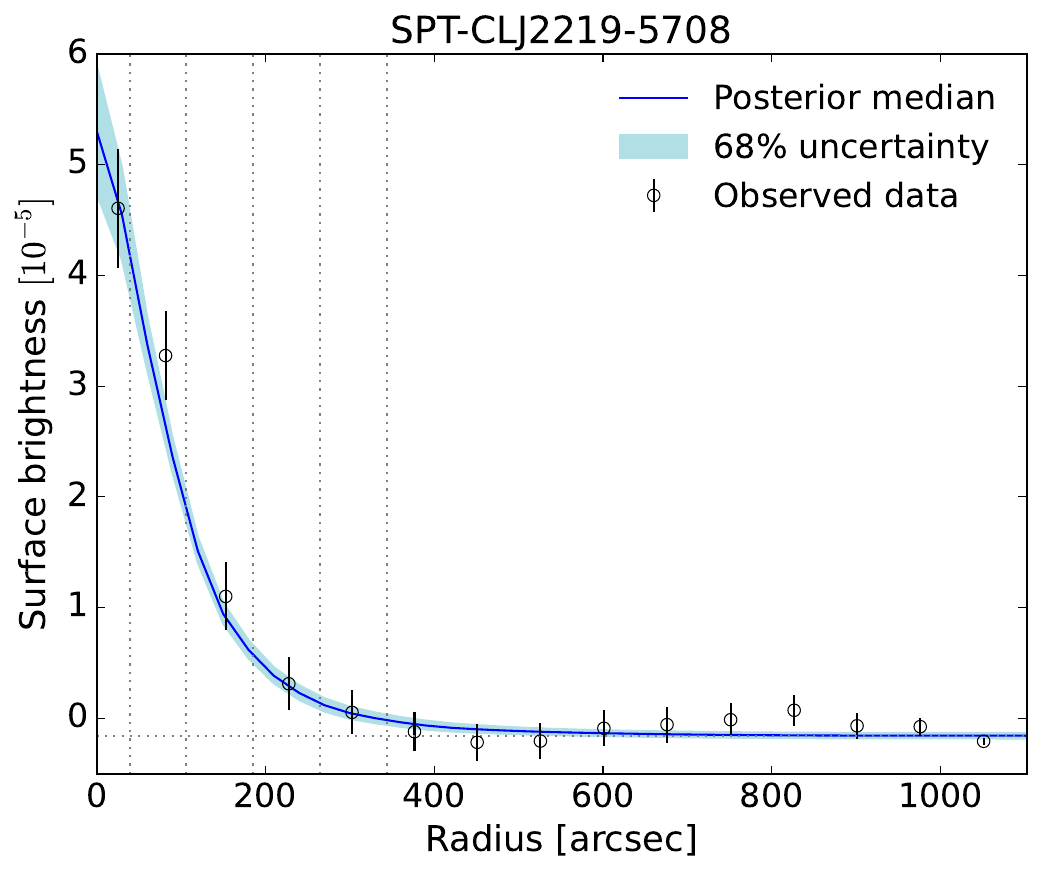} 
\includegraphics[width=.25\textwidth]{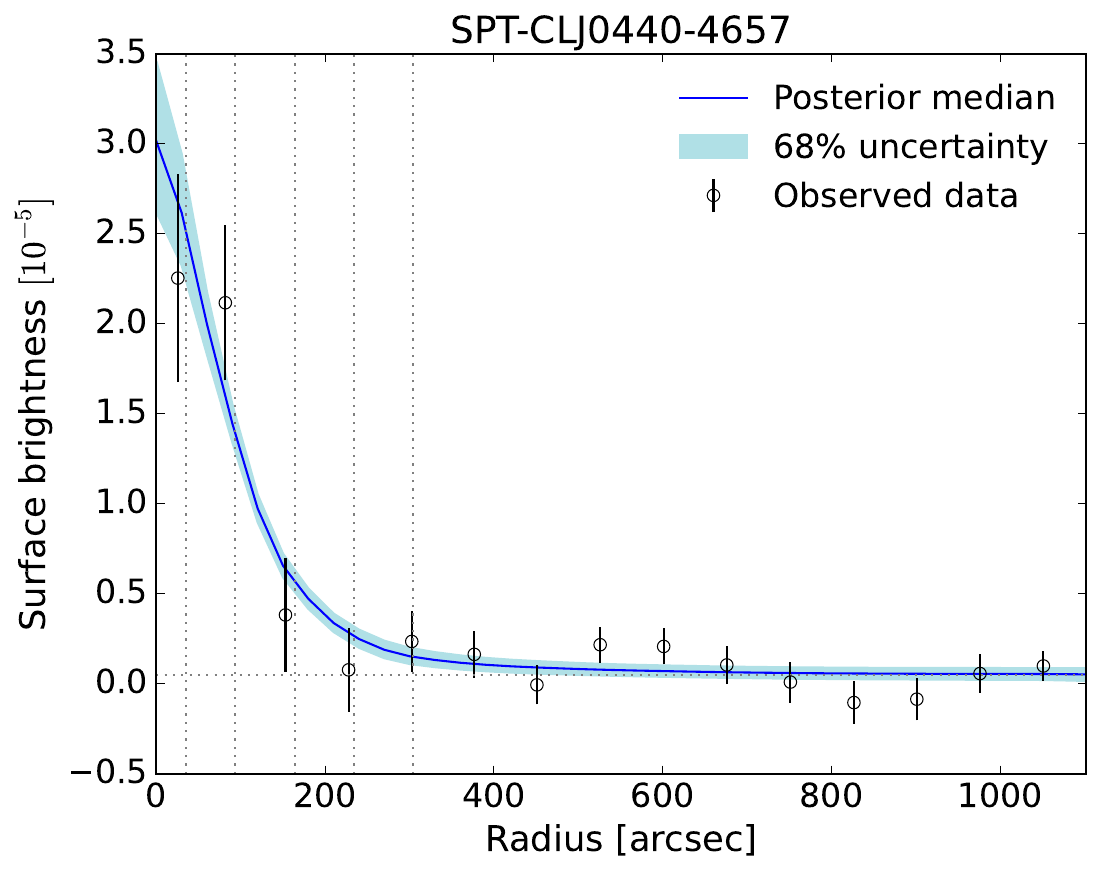}} 
\centerline{\includegraphics[width=.25\textwidth]{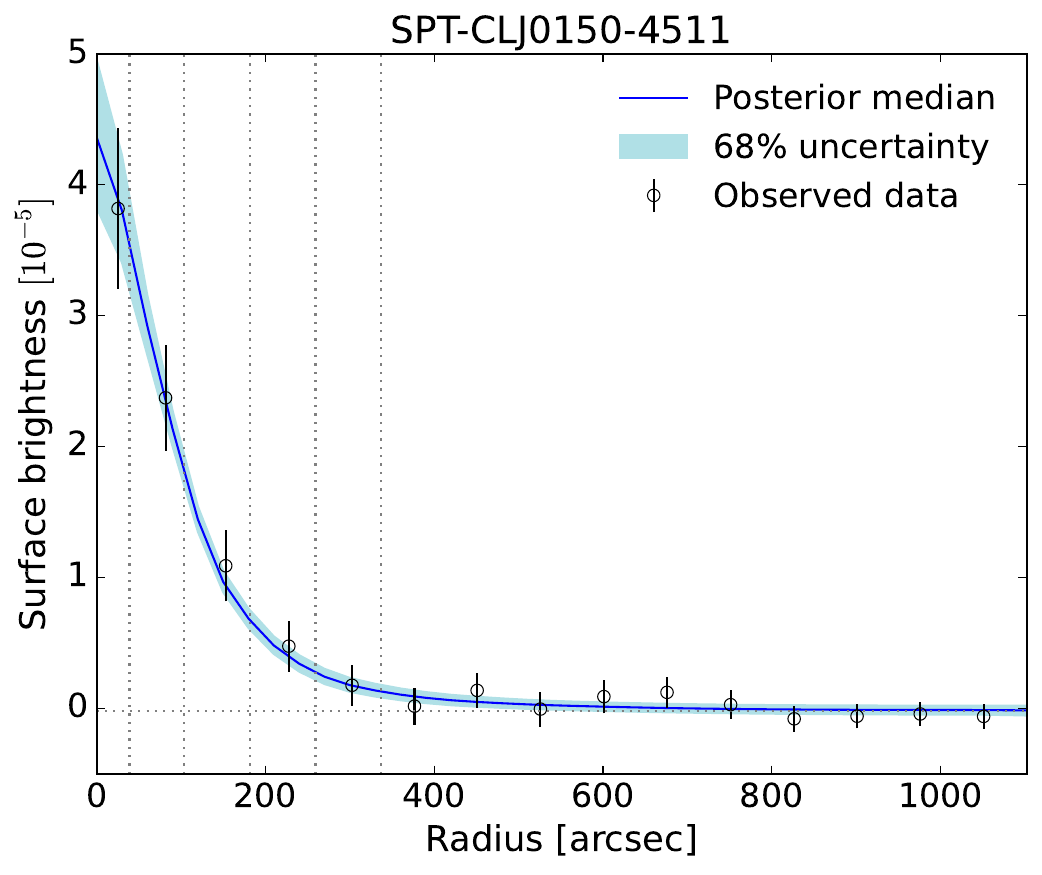} 
\includegraphics[width=.25\textwidth]{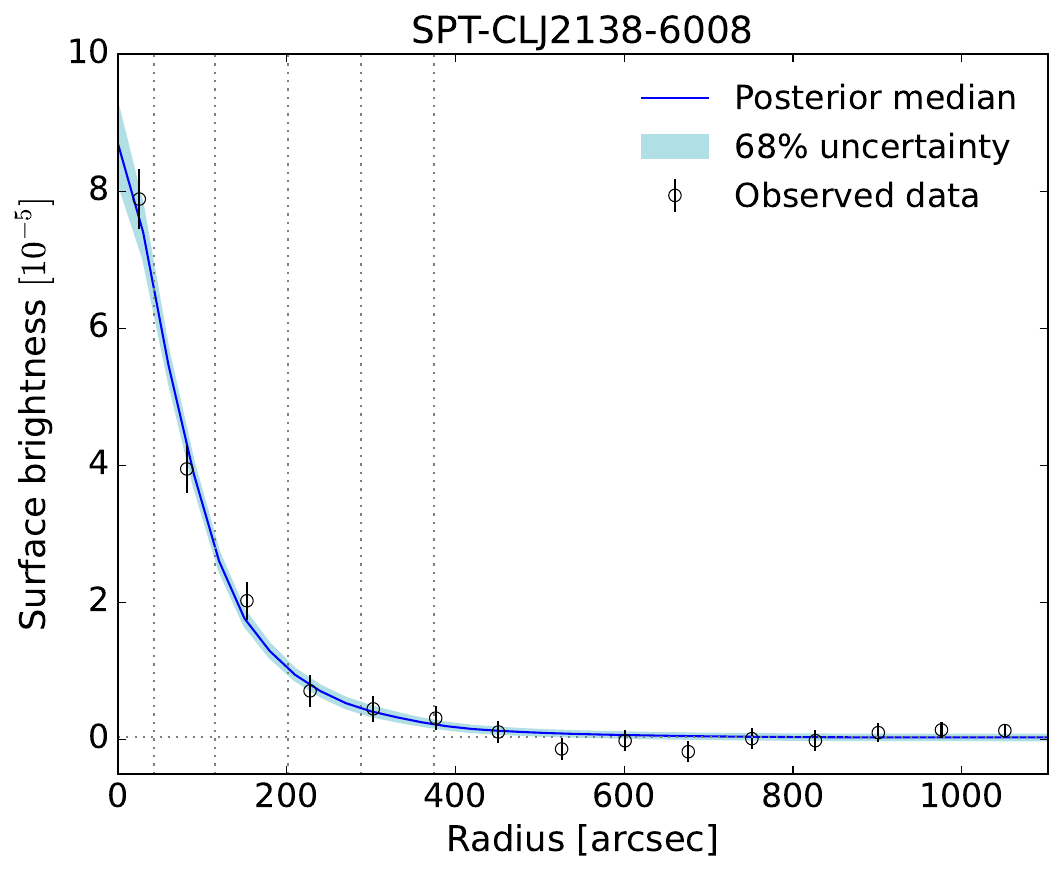} 
\includegraphics[width=.25\textwidth]{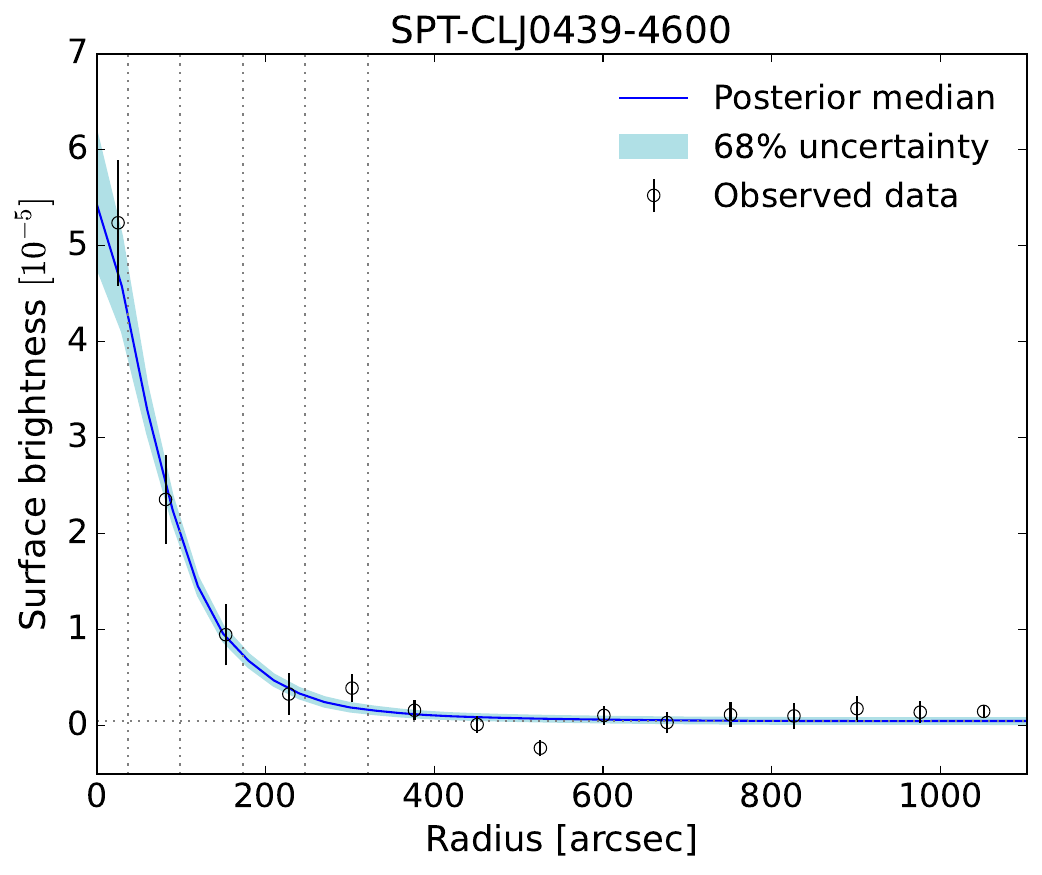} 
\includegraphics[width=.25\textwidth]{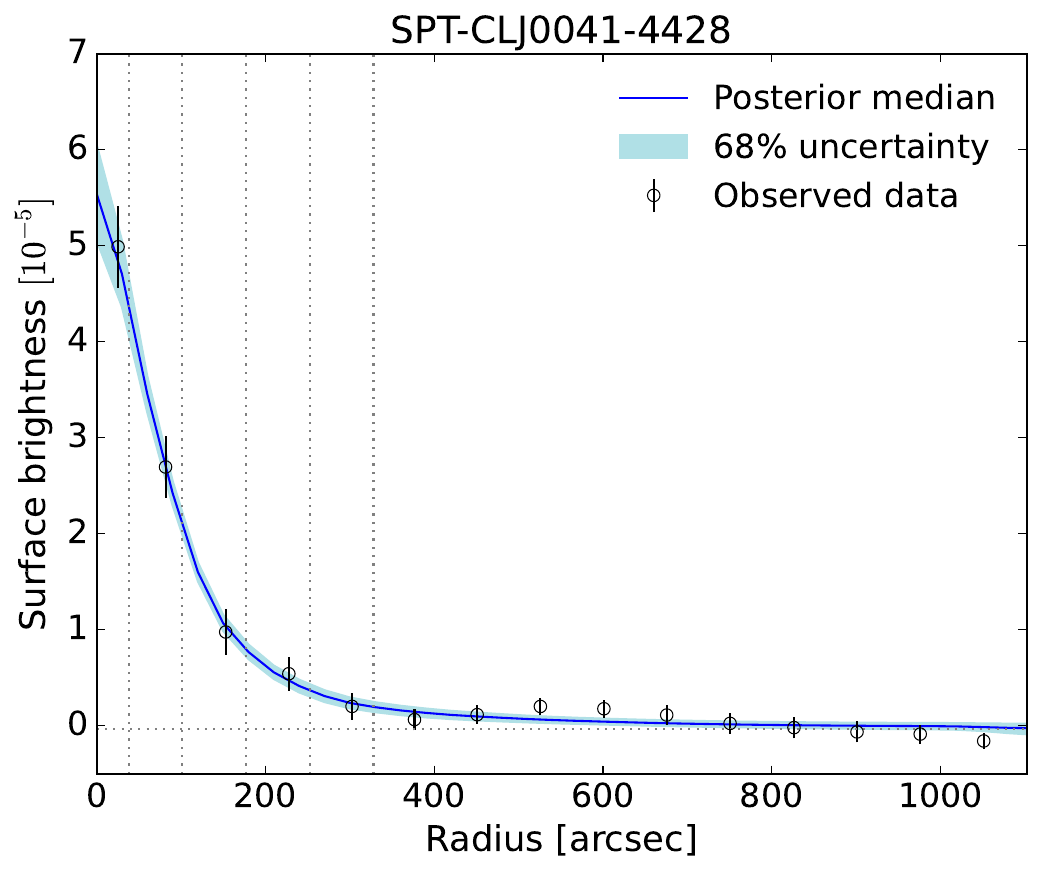}} 
\centerline{\includegraphics[width=.25\textwidth]{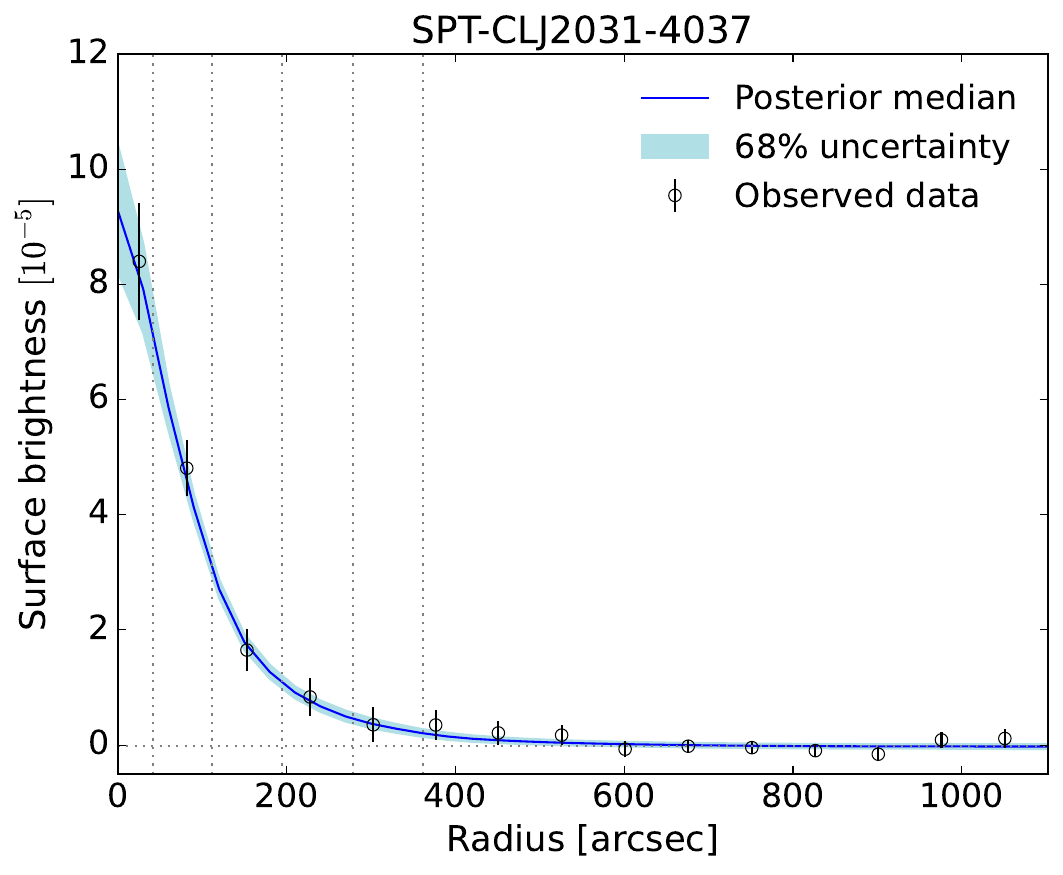} 
\includegraphics[width=.25\textwidth]{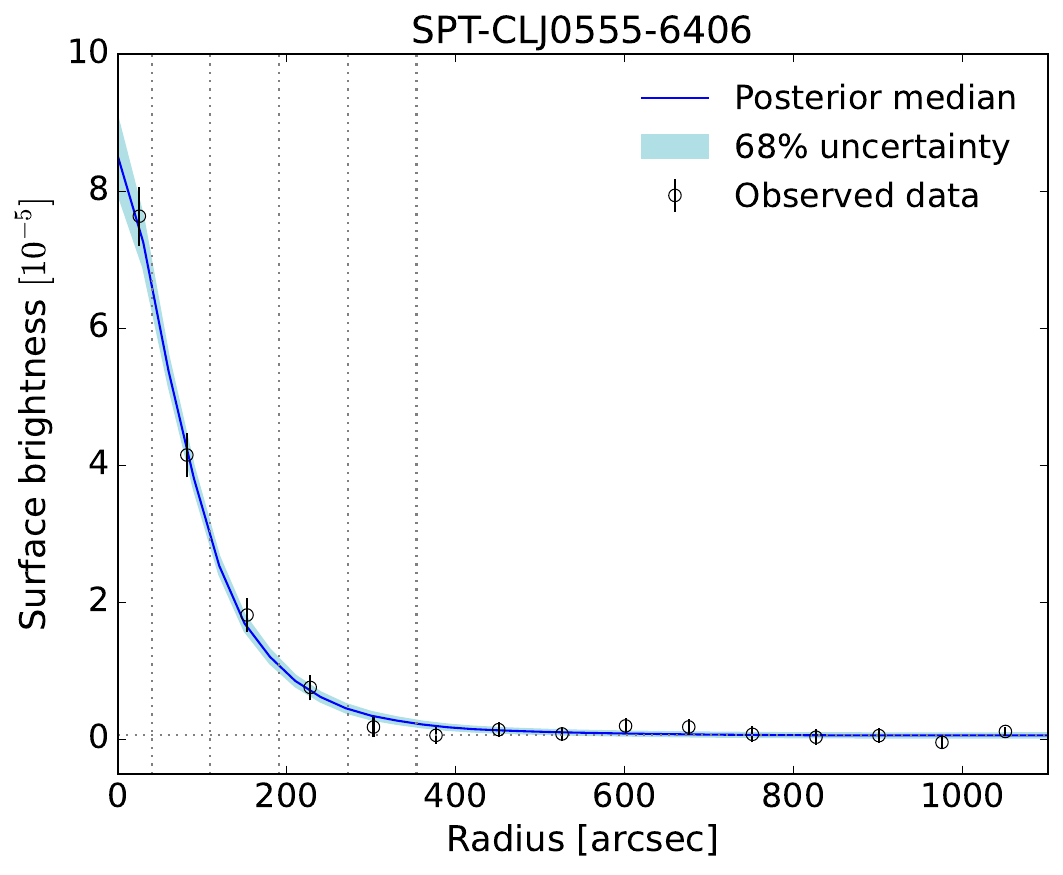} 
\includegraphics[width=.25\textwidth]{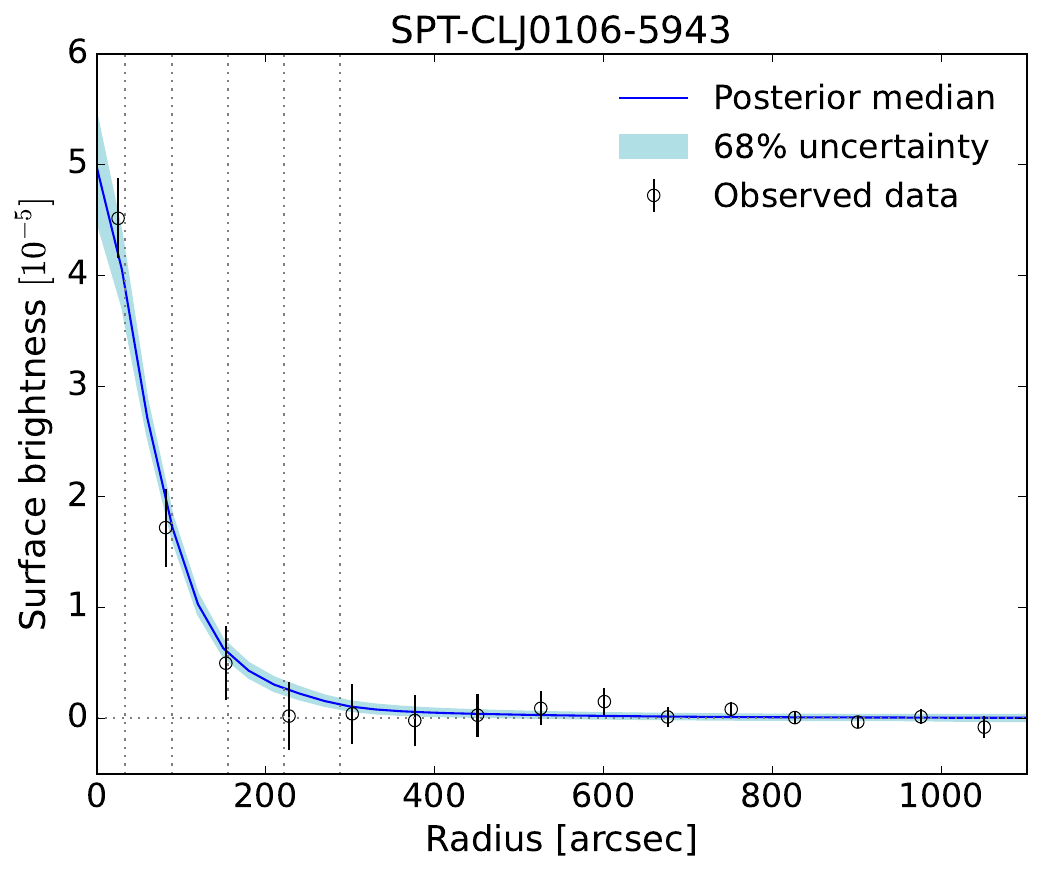} 
\includegraphics[width=.25\textwidth]{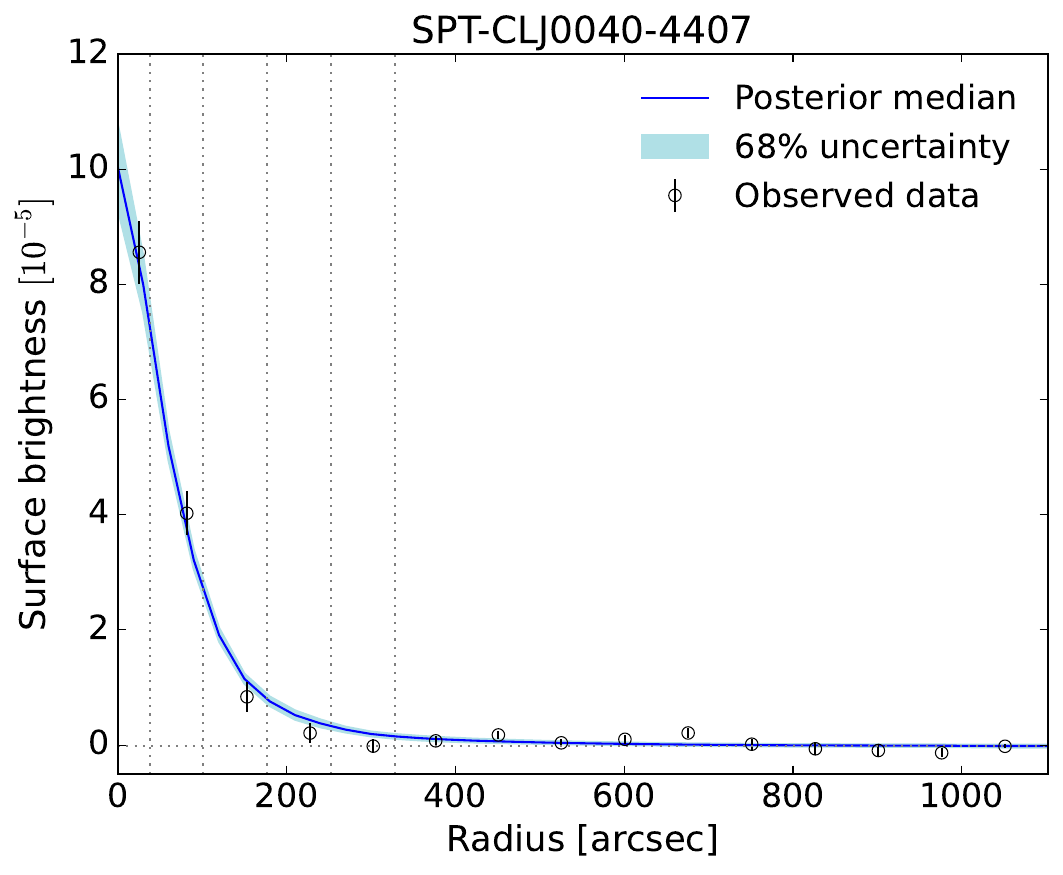}}
\centerline{\includegraphics[width=.25\textwidth]{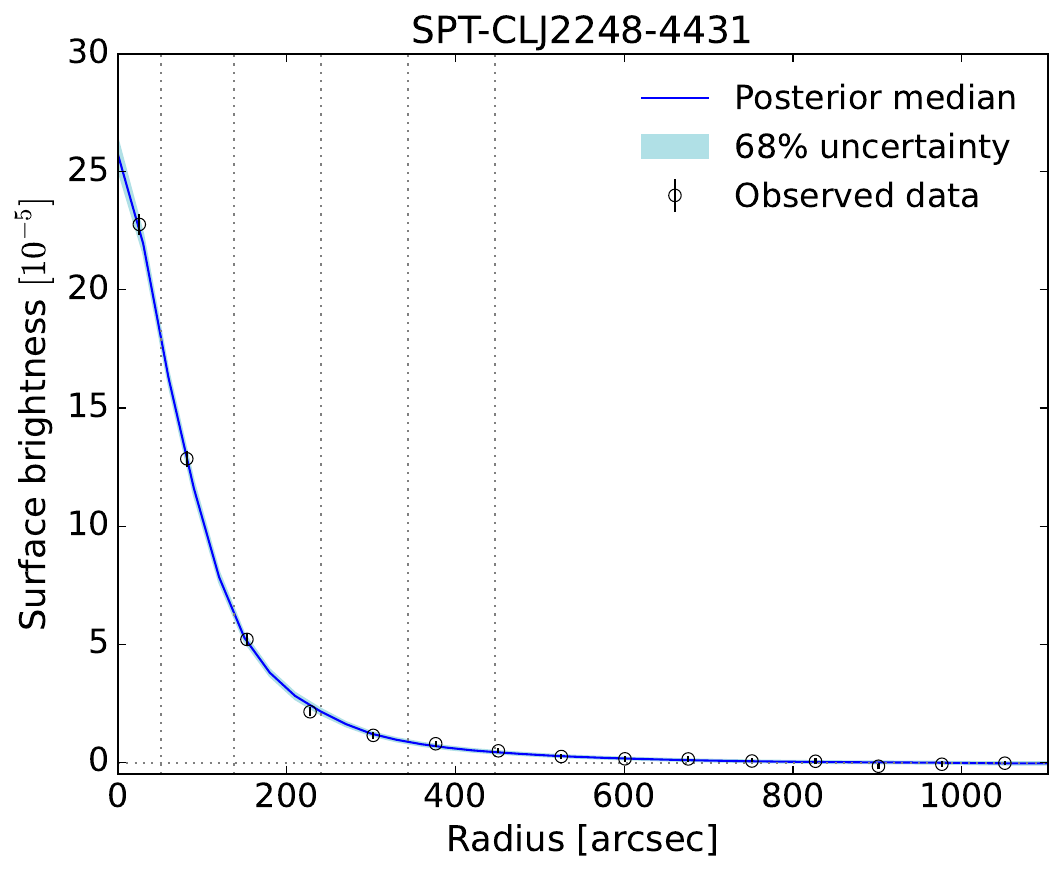} 
\includegraphics[width=.25\textwidth]{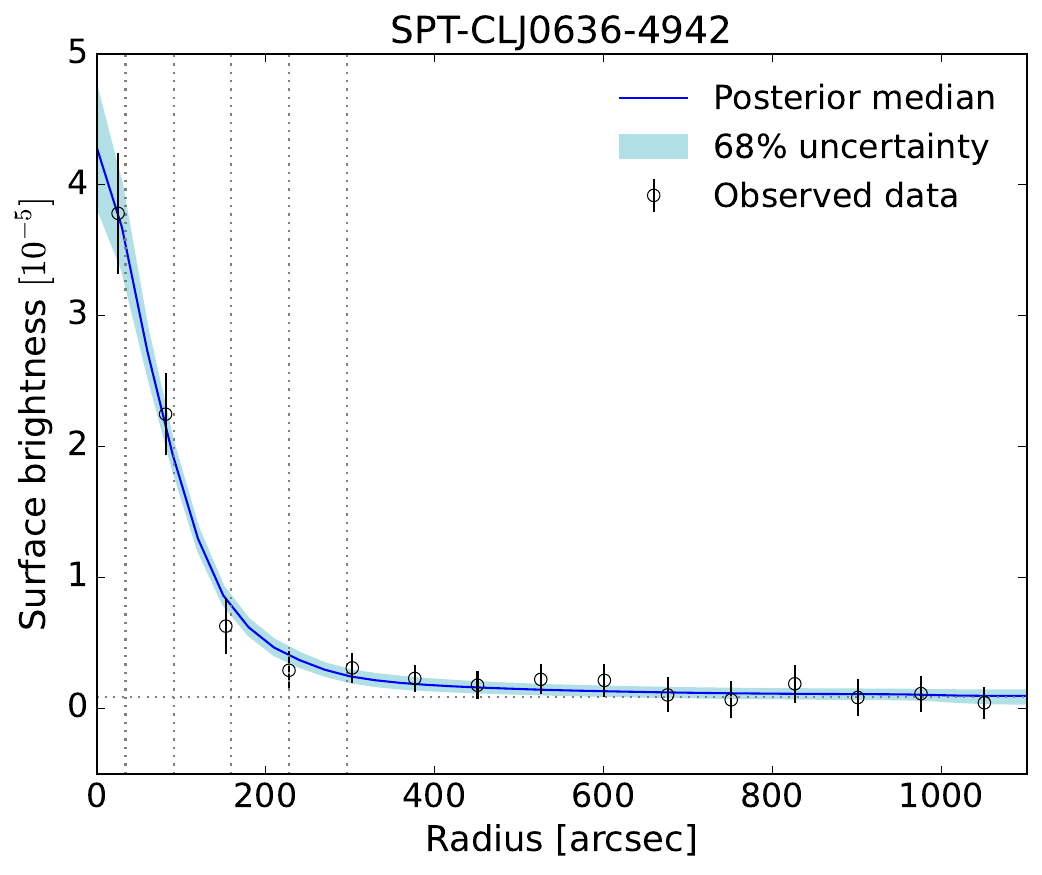} 
\includegraphics[width=.25\textwidth]{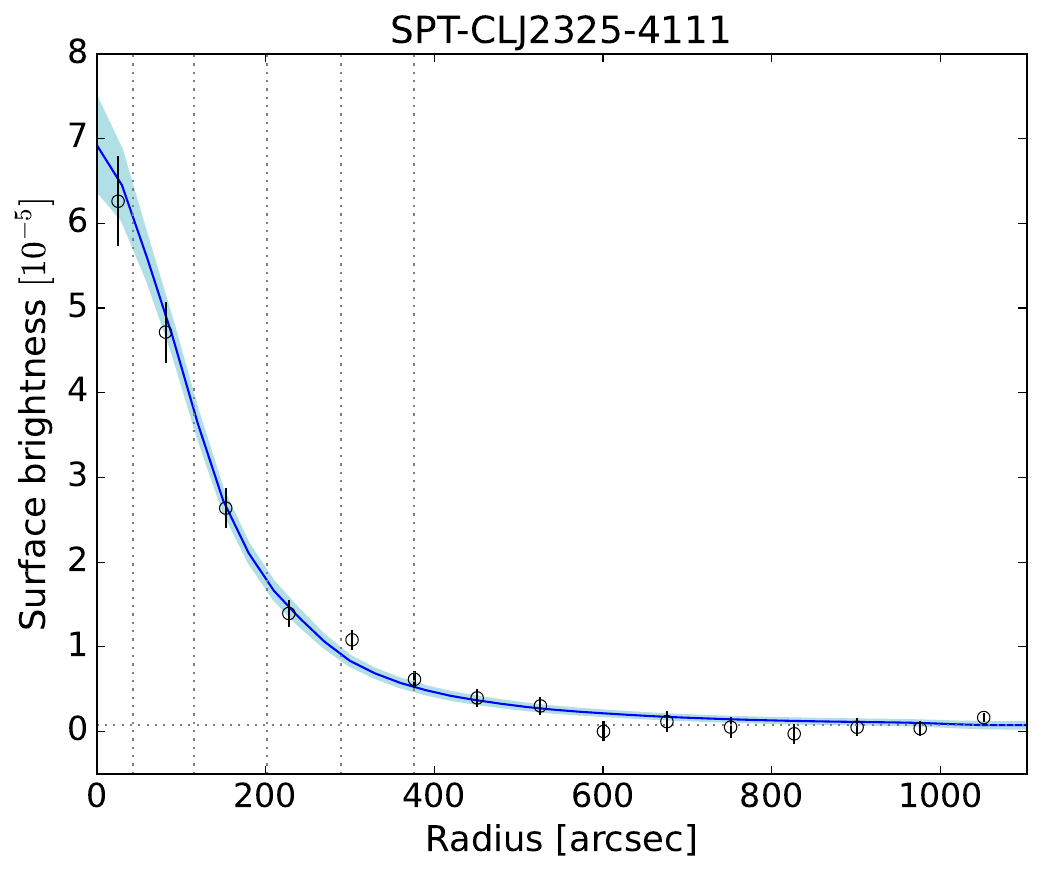} 
\includegraphics[width=.25\textwidth]{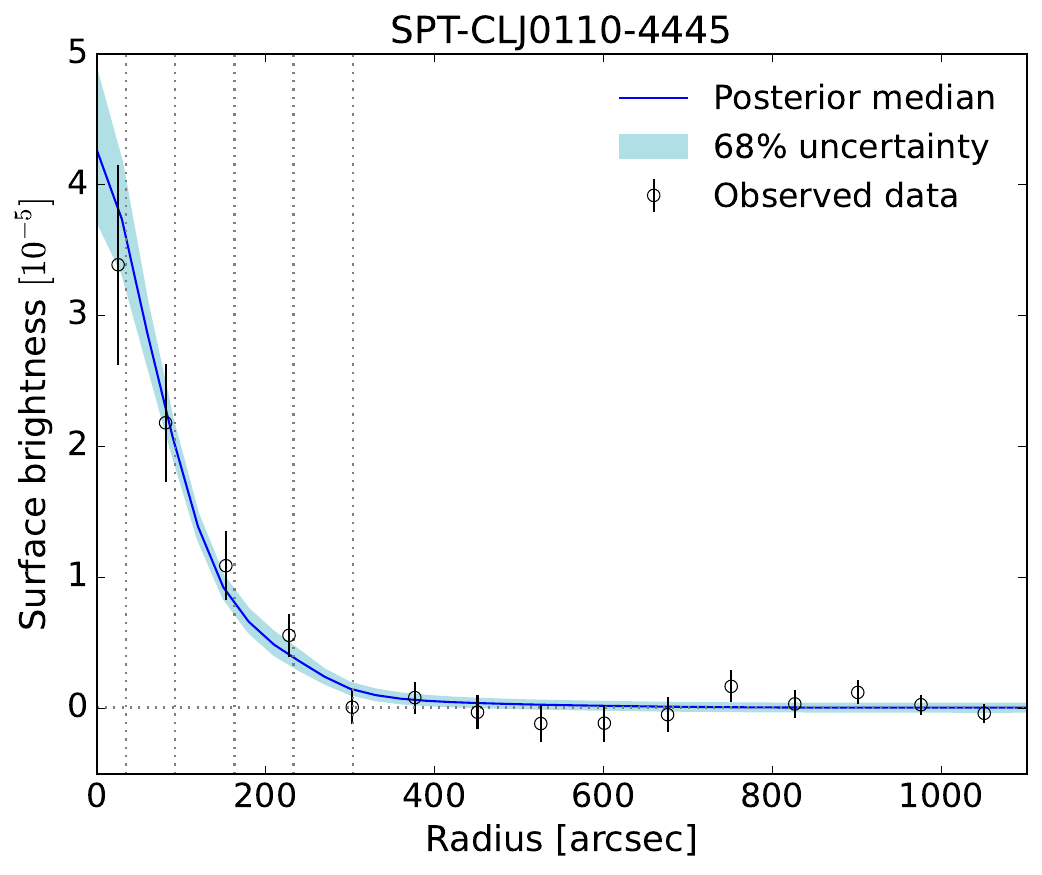}} 
\caption{As previous figure, but for the next 20 clusters }
\label{fig:Compton_prof2}
\end{figure} 

\begin{figure}[h]
\centerline{\includegraphics[width=.25\textwidth]{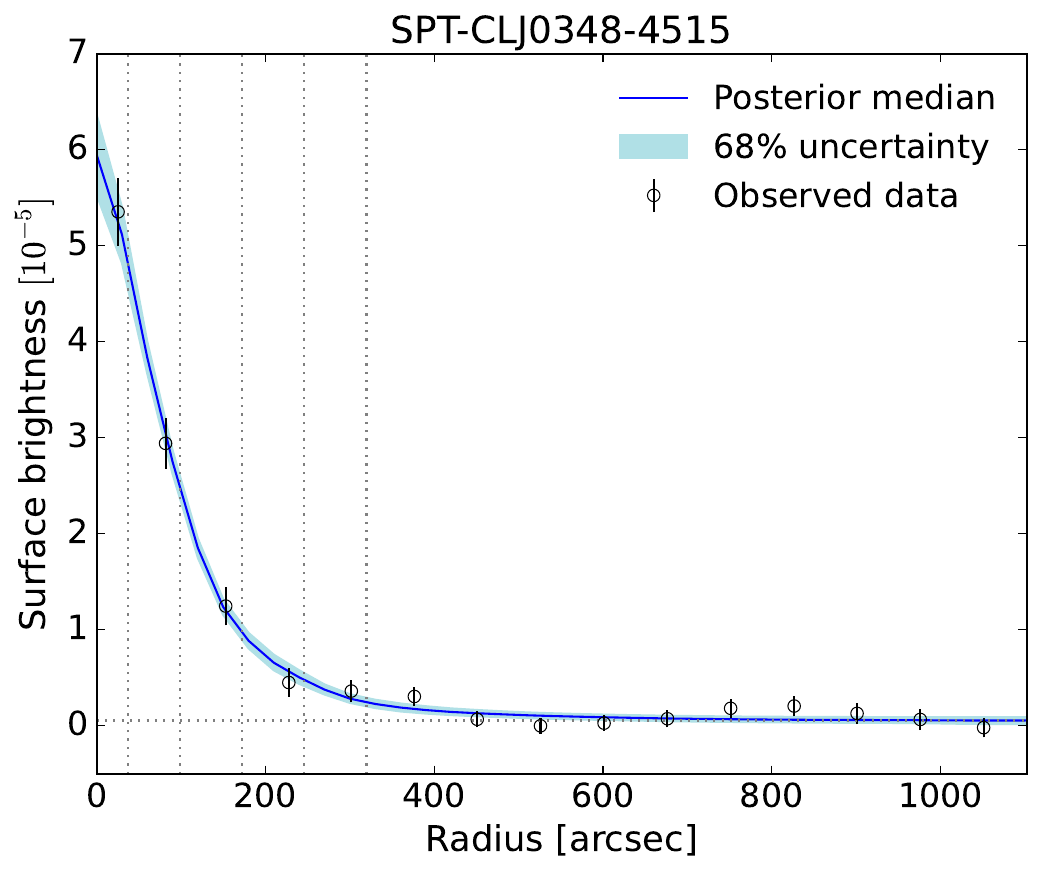} 
\includegraphics[width=.25\textwidth]{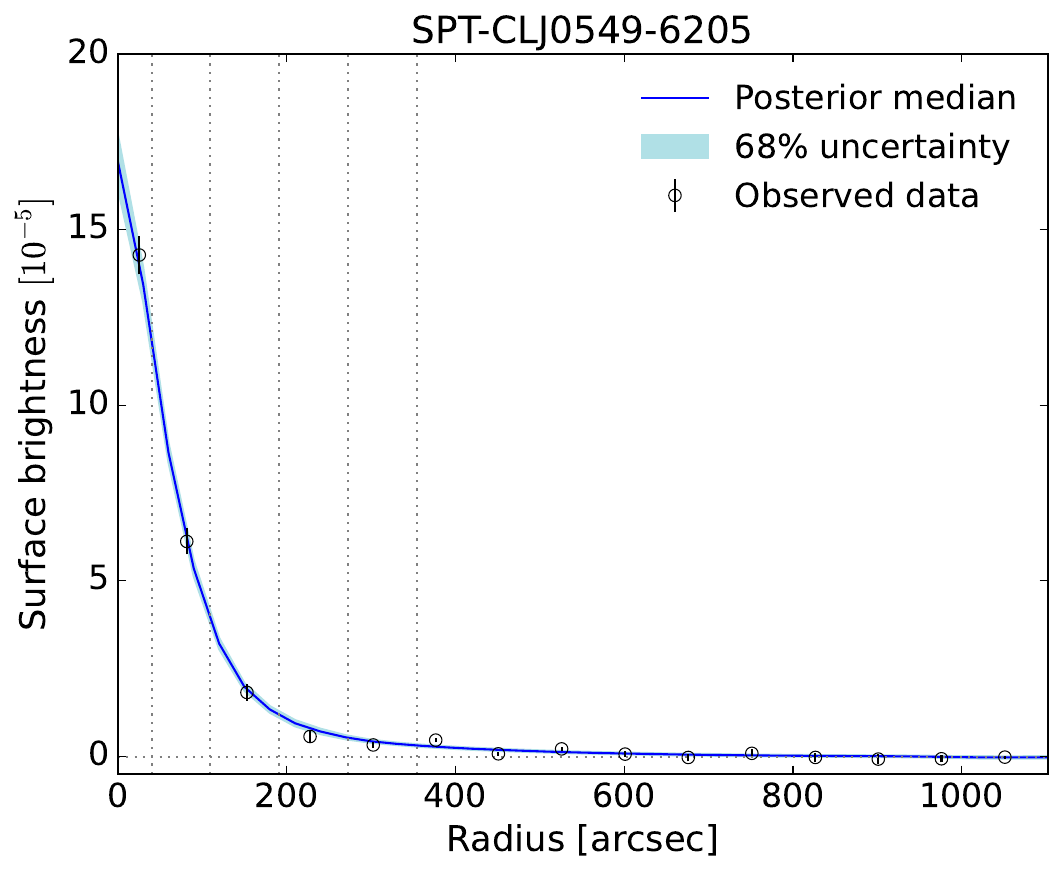} 
\includegraphics[width=.25\textwidth]{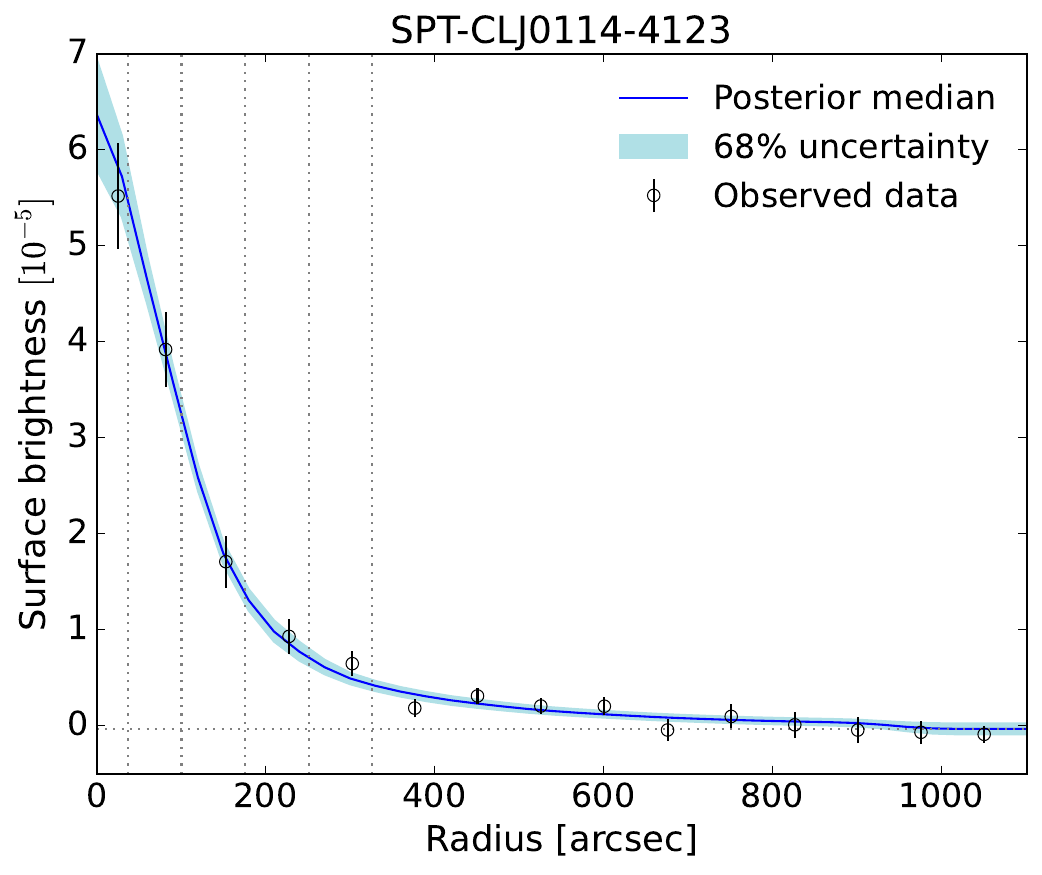} 
\includegraphics[width=.25\textwidth]{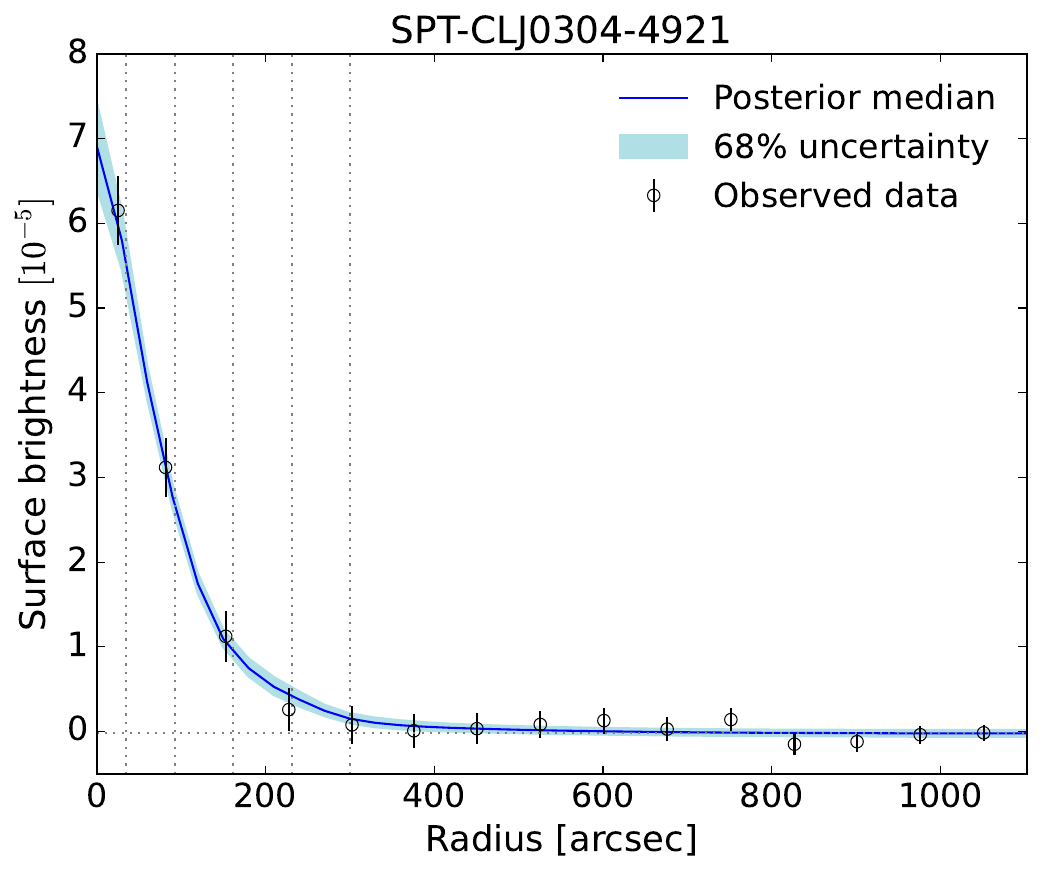}} 
\centerline{\includegraphics[width=.25\textwidth]{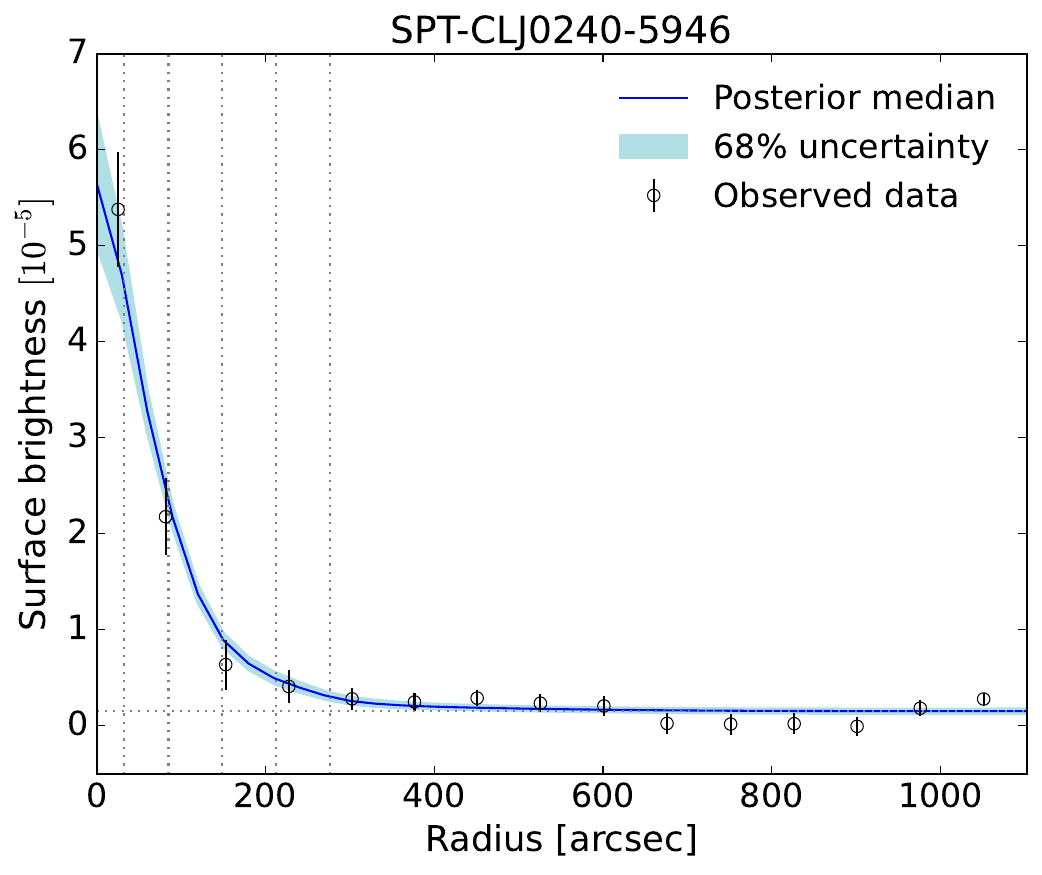} 
\includegraphics[width=.25\textwidth]{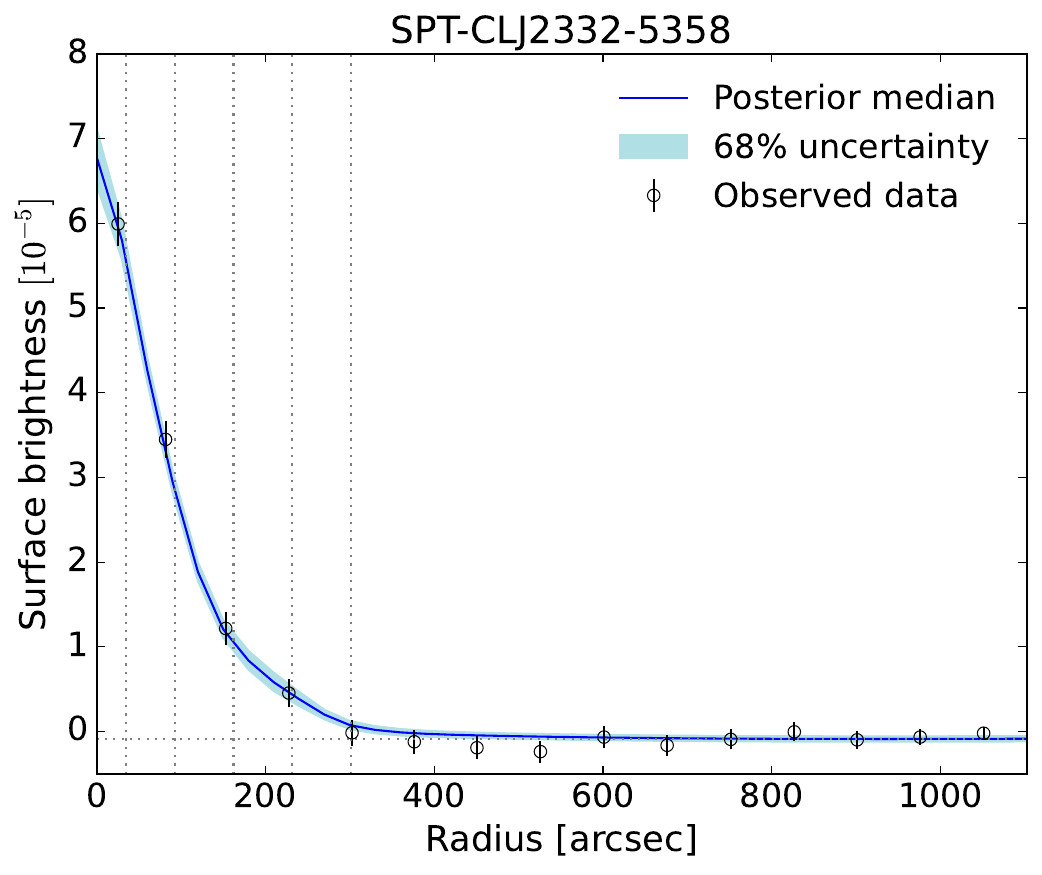} 
\includegraphics[width=.25\textwidth]{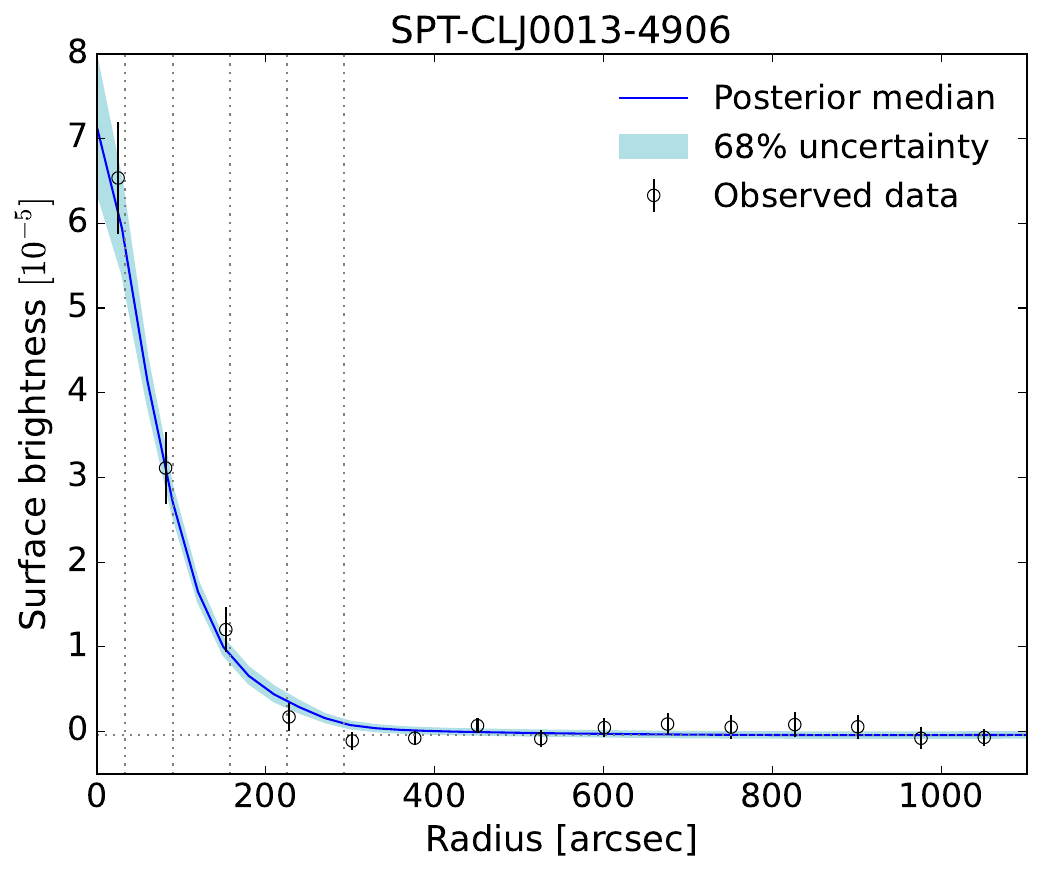} 
\includegraphics[width=.25\textwidth]{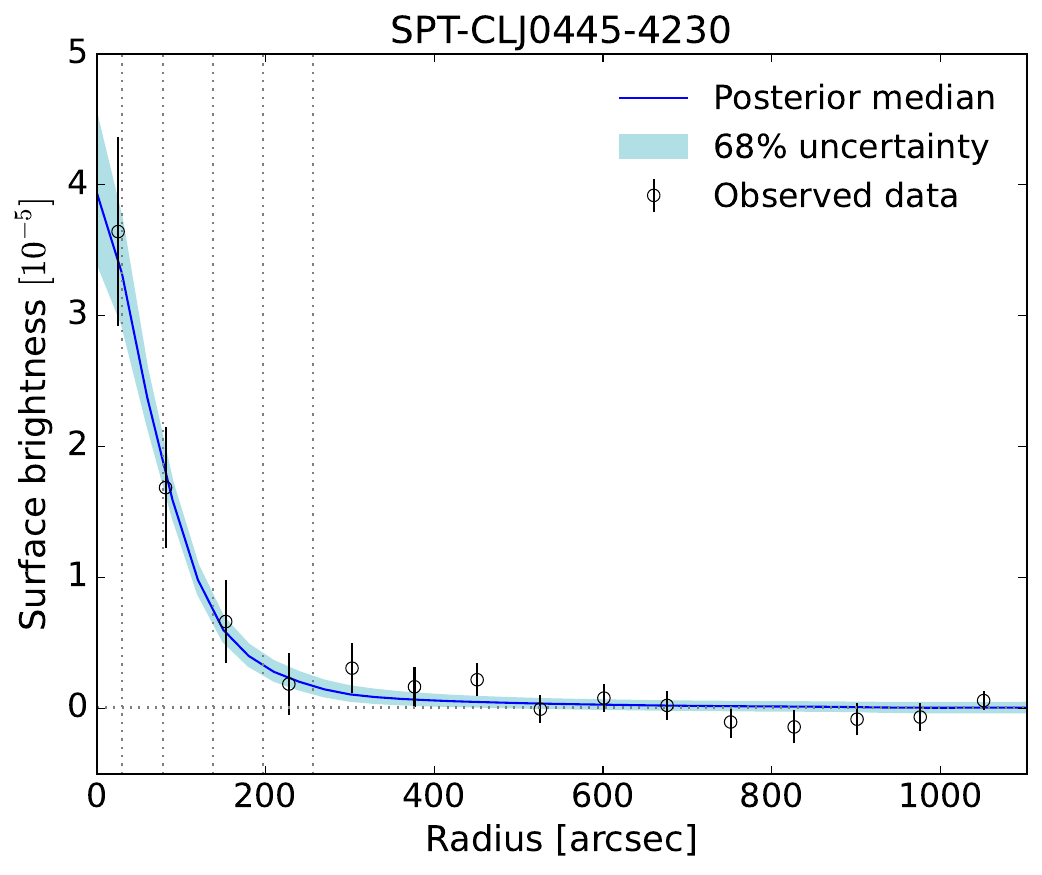}} 
\centerline{\includegraphics[width=.25\textwidth]{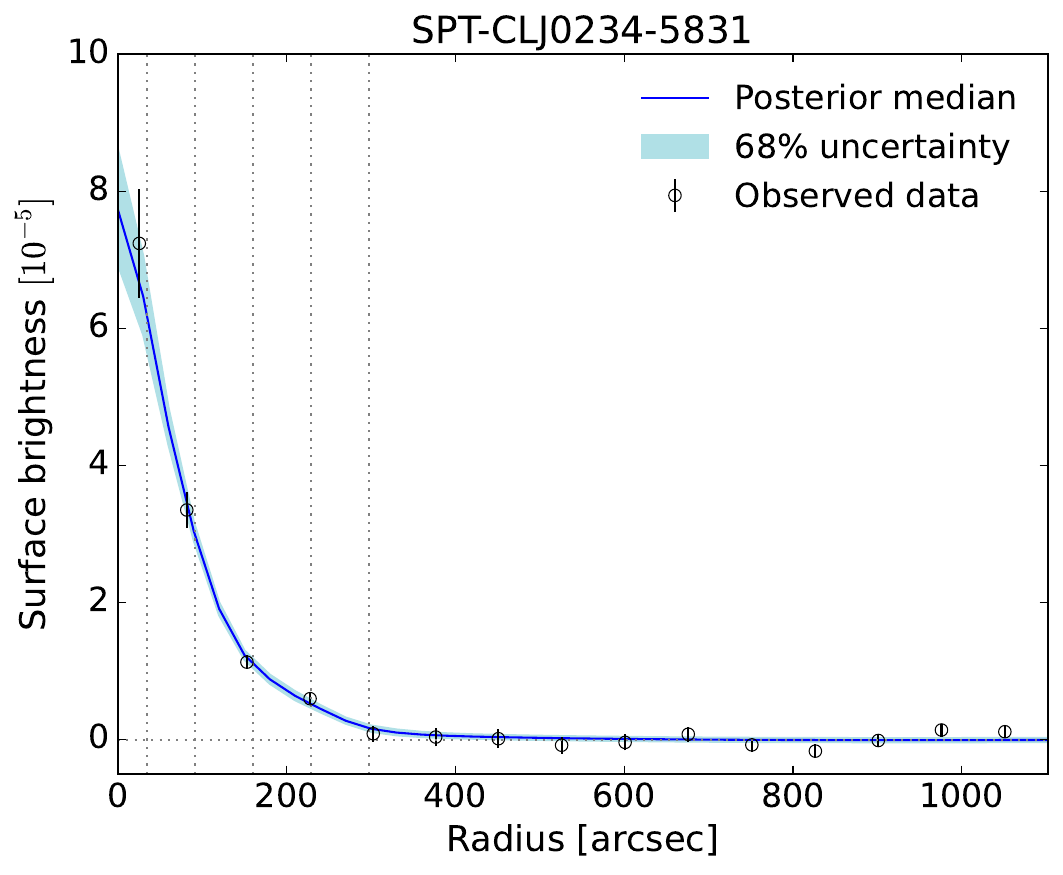} 
\includegraphics[width=.25\textwidth]{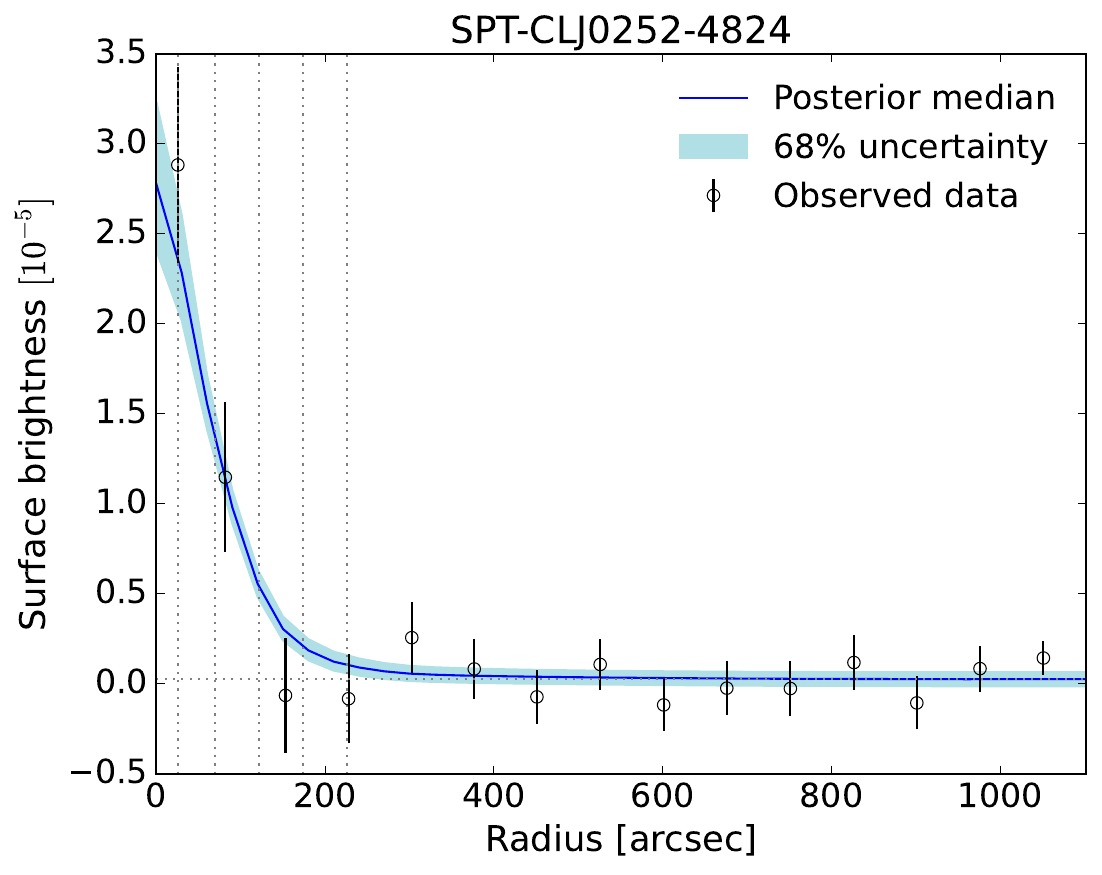} 
\includegraphics[width=.25\textwidth]{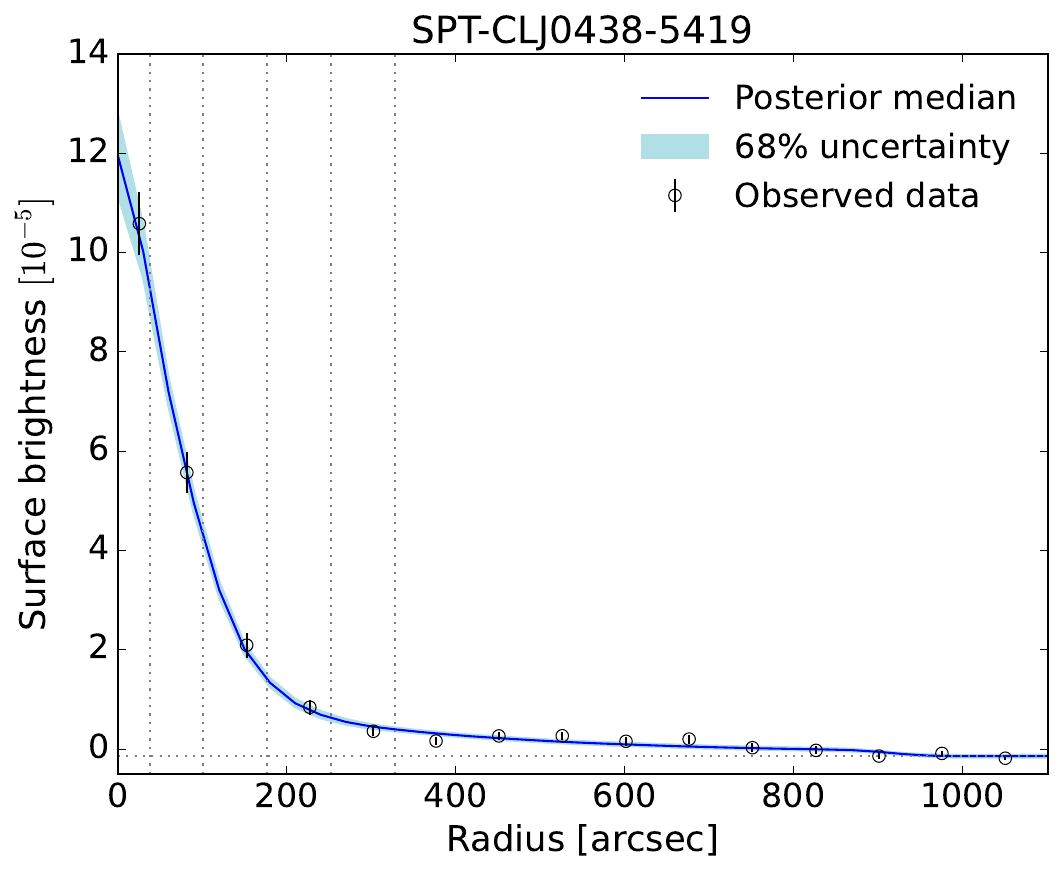} 
\includegraphics[width=.25\textwidth]{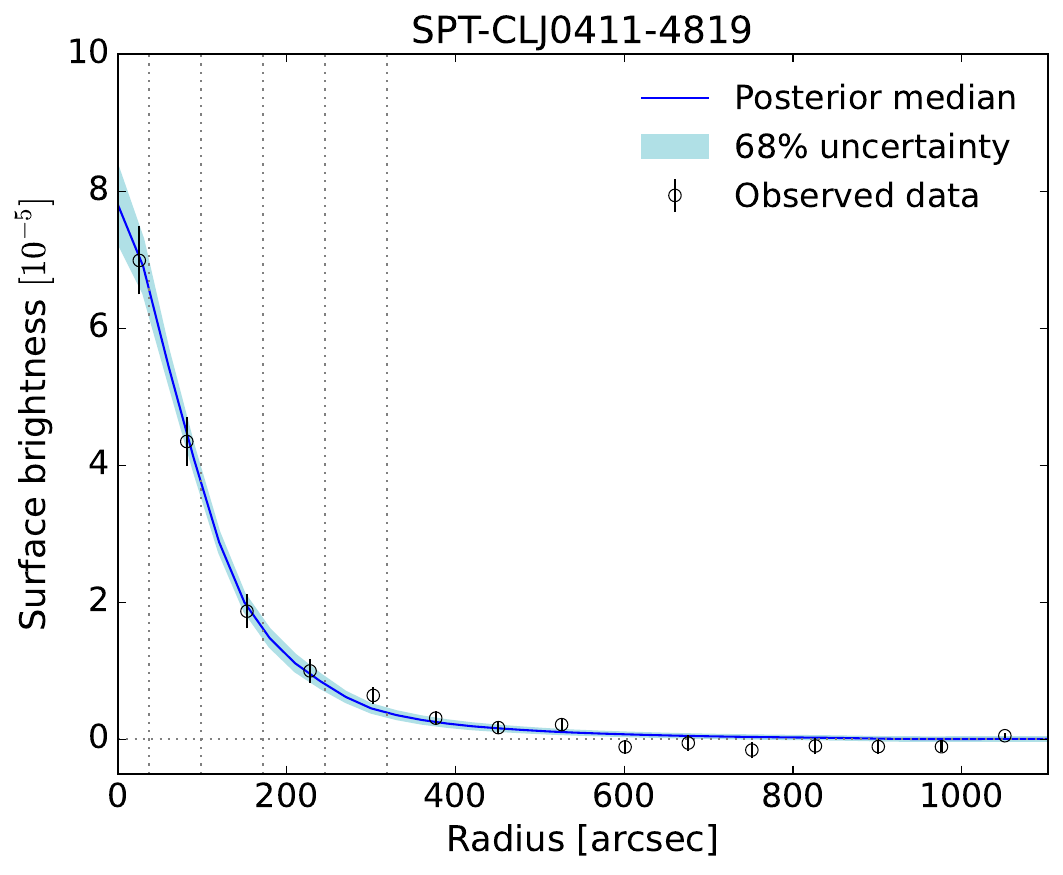}} 
\centerline{\includegraphics[width=.25\textwidth]{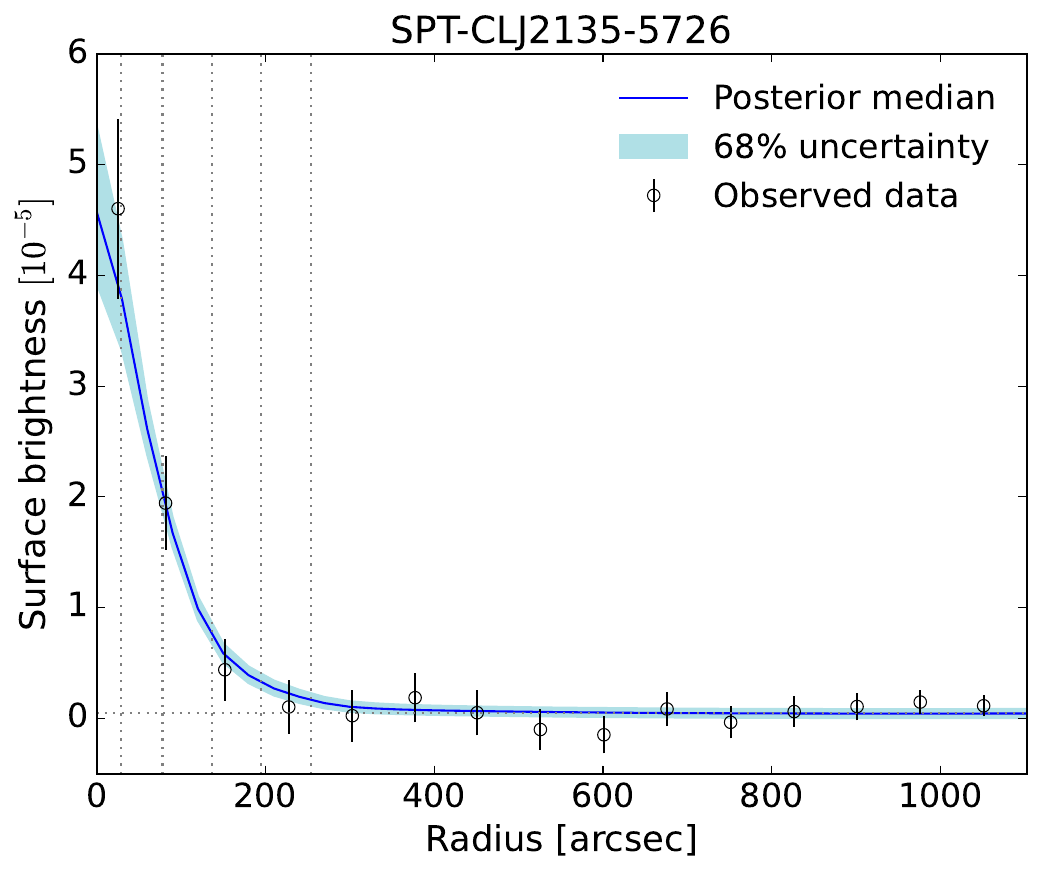} 
\includegraphics[width=.25\textwidth]{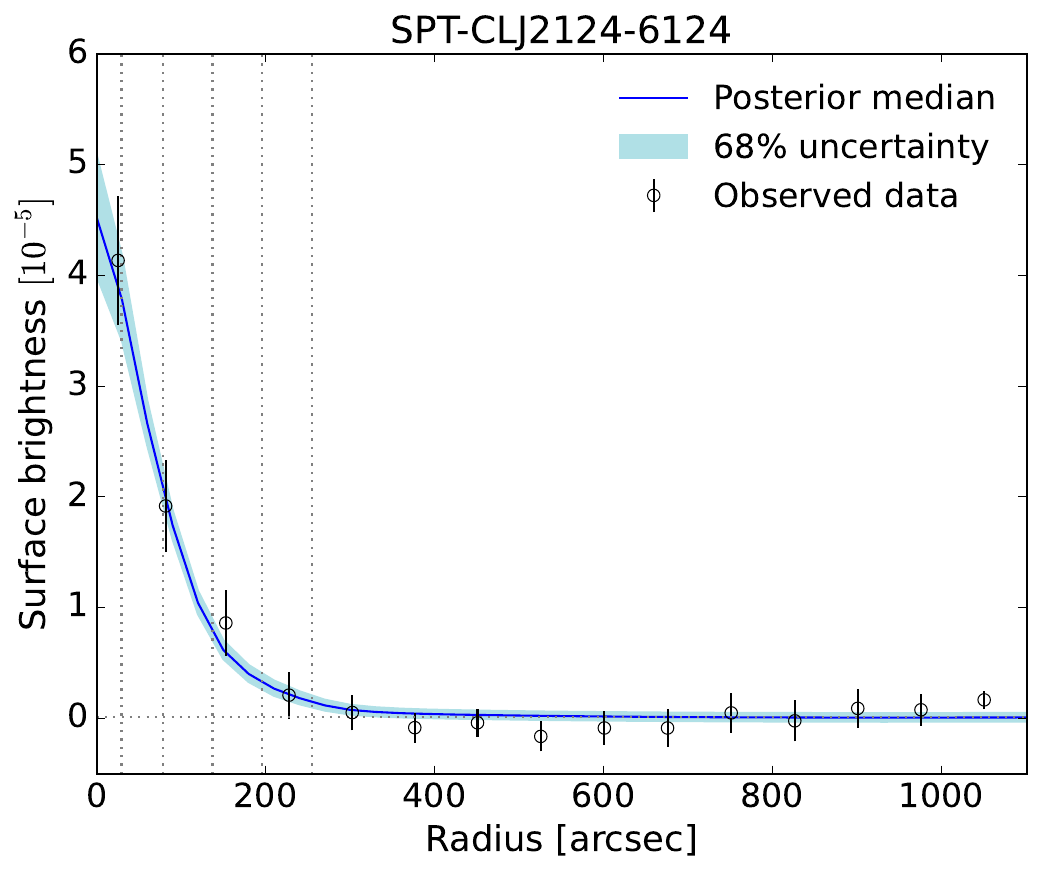} 
\includegraphics[width=.25\textwidth]{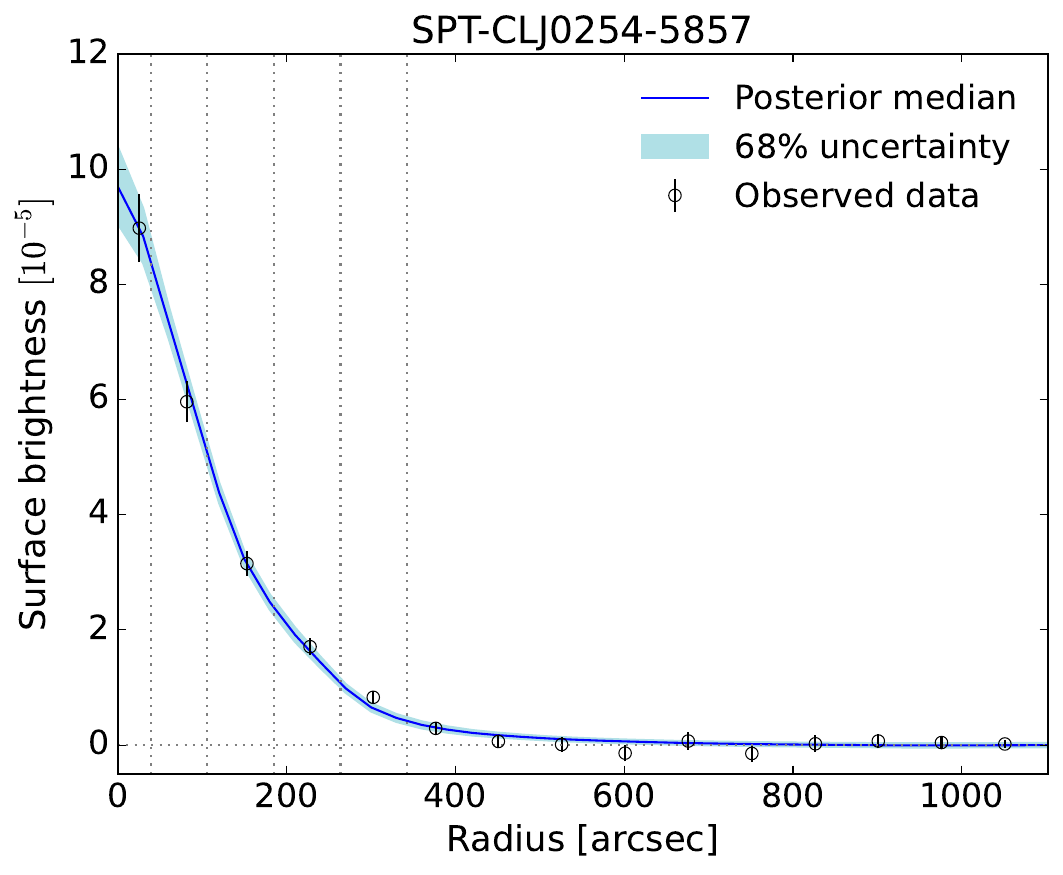} 
\includegraphics[width=.25\textwidth]{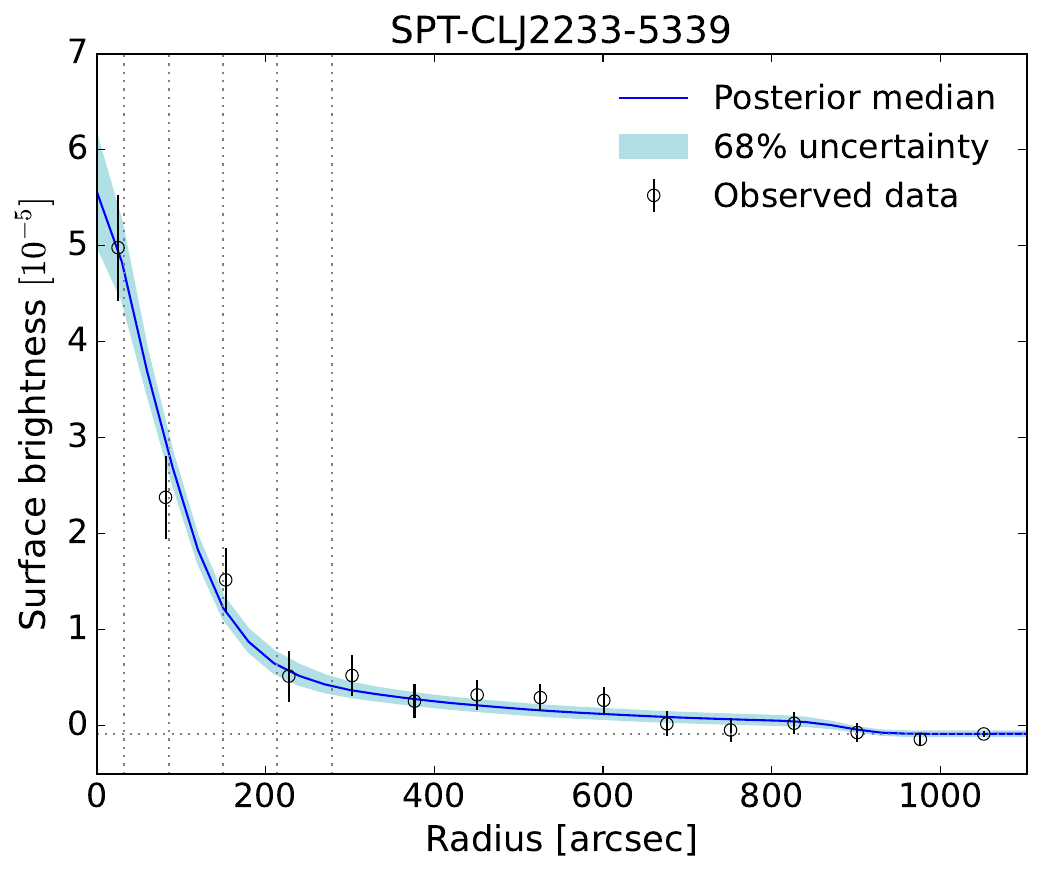}}
\centerline{\includegraphics[width=.25\textwidth]{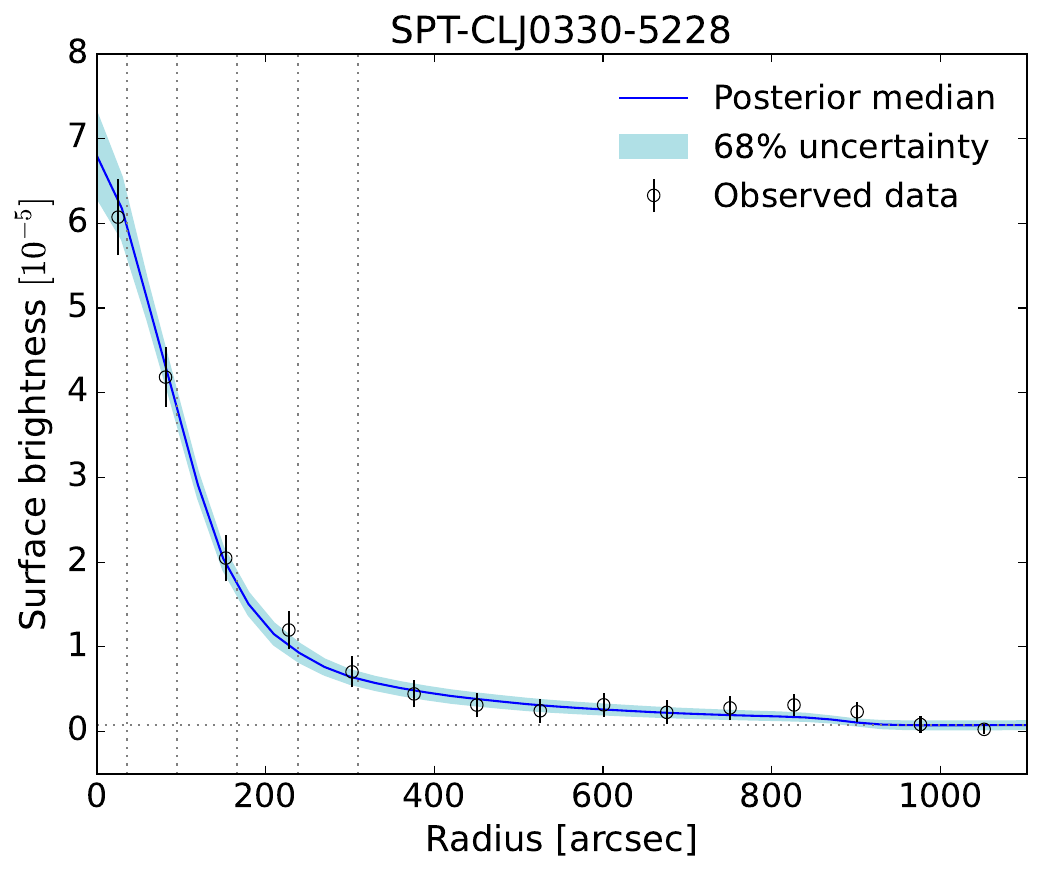} 
\includegraphics[width=.25\textwidth]{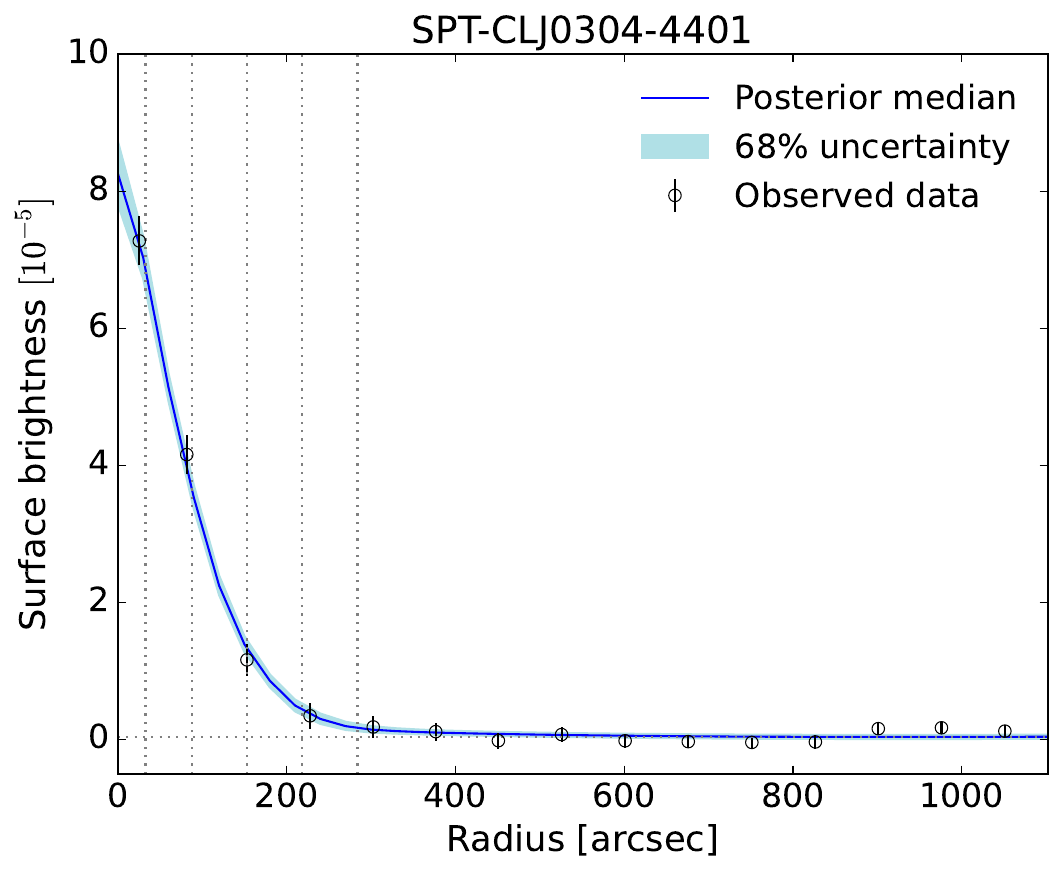} 
\includegraphics[width=.25\textwidth]{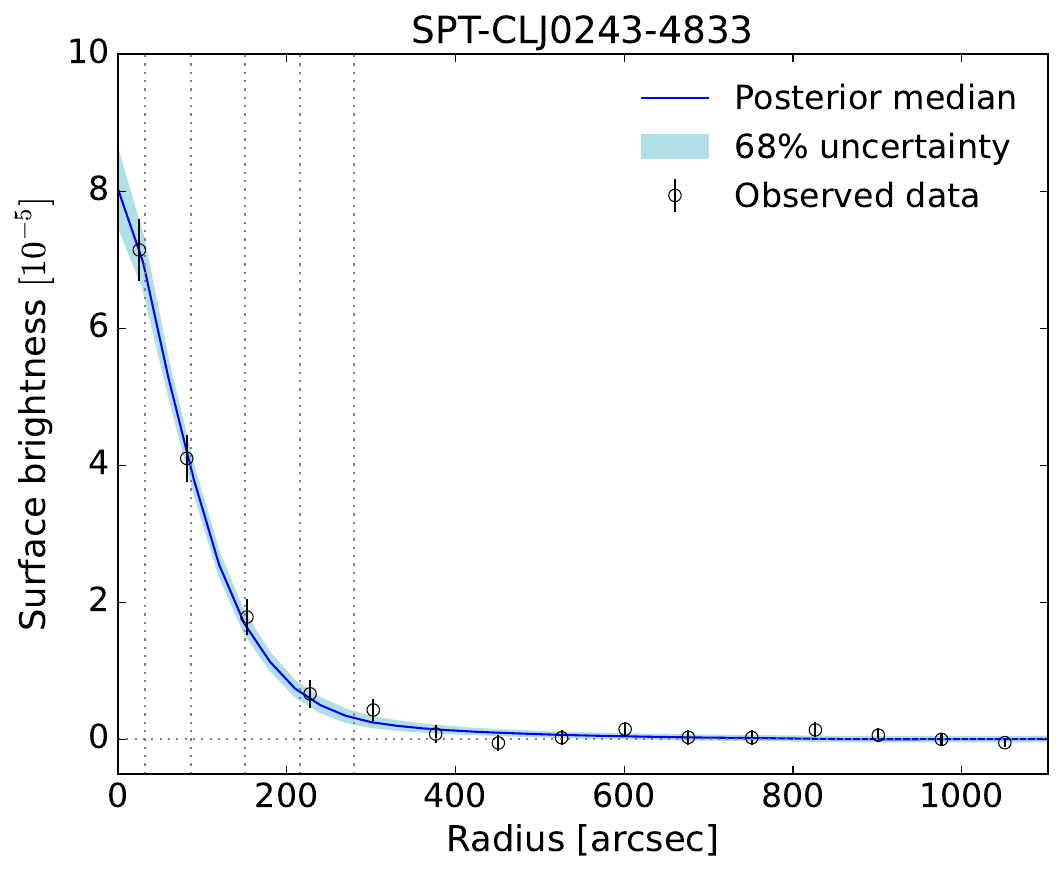} 
\includegraphics[width=.25\textwidth]{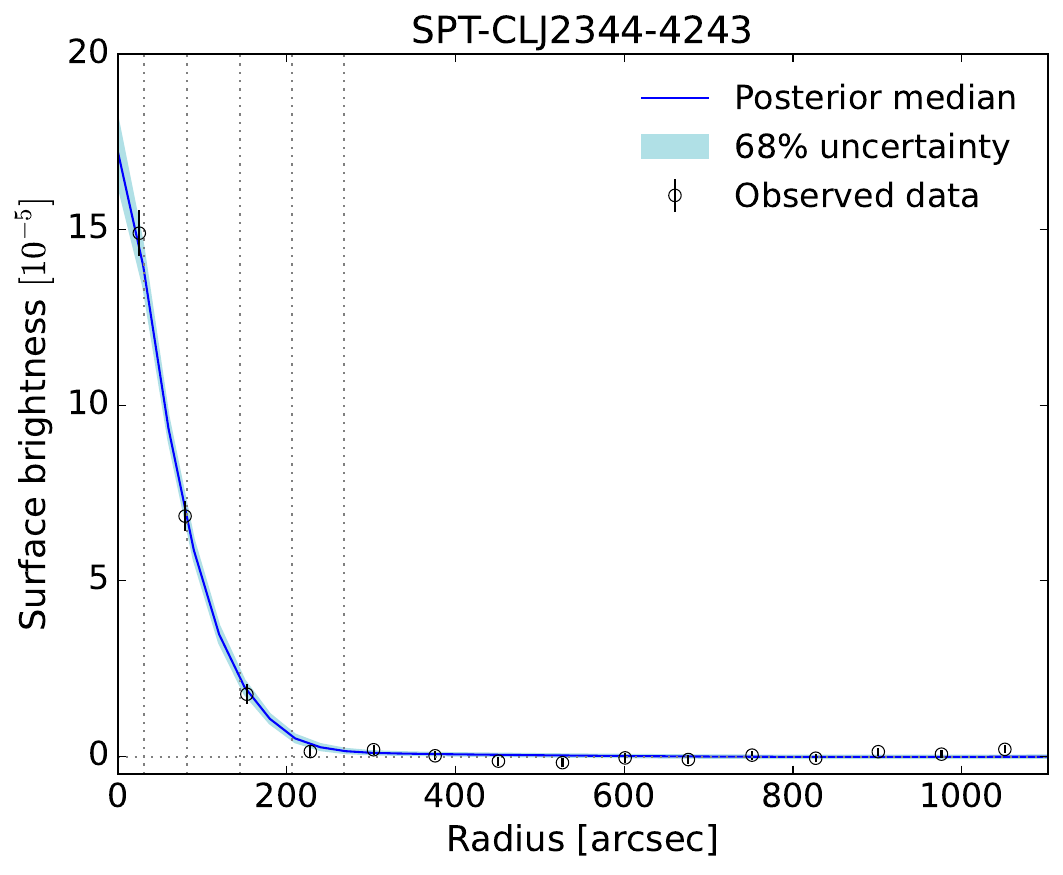}} 
\caption{As previous figure, but for the next 20 clusters }
\label{fig:Compton_prof3}
\end{figure} 

\clearpage
\section{Joint and marginal posterior distributions}
Fig.~\ref{fig:cornerplot} shows joint and marginal posterior distributions of the population parameters.

\begin{figure}[h]
\centering  
\begin{center}
\includegraphics[width=.98\textwidth]{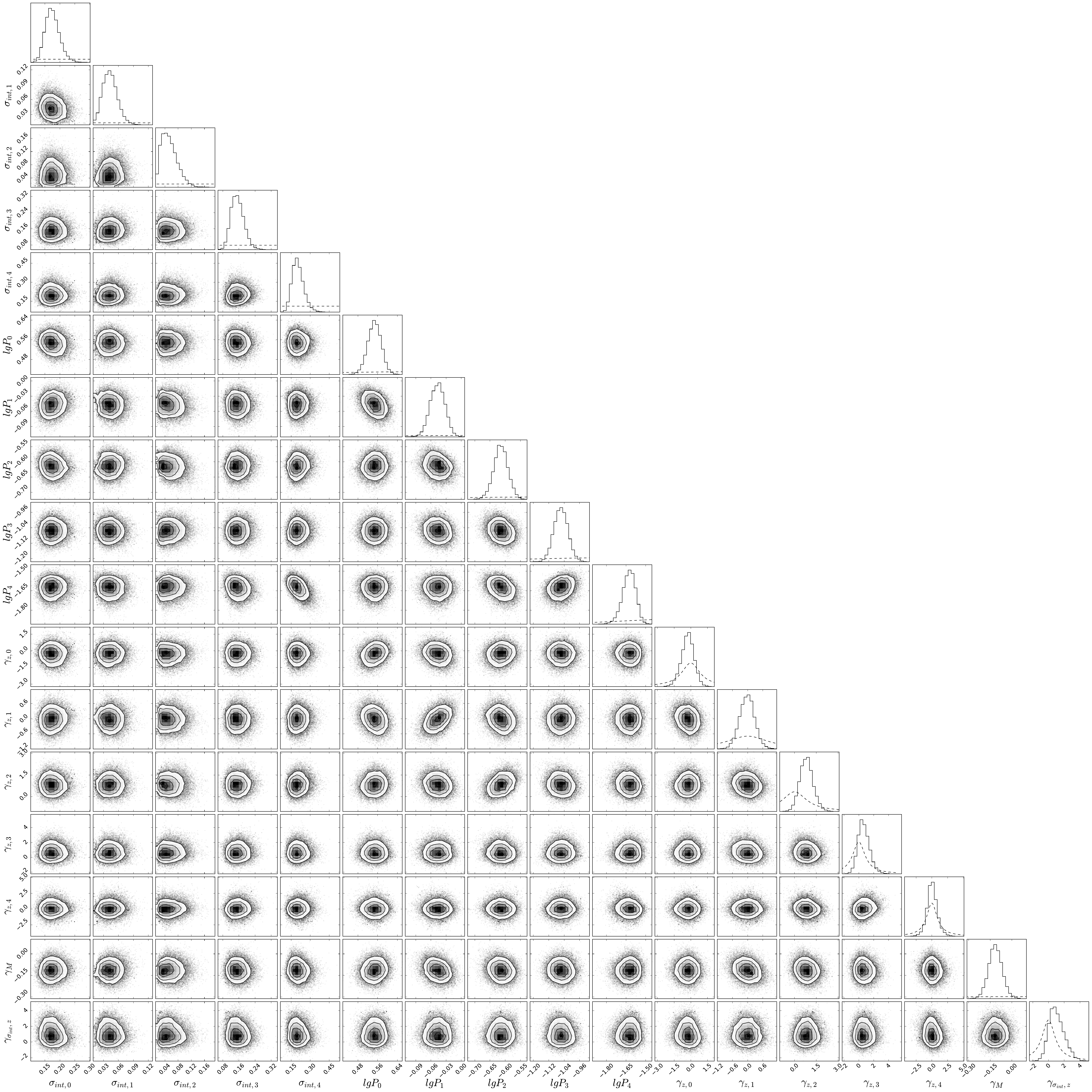} 
\caption{Joint and marginal posterior distributions for the intrinsic scatters $\sigma_{int,k}$, the population mean scaled pressure parameters $lgP_k$ at the five radii (interpolation knots), the five redshift-dependent $\gamma_{z,k}$
terms, the mass-dependent $\gamma_{M}$ term and the redshift-dependent term of the intrinsic scatter $\gamma_{\sigma_{intr,z}}$. Almost all parameters show little correlation, due to our choice of the radial profile modelization and the knots spacing. Prior distributions (dashed lines in the diagonal panels) have a negligible impact on the posterior distributions, except for $\gamma_{z}$ at large radii, where data are inconclusive.}
\label{fig:cornerplot}
\end{center}
\end{figure}

\end{appendix}
\end{document}